\documentclass[11pt,a4paper,english,twoside]{article}

\usepackage{a4wide}
\usepackage{amssymb, amsmath, amsthm}
\usepackage{graphicx}
\usepackage{subcaption}
\usepackage[all]{xy}
\usepackage{enumerate}
\usepackage[pdftex,hyperref,svgnames]{xcolor}
\usepackage[pdftex,colorlinks=true,
pdfstartview=FitV,
pdfnewwindow=true,
linktoc = page,
linkcolor= blue,
citecolor= red,
urlcolor= blue,
hyperindex=true,
hyperfigures=false]{hyperref}
\hypersetup{linktocpage}
\usepackage{dsfont}
\usepackage{empheq}
\usepackage{cite}
\usepackage{float}
\usepackage{cancel}
\usepackage{relsize}
\usepackage{soul}
\usepackage{enumitem}
\usepackage{hhline}
\usepackage{multirow}

\usepackage{pdfpages}

\usepackage{setspace}

\usepackage{ upgreek }

\newcommand{\beq}{\begin{equation}}
\newcommand{\eeq}{\end{equation}}
\def\bea#1\eea{\begin{align}#1\end{align}}
\def\beal#1\eeal{\begin{subequations}\begin{align}#1\end{align}\end{subequations}}
\newcommand{\nn}{\nonumber}
\newcommand{\w}{\wedge}
\newcommand{\R}{\mathcal{R}}
\renewcommand{\i}{\ensuremath{\textnormal{i}}}

\newcommand{\ssum}{{\, \textstyle\sum\, }}
\newcommand{\sprod}{{\, \textstyle\prod\, }}

\def\del {\partial}

\def\d {{\rm d}}

\def\ln {\text{ln}}

\begin{document}
\numberwithin{equation}{section}

\begin{titlepage}
\begin{center}

\phantom{DRAFT}

\vspace{1.3cm}

{\LARGE \bf{Extensions of a scale-separated AdS$_4$ solution}\\ \vspace{0.1in}\bf{and their mass spectrum}}\\

\vspace{2.2 cm} {\Large David Andriot and George Tringas}\\
\vspace{0.9 cm} {\small\slshape Laboratoire d'Annecy-le-Vieux de Physique Th\'eorique (LAPTh),\\
CNRS, Universit\'e Savoie Mont Blanc (USMB), UMR 5108,\\
9 Chemin de Bellevue, 74940 Annecy, France}\\
\vspace{0.5cm} {\upshape\ttfamily andriot@lapth.cnrs.fr; tringas@lapth.cnrs.fr}\\

\vspace{2.8cm}

{\bf Abstract}
\vspace{0.1cm}
\end{center}
%\begin{quotation}
We consider two extensions of the so-called DGKT solution, a 4d scale-separated anti-de Sitter (AdS) solution obtained as a compactification on a 6d torus orbifold. Each extension consists in a specific large $n$ expansion beyond the DGKT solution, where $n$ is the unbounded $F_4$-flux parameter. One of the extensions considered generalizes the known warped, partially backreacted solution. We analyse the two extensions in 10d massive type IIA supergravity as well as in a 4d effective theory, using a general warped compactification formalism, including axions. On top of known corrections to DGKT, we mainly get new ones from $F_4$; other fluxes are very constrained by flux quantization. In each extension, one would expect corresponding corrections to the mass spectrum, before reaching contributions from $\alpha'$-corrections. But the mass spectrum turns out to be robust, and conformal dimensions remain unchanged. 
%\end{quotation}

\end{titlepage}

\tableofcontents

\newpage

\section{Introduction and results summary}

String theory or supergravity solutions on a maximally symmetric spacetime, with cosmological constant $\Lambda$, times compact extra dimensions, may or may not exhibit scale separation. This property, important for phenomenology, has recently received a revived interest. A scale-separated solution admits a large hierarchy between $\Lambda$ and the typical mass scale $m_{{\rm KK}}$ of Kaluza--Klein states coming from the extra dimensions: $m_{{\rm KK}}^2 / |\Lambda| \gg 1$. This hierarchy allows to have an effective description of physics at low energy in the maximally symmetric spacetime only, as in our universe. Scale-separated solutions are however rare, especially when requiring in addition a parametric control (i.e.~an arbitrarily good control on the scale separation in terms of a parameter): until recently, the only such example was the so-called DGKT solution \cite{DeWolfe:2005uu} (see also \cite{Lust:2004ig, Camara:2005dc, Acharya:2006ne}). This is a solution of 10d massive type IIA supergravity on a $4$-dimensional (4d) anti-de Sitter spacetime, times a torus orbifold $T^6/\mathbb{Z}_3^2$, together with a (smeared) orientifold $O_6$-plane. In this solution, the parameter of interest, denoted $n$, is related to an $F_4$ flux integer, and it is unbounded as $F_4$ does not enter a tadpole condition. Interestingly, this parameter $n$ that controls scale separation also governs the classicality or supergravity approximation of the solution, namely, the string coupling becomes small and the toroidal volume becomes large in the limit of large $n$. In this limit, one gets both a (very) scale-separated solution and a classical string theory background. Last but not least, this anti-de Sitter solution is perturbatively stable, and can be a vacuum for the 4d scalar fields. Scale separation, classicality and perturbative stability are rare properties; see e.g.~\cite{Andriot:2022yyj} for the situation of Minkowski and de Sitter supergravity solutions. Scale separation in particular is so uncommon that it has been conjectured not to exist, within the swampland program \cite{Gautason:2018gln, Lust:2019zwm}. This has led to many discussions \cite{Blumenhagen:2019vgj, Apruzzi:2019ecr, Font:2019uva, Emelin:2020buq, Buratti:2020kda, Emelin:2021gzx, DeLuca:2021mcj, DeLuca:2021ojx, Cribiori:2022trc, Lust:2022lfc, Andriot:2022yyj, Andriot:2022brg, DeLuca:2023kjj} (see also \cite{Caviezel:2008ik, Tsimpis:2012tu, McOrist:2012yc, Petrini:2013ika, Gautason:2015tig} for early works). Constructions of similar examples in 4d \cite{Marchesano:2019hfb, Ishiguro:2021csu, Cribiori:2021djm, Bardzell:2022jfh, Becker:2022hse, Carrasco:2023hta, Tringas:2023vzn} and 3d \cite{Farakos:2020phe, VanHemelryck:2022ynr, Farakos:2023nms} have then been performed. These examples are usually related to the original DGKT, e.g.~by T-duality or by considering slight generalizations such as anisotropy. One observation has been that the conformal dimensions in a would-be holographically dual CFT take integer values \cite{Conlon:2020wmc, Conlon:2021cjk, Apers:2022zjx, Apers:2022tfm, Ning:2022zqx, Apers:2022vfp}. This is also rare enough to be noticed, and it raised the question whether the peculiarity of scale separation could be related to this CFT property. It seems not to be the case, as some generalisations of DGKT solutions were found not to provide integer conformal dimensions, in 4d \cite{Quirant:2022fpn} and in 3d \cite{Apers:2022zjx}, but it remains an important property of the original DGKT. We will come back to it in this paper, since we will be interested in its mass spectrum.

One criticism against the DGKT solution is that the $O_6$ source is smeared. On the covering torus $T^6$, the corresponding 9 intersecting sources are not localized or backreacted; only the integrals of their contributions appear in the 10d equations. This is a general issue of supergravity solutions, whenever facing intersecting sources; it goes beyond this 4d anti-de Sitter setting. In contrast, a localized solution on Minkowski times a torus for parallel sources was given in \cite{Andriot:2019hay}. For the DGKT solution, an approximate localized solution has however been recently obtained in \cite{Junghans:2020acz, Marchesano:2020qvg}. To reach this result, the idea there has been to use the parameter $n$ to construct an extension beyond the original DGKT solution, that would capture (part of) the source backreaction. It takes the form of a perturbative expansion in large $n$, where $0^{{\rm th}}$ order fields correspond to the DGKT (smeared) solution, and $1^{{\rm st}}$ order ones are a correction to the former that provides the backreaction; the $1^{{\rm st}}$ order coming as a $1/n$ power. In this paper, we are going to consider two extensions of DGKT in the form of such a $n$-expansion. We will perform the expansion both at a 10d level, to get a corrected solution to the 10d equations, but also at the 4d level, to get corrections to the 4d theory, critical point, and mass spectrum.

Of particular interest to us are the corrections to the mass spectrum that one can obtain via such extensions of DGKT. As emphasized, the DGKT solution possesses rare properties, and whether we can construct more general solutions (here extensions) that share some of these properties is a first question that motivates our work. We mentioned in particular the property of the integer conformal dimensions. A natural guess would be that corrections to the DGKT solution would modify the mass spectrum, thus altering this latter property (see e.g.~\cite{Plauschinn:2022ztd} along these lines, in a dual framework). After considering two extensions of DGKT, including one generalisation of the backreacted solution of \cite{Junghans:2020acz}, we will however conclude that the mass spectrum is very robust, and the integer property is maintained. Another motivation for considering the mass spectrum is the recent observation of a mass bound \cite{Andriot:2022brg}: almost all supersymmetric and many non-supersymmetric anti-de Sitter solutions (in 4d or higher), with radius $l$, admit a scalar whose mass $m$ obeys the bound $m^2 l^2 \leq -2$. In DGKT, depending how some flux signs are chosen (related to supersymmetry), either this claim is verified, saturating this $-2$ bound with some axions, or all scalar masses verify $m^2 > 0$; we detail the mass spectrum in Section \ref{sec:4dkinmass}. The latter case is one of the few known counter examples to the previous observation. Therefore, we would like to know whether extensions of DGKT would alter the mass spectrum,  with respect to the $-2$ bound in the former case, and with respect to positive or negative $m^2$ in the latter case. We will conclude once again on the robustness of the spectrum within the two extensions considered, preventing us to observe any evolution with respect to this mass bound.\\

In Section \ref{sec:DGKT}, we start by rederiving the 10d smeared DGKT solution, and then reproduce it as the critical point of a scalar potential within a corresponding 4d theory. We then compute the mass spectrum. The completeness of the derivation and consistency of the conventions is useful for later sections. To consider extensions beyond this DGKT solution, including a backreacted version, one needs a more general compactification setting with warp factor, dilaton and localized source contributions. We detail such a general warped setting, with 10d conventions and equations in Appendix \ref{ap:conv}. Building on the localized results of \cite{Junghans:2020acz}, we verify explicitly in Appendix \ref{ap:smear} how a smearing procedure can reproduce the smeared source contributions of the 10d solution of Section \ref{sec:DGKT}. In Appendix \ref{ap:4dwarp}, we derive a general 4d theory corresponding to such a warped compactification, namely its kinetic terms and scalar potential. This derivation does not make use of ${\cal N}=1$ supergravity formalism, but is performed as a direct dimensional reduction, building on \cite{Andriot:2022bnb}; it would be interesting to compare this derived warped 4d effective theory to the results of \cite{Giddings:2005ff, Shiu:2008ry, Douglas:2008jx, Frey:2008xw, Martucci:2009sf, Martucci:2014ska, Grimm:2014efa, Grimm:2015mua}. We believe that this theory could find more applications beyond this paper. Having available both a 10d and 4d description of a warped compactification, we can consider $n$-expansions of the fields, including of the warp factor.

\begin{figure}[H]
\centering
\vspace{-0.7in}
\includegraphics[width=0.7\textwidth]{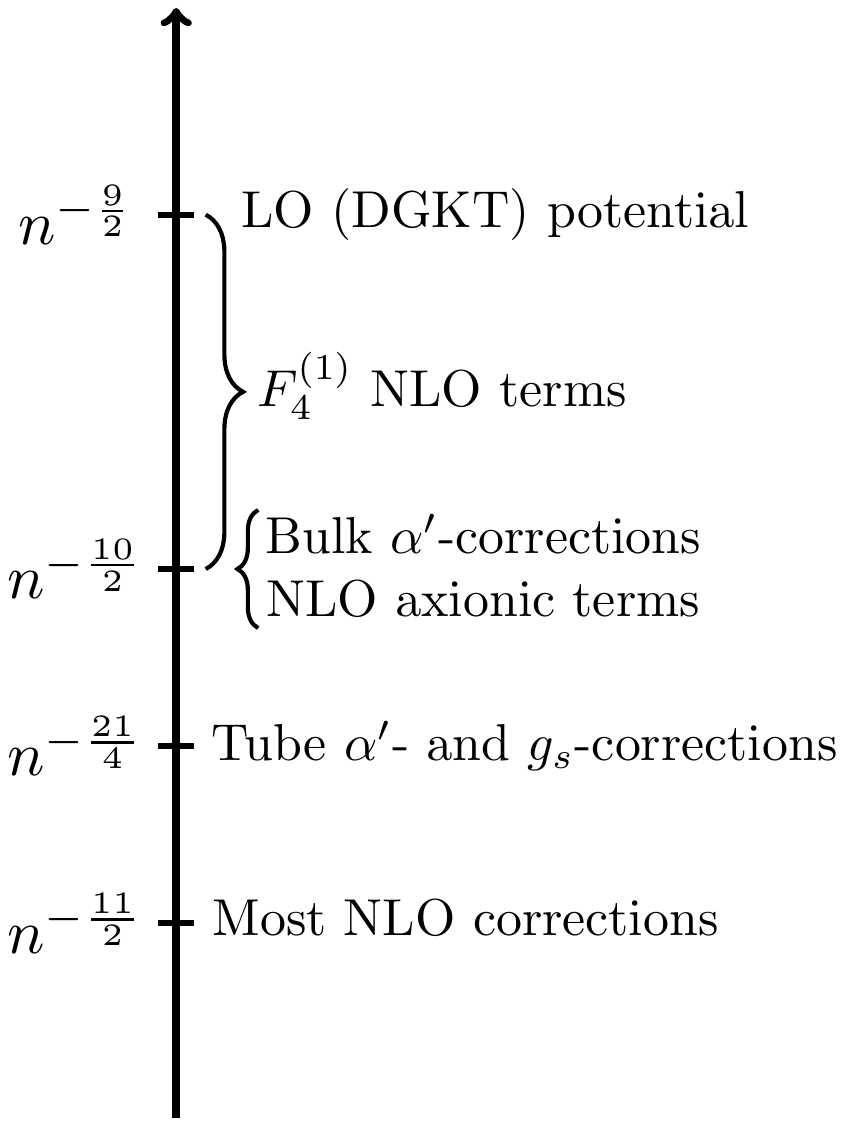}
\vspace{-0.6in}
\caption{Scaling with the parameter $n$ of the various terms in the scalar potential of the 4d theory. The leading order (LO) terms scale at $n^{-9/2}$ and correspond to the 4d DGKT scalar potential. The corrections in powers of $1/n$, due to the perturbative $n$-expansions considered in this paper, are broadly referred to as next-to-leading order (NLO) terms. They appear at various subdominant scaling levels. They compete with $\alpha'$- and $g_s$-corrections, whose scaling given here follows \cite{Junghans:2020acz}. We refer to the main text for more details.}\label{fig:scalingintro}
\end{figure}

In Section \ref{sec:warpDGKT}, we generalise the $n$-expansion of \cite{Junghans:2020acz}, for which an approximate backreacted solution was obtained. We are especially interested in the fluxes $H,F_4,F_6$ at $1^{{\rm st}}$ order, for which we allow for general scalings with $n$. In Section \ref{sec:newextension}, we consider another $n$-expansion, where on top of the freedom for $H,F_4,F_6$, the setting differs in the $1^{{\rm st}}$ order scaling of the metric, the dilaton, and a certain compactification ansatz. The motivation for giving more freedom in $H,F_4,F_6$ is the following. The 4d theory at leading order (LO) in the $n$-expansion, corresponding to the DGKT 4d theory, scales with $n$ as $n^{-9/2}$. When looking for next-to-leading order (NLO) corrections, one finds most contributions at $n^{-11/2}$, and a few corrected axionic terms (not considered in \cite{Junghans:2020acz}) at $n^{-10/2}$ or lower. It is however argued in \cite{Junghans:2020acz} that bulk $\alpha'$-corrections could appear at the level $n^{-10/2}$, while tube $\alpha'$- and $g_s$-corrections (referring to the corrections in the region close to the sources) could appear almost at the same level, namely at $n^{-21/4}$. This implies that computing corrections to the mass spectrum due to the backreaction or localization is meaningless, since those would mix with $\alpha'$-corrections. In turn, it is hard to conclude on the impact of $\alpha'$-corrections on the mass spectrum, as one could be tempted from \cite{Plauschinn:2022ztd}, since those may mix with backreaction contributions. Our strategy is then to consider more general extensions including $H,F_4,F_6$ $1^{{\rm st}}$ order contributions. To our surprise, those turn out to be very constrained, by flux quantization (setting $H^{(1)}=0$) and then equations of motion (setting $F_6^{(1)}=0$). Nevertheless, the $1^{{\rm st}}$ order correction $F_4^{(1)}$ can still contribute and by adjusting its scaling, it would generate a scalar potential scaling as $n^{9/2-g}$ with $0< g < 1/2$, therefore higher than the problematic level $n^{-10/2}$. We summarize this situation in Figure \ref{fig:scalingintro}.

In our two extensions, things however conspire such that the NLO potential due to $F_4^{(1)}$ does not really alter the mass spectrum! For the first $n$-expansion considered in Section \ref{sec:warpDGKT}, $F_4^{(1)}$ is forced to vanish at the critical point, as no field can balance this correction. We show this both with 10d equations and the 4d theory. The spectrum therefore remains the same until its hits $\alpha'$-corrections. In Section \ref{sec:alpha}, we still discuss possible refinements of this claim, by re-analysing possible bulk $\alpha'$-corrections and arguing that they could actually appear at a much lower level; this situation deserves more investigation. For the second $n$-expansion considered in Section \ref{sec:newextension}, $F_4^{(1)}$ provides an actual corrected solution at $1^{{\rm st}}$ order, thanks to the metric and dilaton corrections appearing at the same scaling. This solution is obtained both by solving 10d equations and 4d NLO ones. As is however realised in 4d, this correction can eventually be interpreted (and repackaged) as a redefinition of the parameter $n \rightarrow n' = n + e^{(1)}\, n^{1-g} $, with a constant $e^{(1)}$. This parameter can still obey the $F_4$ quantization condition, so this $n^{1-g}$ correction is admissible. Viewed as a redefinition of the parameter, we understand why, as we eventually find out, the mass spectrum is only altered via a correction of the cosmological constant. As a consequence, $m^2 l^2$ or the conformal dimensions are left unchanged by this correction. This shows again the robustness of the spectrum against extensions.

Without a better understanding of the $\alpha'$- and $g_s$-corrections, or considering different extensions of DGKT, we conclude that this solution seems to remain robust and very specific, in its properties and mass spectrum. One aspect we left aside are the blow-up modes, that could be worth investigating. Indeed, some of the peculiarities of DGKT are due to its orbifold, but the latter comes at the price of having blow-up modes. Surprisingly still, those are K\"ahler moduli and admit $m^2 >0$. It would be interesting to see if any of the extensions or corrections mentioned above could affect this spectrum. To that end, one would first need to establish on firm grounds the scalar potential for those modes, with a clear 10d origin, and then see how it gets affected by corrections in $n$, e.g.~to the metric; this goes beyond the scope of this work.

\section{DGKT}\label{sec:DGKT}

In this section, we review the DGKT anti-de Sitter solution on a torus orbifold \cite{DeWolfe:2005uu}, first in 10d type IIA massive supergravity and then as the critical point of a scalar potential in a 4d theory. We finally derive the 4d mass spectrum. This provides us with a starting point for the extensions considered in the next sections.

\subsection{10d DGKT solution}

We describe here the DGKT solution in 10d type IIA massive supergravity. Conventions are detailed in Appendix \ref{ap:conv}. The solution requires a certain amount of notations that we try to keep minimal.

\subsubsection{Compactification ansatz}\label{sec:ansatz}

The solution of 10d type IIA massive supergravity presented in \cite{DeWolfe:2005uu} has a 10d spacetime split as 4d anti-de Sitter times a 6d compact space being a torus orbifold, together with a space-filling orientifold $O_6$-plane. Let us start with the torus $T^6=T^2 \times T^2 \times T^2$ and introduce for $i=1,2,3$ the coordinates $z^i = y^{2i-1} + \i\, y^{2i}$, i.e. $z^1 = y^1 + \i\, y^2$ etc. Each $T^2$ is made compact with the following discrete identifications
\beq
z^i \sim z^i + 1 \sim z^i + e^{\i \frac{\pi}{3}} \ .
\eeq
From there, we consider the orbifold $T^6/\mathbb{Z}_3^2$, and we refer to \cite{DeWolfe:2005uu} for the two $\mathbb{Z}_3$ actions and details on the resulting space. Finally, one has the orientifold reflection: $z^i \rightarrow - \overline{z}^i$. On the covering torus $T^6$, the $O_6$-plane can be viewed as 9 localized sources, as detailed e.g.~in \cite{Junghans:2020acz}.

The orbifold and orientifold project out many field components of type IIA supergravity, and the solution ansatz has to respect these projections. To start with, the 6d metric is given by
\beq
\d s^2 = g_{mn} \d y^m \d y^n = 2 (\kappa \sqrt{3} )^{1/3} \sum_{i=1}^{3} v_i\,  ((\d y^{2i-1})^2 + (\d y^{2i})^2 ) \ . \label{6dmetric}
\eeq
The overall normalisation is for later convenience; we take $\kappa$ to carry the physical dimension of a $\text{{\it length}}^6$, while $y^i$ and $v_i$ are dimensionless. The freedom in the metric ansatz is carried by the $v_i$, which will be fixed in the 10d solution, and appear as scalar fluctuations in 4d.

On this 6d space, only few forms can be defined, as they need to be invariant under the two $\mathbb{Z}_3$; they will also appear in the flux ansatz. The even and odd forms are respectively given by
\bea
& w^i = (\kappa \sqrt{3} )^{1/3}\, \i\, \d z^i \w \d \overline{z}^i \ , \quad \tilde{w}^i  = \frac{1}{\kappa}\, w^j \w w^k \ \ \text{with}\ \{ i,j,k\} =\{1,2,3\} \ ,\\
& {\rm vol}_6 = v_1v_2v_3\, w^1 \w w^2 \w w^3 = \sqrt{|g_6|}\ \d y^1 \w ... \w \d y^6 \ ,\\
& \Omega = 3^{\frac{1}{4}} \, \i \, \d z^1 \w \d z^2 \w \d z^3 = \frac{1}{\sqrt{2}} (\alpha_0 + \i \, \beta_0) \ ,
\eea
where the normalisations are chosen for future convenience. It will be useful to have the explicit expressions for some of these forms, namely
\bea
& \alpha_0 =\sqrt{2}\,3^{\frac{1}{4}}\left(\text{d}y^2\wedge\text{d}y^4\wedge\text{d}y^6 -\text{d}y^2\wedge\text{d}y^3\wedge\text{d}y^5-\text{d}y^1\wedge\text{d}y^3\wedge\text{d}y^6
    -\text{d}y^1\wedge\text{d}y^4\wedge\text{d}y^5 \right)\\
& \beta_0 =\sqrt{2}\,3^{\frac{1}{4}}\left(\text{d}y^1\wedge\text{d}y^3\wedge\text{d}y^5
    -\text{d}y^1\wedge\text{d}y^4\wedge\text{d}y^6
    -\text{d}y^2\wedge\text{d}y^3\wedge\text{d}y^6
    -\text{d}y^2\wedge\text{d}y^4\wedge\text{d}y^5\right) \label{beta0}\\
& \tilde{w}^1= 4 (3\kappa^{-1})^{\frac{1}{3}}\,  \d y^3 \w \d y^4 \w \d y^5 \w \d y^6 \ ,\ \tilde{w}^2= 4 (3\kappa^{-1})^{\frac{1}{3}}\,  \d y^1 \w \d y^2 \w \d y^5 \w \d y^6 \ ,\\
& \tilde{w}^3= 4 (3\kappa^{-1})^{\frac{1}{3}}\,   \d y^1 \w \d y^2 \w \d y^3 \w \d y^4 \ .\nn
\eea
The real forms $\alpha_0$ and $\beta_0$ are even, respectively odd, under the orientifold reflection.

Let us finally give few integrals of these forms. Given the coordinate identifications, each $T^2$ corresponds to a parallelogram, whose area is the same as the rectangle defined by $y^{2i-1} \sim y^{2i-1} +1$ and $\, y^{2i} \sim y^{2i} + \frac{\sqrt{3}}{2}$. One therefore gets the following integral
\beq
\int_{T^2} \i \, \d z^i \w \d \overline{z}^i = \int_{T^2} 2 \d y^{2i-1} \w \d y^{2i} = 2 \, \frac{\sqrt{3}}{2} = \sqrt{3} \ . \label{inty}
\eeq
Modding out the $T^6$ by the two $\mathbb{Z}_3$ divides the 6d volume by 9. This leads to the following 6d integrals
\bea
& \int_{T^6/\mathbb{Z}_3^2} w^1 \w w^2 \w w^3 = \kappa \ ,\quad \int_{T^6/\mathbb{Z}_3^2} w^i \w \tilde{w}^j = \delta^{ij} \ ,\\
& \int_{T^6/\mathbb{Z}_3^2} {\rm vol}_6 = \kappa\, v_1v_2v_3 \equiv {\rm vol} \ , \label{vol}\\
& \int_{T^6/\mathbb{Z}_3^2} \i \Omega \w \overline{\Omega} = \int_{T^6/\mathbb{Z}_3^2} \alpha_0 \w \beta_0 = 1 \ ,
\eea
where \eqref{vol} defines the scalar quantity ${\rm vol}$.

We can now present the ansatz for the 6d background fluxes, allowed by the projections. We start with the following ones
\beq
F_4= \sqrt{2}\, e_i \tilde{w}^i \,, \quad
H = -p \beta_0 \,, \quad
F_0 = -\sqrt{2}\, m_0 \,,
\eeq
where the normalisations and signs are due to our conventions, detailed in Appendix \ref{ap:conv}. The quantities $m_0, p, e_i$, referred to as flux numbers in the following, will be related to integers thanks to flux quantization; we can already specify their physical dimension
\beq
m_0 \sim \text{{\it length}}^{-1} \,,\quad
p \sim \text{{\it length}}^2 \,,\quad
e_i \sim \text{{\it length}}^5 \,. \label{dim}
\eeq
We take in addition for this solution
\beq
F_2=0 \,,
\eeq
as is standard for a smeared solution on a toroidal manifold. $F_6$ can be left free in the ansatz, but the equations of motion will set $F_6=0$.

Axions do not enter the 10d solution, but as fluctuations, they are still constrained by the geometry and the allowed forms. They are parameterized as follows
\beq
B_2 = \sum_{i=1}^{3} b_i\, w^i \,,\quad
C_3 = \sqrt{2}\, \xi\,  \alpha_0 \,,\label{axions}
\eeq
where $b_i, \xi$ will appear as 4d scalars, and one has $C_1=0$.

Finally, the dilaton $\phi$ will be fixed to a constant value in the 10d solution, and will be considered as a scalar fluctuation in the 4d theory. We could also give more detail on the $O_6$-plane whose contribution appears in the 10d equations; the fact it will taken smeared requires however a longer discussion that we leave to the next subsection. Having specified the 10d geometry and the compactification ansatz of the fields, we now show how the 10d equations are solved, fixing the free parameters and expressing the whole solution in terms of the flux numbers only.

\subsubsection{Solution}\label{sec:sol}

We show here how the previous ansatz leads to a solution of 10d equations. As we will see, the solution is said to be smeared: the first consequence is that the dilaton $\phi$ is constant and there is no warp factor, i.e.~the spacetime is a direct product of a 4d anti-de Sitter spacetime and the 6d manifold. With respect to the more general ansatz of Appendix \ref{ap:conv}, this sets $e^A=1$. This, together with the ansatz previously presented, simplify the 10d equations of Appendix \ref{ap:conv}.

We start with the flux equations of motion \eqref{fluxeq0} - \eqref{fluxeq}. All exterior derivatives on internal forms vanish. Provided that either $H$ or $F_4$ is non-vanishing, as desired here, we deduce the requirement
\beq
F_6=0 \ .
\eeq
All those equations are then satisfied. Among the others, \eqref{BIs} and \eqref{eq1} - \eqref{eq3}, the non-trivial equations boil down to
\bea
\text{$F_2$ Bianchi identity:}\quad & - H F_0 = \frac{ T_{10} }{7}\, {\rm vol}_{\bot} \label{BI}\\
\text{dilaton e.o.m.:}\quad & 2 {\cal R}^S_4 + e^{\phi} \frac{ T_{10} }{7} - |H|^2 = 0 \label{dilatoneom}\\
\text{4d Einstein equation:}\quad & 4 {\cal R}^S_4 = e^{\phi} \frac{ T_{10} }{7} - 2|H|^2 +  e^{2 \phi} ( |F_0|^2 - 3 |F_4|^2 ) \label{4dEinstein} \\
\text{6d Einstein:}\quad & 0  = e^{2\phi} \frac{1}{3!} F_{4\ mpqr}F_{4\ n}^{\ \ \ pqr} + \frac{1}{2} H_{mpq}H_n^{\ pq}  \label{6dEinstein}\\
& \ + e^{\phi} T_{mn} + \frac{g_{mn}}{8} \left( - e^{\phi} T_{10} - 2|H|^2 + e^{2 \phi} ( |F_0|^2 - 3 |F_4|^2 ) \right)  \ ,\nn
\eea
with $m= 1,..., 6$ and indices are lifted with the 6d metric $g_{mn}$. ${\cal R}^S_4$ denotes the Ricci scalar associated to the 4d metric $g_{\mu\nu}^S$ in \eqref{metric10d}, and we refer to Appendix \ref{ap:conv} for further notations. The source contributions $T_{mn}$ will be defined below.\\

Before solving these equations, we need to clarify the situation of the sources, namely the $O_6$-plane. The contributions $T_{10}^I$ and alike are defined in Appendix \ref{ap:conv} with localized and backreacted sources. The DGKT solution is however smeared, meaning that the $\delta$-functions appearing in those quantities are replaced by $1$; similarly, the warp factor that plays the role of the Green's function for these $\delta$-functions (see e.g.~\cite{Andriot:2019hay}) is set to a constant so that its derivatives vanish. Overall, smearing amounts to consider the integral of equations of motion, instead of their localized version. If one would know the localized solution, this smearing procedure could in principle be performed to get the corresponding constant value for $T_{10}^I$, and other smeared quantities. This is done to some extent in \cite{Junghans:2020acz} starting with the 9 localized sources on the torus covering. We revisit this smearing procedure in Appendix \ref{ap:smear} to match the smeared DGKT solution. In this subsection, we take a different path: we rather consider the torus orbifold, on which there is only one $O_6$. This corresponds to having only one source set $I=1$, so we drop the label $I$ in above equations. We do not know the localized solution for it, but we follow \cite{DeWolfe:2005uu} considering that there exists a smeared version, described by a constant quantity $T_{10}$. Instead of fixing $T_{10}$ by its localized definition together with a smearing procedure, it gets fixed here by solving the various equations.

A related matter is to determine the volume forms ${\rm vol}_{||}$ and ${\rm vol}_{\bot}$. The $O_6$ is said in \cite{DeWolfe:2005uu} to wrap the cycle along $\alpha_0$. We thus take ${\rm vol}_{||}$ to be proportional to $\alpha_0$, and ${\rm vol}_{\bot}$ proportional to $\beta_0$, obeying the standard constraint ${\rm vol}_{||} \w {\rm vol}_{\bot} = {\rm vol}_{6}$. It is not straightforward to understand what are here the corresponding 3 dimensions wrapped, since both $\alpha_0$ and $\beta_0$ are a sum of four 3-forms along different directions. In both cases however, the directions in each 3-form share the same metric volume $\sqrt{8\sqrt{3}\,\kappa\, v_1 v_2 v_3}$. The two volume forms should then be proportional to this factor, and what remains to be determined is an overall numerical factor. The constraint ${\rm vol}_{||} \w {\rm vol}_{\bot} = {\rm vol}_{6}$ leaves one possible relative factor between the two forms. The Bianchi identity involves $\frac{ T_{10} }{7}\, {\rm vol}_{\bot}$, so this normalisation ambiguity is also present in $\frac{ T_{10} }{7}$ (since the other side, $-H F_0$, is fixed). The quantity $\frac{ T_{10} }{7}$ also appears alone in further independent equations, so this ambiguity is eventually fixed. With this knowledge, we finally pick
\beq
{\rm vol}_{\bot} = \frac{1}{2} \sqrt{\kappa\, v_1v_2v_3}\ \beta_0 \ , \quad
{\rm vol}_{||} = 2 \sqrt{\kappa\, v_1v_2v_3}\ \alpha_0 \ . \label{volbp}
\eeq
The Bianchi identity is then solved by
\beq
 \frac{ T_{10} }{7} = -\frac{p m_0 2\sqrt{2}}{\sqrt{\kappa\, v_1v_2v_3}} \ . \label{tad}
\eeq
Having only an $O_6$-plane, we must have $T_{10} >0$, so $pm_0 <0$.

Finally, we will also use an alternative, or effective, description of the above smeared source. One may effectively consider each of the four 3-forms in $\alpha_0$ as a subspace wrapped by an $O_6$, carrying $1/4$ of the total charge. Doing so, we then consider four sets of sources, $I=1,...,4$, along these 4 subspaces. We introduce the corresponding volume forms ${\rm vol}_{||_I}, {\rm vol}_{\bot_I}$, with
\beq
{\rm vol}_{||} = \ssum_I {\rm vol}_{||_I} \ ,\quad {\rm vol}_{\bot} = \frac{1}{4} \ssum_I {\rm vol}_{\bot_I} \ ,\quad {\rm vol}_{||_I} \w {\rm vol}_{\bot_I} = {\rm vol}_{6} \ ,
\eeq
together with the (smeared) charge contributions for each set $T_{10}^1=T_{10}^2=T_{10}^3=T_{10}^4 = \frac{1}{4} T_{10}$, and $T_{10}= \ssum_I  T_{10}^I$. The Bianchi identity \eqref{BI} can be rewritten as
\beq
- H F_0 = \sum_{I=1}^4 \frac{ T_{10}^I }{7}\, {\rm vol}_{\bot_I} \ .
\eeq
This alternative, effective description is useful to solve the Einstein equation, where $T_{mn}$ defined as follows \cite{Andriot:2017jhf} appears
\beq
T_{mn} = \sum_I \delta^{m_{||_I}}_{m} \delta^{n_{||_I}}_{n} g_{m_{||_I}n_{||_I}} \frac{T_{10}^I}{7} \ . \label{Tijdef}
\eeq
Here, $m_{||_I}$ refers to the directions along the $O_6$ in set $I$; those are precisely easier to identify when considering the four sets, rather than in the full $\alpha_0$. We get that 2 sets $I$ contribute to each direction $m$, giving thus half a charge of $T_{10}$
\beq
T_{mn}= g_{mn} \frac{1}{2} \frac{T_{10}}{7} \ . \label{Tmntrick}
\eeq
A cross-check of this result is that $g^{mn} T_{mn} = \tfrac{3}{7} T_{10}$ as it should. Having explicitly determined the source contributions, we are ready to solve the remaining equations.\\

The solution, as given in \cite{DeWolfe:2005uu}, is essentially captured by the background values for the 4d fluctuations in terms of the flux numbers: those are
\beq
v_i = \frac{v}{|e_i|} = \frac{1}{|e_i|} \sqrt{\frac{5}{3} \left|\frac{e_1e_2e_3}{\kappa m_0} \right|} \,,\quad e^{\phi} = \frac{3}{4} |p| \left(\frac{5}{12} \frac{\kappa}{|m_0 e_1e_2e_3|} \right)^{1/4} \,,\quad
b_i = \xi = 0 \ .\label{bcgdfields}
\eeq
The above gives the useful relation
\beq
e^{\phi}  = \frac{5}{4\sqrt{2}} \frac{| p | }{\sqrt{\kappa\, v_1v_2v_3 } | m_0 | }   \ . \label{relphivol}
\eeq
Let us now verify that this is indeed a solution of the above equations. The dilaton and the Einstein equations require the flux squares: those are given in general by
\beq
|H|^2 = \frac{p^2}{\kappa\, v_1v_2v_3} \,,\quad
|F_4|^2 = \frac{\sum_{i=1}^{3} e_i^2 v_i^2}{(\kappa\, v_1v_2v_3)^2 } \ . \label{H2}
\eeq
Using the background values \eqref{bcgdfields} and \eqref{relphivol}, we obtain the convenient expressions
\beq
e^{\phi} \frac{ T_{10} }{7} = \frac{5}{2}  \frac{p^2}{\kappa\, v_1v_2v_3} \,,\quad
\ e^{2\phi} |F_4|^2 = \frac{27}{16}\frac{p^2}{\kappa\, v_1v_2v_3 } \,, \quad
e^{2\phi} |F_0|^2 = \frac{25}{16}\frac{p^2}{\kappa\, v_1v_2v_3 } \ . \label{Fq2}
\eeq
The dilaton e.o.m.~\eqref{dilatoneom} then fixes the 4d Ricci scalar as follows
\beq
{\cal R}^S_4 =  - \frac{3}{4}  \frac{p^2}{\kappa\, v_1v_2v_3} \ , \label{R4s}
\eeq
its sign allowing us to verify that this is an anti-de Sitter solution. The 4d Einstein equation \eqref{4dEinstein} offers the first non-trivial check of the above quantities, and is satisfied. The other non-trivial check is to satisfy the 6d (trace-reversed) Einstein equations \eqref{6dEinstein}. To verify this, we first compute each of the quantities appearing: those are proportional to $v_k$ for $m,n=2k-1,2k-1$ and $2k ,2k$
\bea
& \frac{e^{2\phi}}{3!} F_{4\ mpqr}F_{4\ n}^{\ \ \ pqr} = \frac{9}{4}\ \frac{(\kappa \sqrt{3})^{\tfrac{1}{3}}  p^2}{\kappa\, v_1v_2v_3} \, v_k \ ,\ \frac{1}{2} H_{mpq}H_n^{\ pq}  = \frac{(\kappa \sqrt{3})^{\tfrac{1}{3}}  p^2}{\kappa\, v_1v_2v_3}\, v_k \ ,\ e^{\phi} T_{mn} = \frac{5}{2}\ \frac{(\kappa \sqrt{3})^{\tfrac{1}{3}}  p^2}{\kappa\, v_1v_2v_3}\, v_k  \nn\\
& \frac{g_{mn}}{8} \left( - e^{\phi} T_{10} - 2|H|^2 + e^{2 \phi} ( |F_0|^2 - 3 |F_4|^2 ) \right) = -\frac{23}{4}\ \frac{(\kappa \sqrt{3})^{\tfrac{1}{3}}  p^2}{\kappa\, v_1v_2v_3}\, v_k \ , \label{F4F4}
\eea
where we used \eqref{Tmntrick} for $T_{mn}$. With the above quantities, it is straightforward to verify that the 6d Einstein equations are satisfied. We have then verified that the DGKT solution indeed solves the smeared 10d equations of type IIA supergravity.\\

For completeness, one may consider the source quantization for one $O_6$, following the definitions \eqref{DBIrewrite}-\eqref{BIRHS} together with the flux quantization (see e.g. \cite{Andriot:2020wpp}) to reach the following equalities
\beq
\int \frac{ T_{10} }{7}\, {\rm vol}_{\bot} =  2 \times 2\pi \sqrt{\alpha'} \, \quad
\int H =  (2\pi \sqrt{\alpha'})^2\, h_3 \,, \quad
F_0= (2\pi \sqrt{\alpha'})^{-1}\, f_0 \,,
\eeq
with the flux integers $f_0, h_3 \in \mathbb{Z}$. Applied to the Bianchi identity, the above integrals give a tadpole condition
\beq
 f_0 h_3 = -2 \ .\label{tadpole}
\eeq
This condition is not needed to solve the 10d equations, nor in the 4d analysis; we will still come back to it when considering extensions of the DGKT solution. Expressing the tadpole condition in terms of $p, m_0$, or $T_{10}$, would be interesting but this requires to know $\int \beta_0$. That question is related to the smearing procedure mentioned previously, as it considers the integral over the localized definition of $T_{10}$, with the quantized nature of the source. We thus discuss these subtleties in Appendix \ref{ap:smear}.

In the following, we show that this 10d solution can be obtained as a critical point of a scalar potential in a 4d effective theory.

\subsection{4d description}

We now provide a 4d effective theory for the 10d compactification setting described above. This 4d theory is of the form
\begin{equation}
{\cal S} = \int \d^4 x \sqrt{|g_4|} \left(\frac{M_p^2}{2} \R_4 - \frac{1}{2} \tilde{g}_{ij} \del_{\mu}\varphi^i \del^{\mu}\varphi^j - V \right) \ ,\label{S4dgen}
\end{equation}
where $\tilde{g}_{ij}$ denotes the field space metric, and $M_p$ the Planck mass. We show that the previous 10d solution can be obtained as a critical point of the scalar potential $V$. Having determined the kinetic terms, we then compute the mass spectrum at this anti-de Sitter solution. Most of these results follow \cite{DeWolfe:2005uu}, with few extensions. This presentation will be useful for the next sections.

To derive the 4d theory, in particular the scalar potential with background fluxes, we follow \cite{Andriot:2022bnb}, whose conventions are compatible with the above 10d ones and with the toroidal compactification setting. The 4d theory comes from a direct dimensional reduction of 10d supergravity; this approach is useful in view of the solution extensions considered in the next sections. For the DGKT setting alone, an alternative could be to use 4d ${\cal N}=1$ supergravity, where the 4d real fields, the saxions $\{v_i,\phi \}$ and the axions $\{b_i,\xi \}$, pair up into 3 complex K\"ahler moduli and the axio-dilaton. The interest of this compactification is indeed the absence of complex structure moduli. The orbifold however comes at the price of having 9 extra K\"ahler moduli corresponding to the blow-up modes (or twisted sector); we ignore those in this work.

\subsubsection{4d scalar potential and critical point}\label{sec:4dV}

We start with the general potential derived in \cite{Andriot:2022bnb}. Considering only the fields in the ansatz detailed in Section \ref{sec:ansatz}, this potential first boils down to
\begin{equation}
\begin{split}
\frac{1}{(2\pi \sqrt{\alpha'})^6}\, \frac{2}{M_p^2}\, V=  \frac{e^{2\phi}}{{\rm vol}}  \bigg(\frac{1}{2} |H|^2  - e^{\phi}&\frac{T_{10}}{7}  + \frac{e^{2\phi}}{2} \bigg[ F_0^2 +\big| F_0 B_2\big|^2 +\left|F_4  +\frac{1}{2}F_0\ B_2 \w B_2\right|^2 \label{potgen0}\\
&  +\left|C_3 \w H + F_4 \w B_2 + \frac{1}{6}F_0\ B_2 \w B_2\w B_2\right|^2 \bigg] \bigg)
\end{split}
\end{equation}
where fluxes are given by their background values, and gauge potentials $B_2$ and $C_q$ are the 4d axions. $T_{10}$ has here the off-shell expression of \cite{Andriot:2022bnb},\footnote{The expression for $T_{10}$ in \cite{Andriot:2022bnb} allows for a $B$-field dependence, coming from the DBI action. Here however, there is no $B_2$ component on the world-volume of hypothetical $D_6$ parallel to the $O_6$, as can be seen in \eqref{axions}, so we do not need to include this dependence.} but it can equivalently be given by the tadpole cancellation condition or Bianchi identity that should hold in any case, so we can use here the expression \eqref{tad}. The expression \eqref{potgen0} of the potential is the same as in \cite{Andriot:2022bnb}, except that we gave it in terms of the dimensionful volume ``${\rm vol}$'' defined in \eqref{vol}. In that case, the Planck mass does not depend on the background volume or dilaton (those are fully included in the fields), and is thus simply given here by $M_p = (\pi \alpha')^{-1/2} $, with $\alpha' = l_s^2$. Our $V$ has mass dimension 4; comparing to \cite[(3.20)]{DeWolfe:2005uu} where the potential has mass dimension $-4$, we get the following relation
\beq
\frac{2}{(2\pi)^7 {\alpha'}^4}\ V_{{\rm DGKT}} = V_{{\rm here}} \ .\label{relconvV}
\eeq
We further compute the potential in terms of the 4d fields with the compactification ansatz of Section \ref{sec:ansatz} and obtain
\bea
\frac{2/M_p^2}{(2\pi \sqrt{\alpha'})^6}\, V=  &\, \frac{p^2}{2} \frac{e^{2\phi}}{{\rm vol}^2} - 2 \sqrt{2} |pm_0| \frac{ e^{3\phi} }{{\rm vol}^{\frac{3}{2}}}  + m_0^2 \frac{ e^{4\phi}}{{\rm vol}} + \frac{e^{4\phi}}{{\rm vol}^3} \ssum e_i^2 v_i^2   \label{pot}\\
& + e^{4\phi} \bigg( \frac{ m_0^2}{{\rm vol}} \ssum \frac{b_i^2}{v_i^2} -  \frac{2 \kappa m_0 b_1b_2b_3}{{\rm vol}^3} \ssum \frac{e_i v_i^2}{b_i} +\frac{1}{{\rm vol}^3} \big(-\xi p + \ssum e_i b_i  \big)^2 \bigg) + {\cal O}(b_i^4, \xi b_i^3) \nn
\eea
where the sums are on $i=1,2,3$. This reproduces the potential in \cite{DeWolfe:2005uu}, as can be verified with \eqref{relconvV}, and up to $2\pi \sqrt{\alpha'}$ overall factors that were missed in \cite[(3.21)]{DeWolfe:2005uu}, as can be seen using e.g.~\eqref{dim}.

Introducing $V_{1,2}$ for the first and second line of the potential \eqref{pot} (neglecting higher dependence in the axions than quadratic), we rewrite $V=V_1 +V_2$ using the following simplifying variables:
\begin{equation}
r_i^2= \sqrt{\frac{|m_0|}{E}}|e_i|v_i \,, \quad
g = \frac{e^{\phi}}{\sqrt{{\rm vol}}}\frac{1}{|p|} \sqrt{\frac{E}{|m_0|}}\,, \quad
E=\frac{|e_1e_2e_3|}{\kappa}\,, \quad
\tilde{b}_i=|e_i|\, b_i\,, \quad
\tilde{\xi}=|p|\,\xi \,, \label{var}
\end{equation}
where one verifies that $r_i, g$ are dimensionless. We rewrite the potential as follows
\begin{equation}
\frac{2/M_p^2}{(2\pi \sqrt{\alpha'})^6}\, V
=\frac{2/M_p^2}{(2\pi \sqrt{\alpha'})^6}\, \Big( V_1 +V_2 \Big)
= \frac{p^4 |m_0|^\frac{5}{2} }{E^\frac{3}{2}}\, \Big( \tilde{V}_1 + \tilde{V}_2 \Big) \ ,
\end{equation}
with
\begin{align}
\tilde{V}_1 & =   \frac{1}{2} \frac{g^2}{\sprod r_i^2} - 2\sqrt{2}\, g^3  + \frac{g^4 \ssum r_i^4}{\sprod r_i^2} + g^4 \sprod r_i^2  \ , \label{V1}\\
\tilde{V}_2 & = \frac{|m_0|}{E}\, \frac{g^4}{\sprod r_i^2} \left( \sprod r_i^4 \ssum \frac{\tilde{b}_i^2}{r_i^4} - 2 \sprod r_i^2 \ssum s_i r_i^2\, \frac{\tilde{b}_j}{r_j^2} \frac{\tilde{b}_k}{r_k^2} + (\tilde{\xi} + \ssum s_i \tilde{b}_i  )^2   \right) \ , \label{V2}
\end{align}
where $s_i = {\rm sign}(m_0 e_i)$, while products and sums are on $i=1,2,3$, and the sum in the second axionic term contains 3 terms, each including the 3 indices. We used that $pm_0 <0$. This expression of the potential shows that the dependence on flux numbers is only an overall factor; changing those will therefore not affect the critical point, nor the mass spectrum (except for the signs $s_i$), up to an overall rescaling. This is a very specific property of DGKT.\\

Having determined the scalar potential, we now find its critical point. Since it is (at least) quadratic in the axions $b_i, \xi$, we consider the critical point solution given by
\beq
b_i = \xi = 0 \ .\label{bx}
\eeq
The first derivative of the potential with respect to the saxions, evaluated at the critical point, thus only involves $V_1$ to which we restrict. We disagree with \cite[(3.26)]{DeWolfe:2005uu} and we thus generalize the procedure of \cite{DeWolfe:2005uu} by keeping three different variables $v_i$ or $r_i$ to determine the critical point; this will however lead to the same final critical point. We compute
\begin{equation}
\begin{split}
& g \del_g \tilde{V}_1 + 2 \ssum r_i \del_{r_i} \tilde{V}_1  = 4 g^4 \sprod r_i^2 \left( 4 - \frac{3}{\sqrt{2}} (g \sprod r_i^2)^{-1} - \frac{5}{4} (g \sprod r_i^2)^{-2} \right) \\
\Rightarrow &\ g \sprod r_i^2 \,{}\big|_0 = \frac{5}{4\sqrt{2}} \ ,\label{gr}
\end{split}
\end{equation}
where the last value is obtained at the critical point, setting the above line to 0. Using then $\del_g V = 0$, one gets the relation at the critical point
\beq
\frac{1}{3} \ssum r_i^4 \,{}\big|_0 = \frac{9}{25} \sprod r_i^4 \,{}\big|_0 \ . \label{sp}
\eeq
Using finally $\del_{r_i} V= 0$, one concludes on the values
\beq
r_1^2 = r_2^2 =r_3^2 \,{}\big|_0\equiv r^2 \ \Rightarrow \ \eqref{sp}:\ r^4 = \frac{5}{3} \ ,\ \eqref{gr}:\ g \,{}\big|_0 = \sqrt{\frac{27}{160}} \ , \label{r0g0}
\eeq
at the critical point, as in \cite{DeWolfe:2005uu}. From those, with \eqref{var} and \eqref{bx}, one recovers the background values given in 10d in \eqref{bcgdfields}.

These field values allow us to evaluate the potential at the critical point
\beq
\frac{2/M_p^2}{(2\pi \sqrt{\alpha'})^6}\, V|_0 =  - \frac{p^4 |m_0|^\frac{5}{2} }{E^\frac{3}{2}}\,  \left( \frac{3}{5} \right)^{\frac{5}{2}} \frac{27}{2^8} \ ,\label{pot0}
\eeq
verifying that we have an anti-de Sitter solution. We can relate the latter to the 4d Ricci scalar in 4d Einstein frame
\beq
{\cal R}_{4} = 4\, \frac{V|_0}{M_p^2} = - (2\pi \sqrt{\alpha'})^6\ \frac{p^4 |m_0|^\frac{5}{2} }{E^\frac{3}{2}}\,  \left( \frac{3}{5} \right)^{\frac{5}{2}} \frac{27}{2^7} \ . \label{R4EV}
\eeq
Following conventions of \cite{Andriot:2022bnb}, the 4d metric in 10d string frame, respectively its Ricci scalar, are related to the 4d metric in 4d Einstein frame, respectively the above Ricci scalar (on the solution), by
\beq
g_{\mu\nu}^S = (2\pi \sqrt{\alpha'})^6\ \frac{e^{2\phi}}{{\rm vol}}\ g_{\mu\nu} \ , \qquad {\cal R}_{4} =  (2\pi \sqrt{\alpha'})^6\ \frac{e^{2\phi}}{{\rm vol}} \ {\cal R}_4^S \ .\label{relR4}
\eeq
We then compute ${\cal R}_4^S$, and verify that it matches the one obtained in the 10d solution \eqref{R4s}.

\subsubsection{4d kinetic terms and masses}\label{sec:4dkinmass}

We now provide the kinetic terms in the 4d theory \eqref{S4dgen}; this will allow us to compute the mass spectrum at the above critical point. The kinetic terms for the saxions are computed directly from 10d supergravity, using known results, in Appendix \ref{ap:kin}. They are given by
\beq
{\cal S}_{{\rm kin}_{sax}} =- \frac{1}{2} \int \d^4 x \sqrt{|g_4|}\ 2 M_p^2\  \left( (\del \ln\, g)^2 + (\del \ln\, r_1 )^2 + (\del \ln \, r_2 )^2 +  (\del \ln\, r_3 )^2 \right) \ .\label{kinsax2}
\eeq
From this expression, we can easily read the canonically normalized fields $\hat{g}, \hat{r}_i$. The kinetic terms for the axions are simpler to obtain and are generically given in \cite[(2.23)]{Andriot:2022bnb}. We deduce here the following kinetic terms, that we rewrite in terms of the different field variables, in particular \eqref{var}
\bea
{\cal S}_{{\rm kin}_{ax}} & =- \frac{M_p^2}{4} \int \d^4 x \sqrt{|g_4|} \left( |\del B_2|^2 + e^{2\phi} |\del C_3|^2  \right) \nn\\
& =- \frac{M_p^2}{4} \int \d^4 x \sqrt{|g_4|} \left( \ssum \frac{1}{v_i^2} (\del b_i)^2 + 2 \frac{e^{2\phi}}{{\rm vol}} (\del \xi)^2 \right)\nn \\
& = - \frac{1}{2} \int \d^4 x \sqrt{|g_4|}\ \frac{M_p^2}{2} \frac{|m_0|}{E}\, \left( \ssum \frac{1}{r_i^4} (\del \tilde{b}_i)^2 + 2 g^2 (\del \tilde{\xi})^2 \right) \ . \label{kinax1}
\eea
Since $g, r_i$ are 4d fields, we refrain from going to a canonical basis for the axions. We now have all ingredients to compute the mass spectrum.

In our conventions, the masses squared are the eigenvalues of the mass matrix $M^i{}_j= \tilde{g}^{ik} \nabla_k \del_j V$. At the critical point and in the canonical basis, they boil down to those of $\hat{M}^i{}_j= \delta^{ik} \del_k \del_j V$. The field space metric is block diagonal in the axions and the saxions. The Hessian of the potential at the critical point is also block diagonal in the axions and the saxions. So we can treat both independently.

We start by the saxions, only having to consider for them the contribution of $\tilde{V}_1$. We get the following evaluation at the critical point
\begin{equation}
\begin{split}
g^2 \del_g^2 \tilde{V}_1|_0 &=  \left( \frac{3}{5} \right)^{\frac{5}{2}} \frac{11\times 9}{2^6} \,,\quad
r_i^2 \del_{r_i}^2 \tilde{V}_1|_0 =  \left( \frac{3}{5} \right)^{\frac{5}{2}} \frac{9\times 17}{2^6} \,\\
g r_i \del_g \del_{r_i}\tilde{V}_1|_0&=\left( \frac{3}{5} \right)^{\frac{5}{2}}\frac{9}{2^4}\,,\quad
r_j r_i \del_{r_j} \del_{r_i} \tilde{V}_1|_0 = \left( \frac{3}{5} \right)^{\frac{5}{2}} \frac{9}{8} \ . 
\end{split}
\end{equation}
We deduce the mass matrix in the canonical basis $\{\hat{g},\hat{r_i}\}$
\bea
\hat{M} & = \frac{(2\pi \sqrt{\alpha'})^6}{2/M_p^2} \frac{p^4 |m_0|^\frac{5}{2} }{E^\frac{3}{2}}\, \delta^{ik} \del_k \del_j \tilde{V}_1|_0 \nn\\
& = \frac{(2\pi \sqrt{\alpha'})^6}{2^6} \frac{p^4 |m_0|^\frac{5}{2} }{E^\frac{3}{2}}\,   \left( \frac{3}{5} \right)^{\frac{5}{2}} 9 \, \left( \begin{array}{cccc}  \frac{11}{4} & 1 & 1 &  1\\
1 &  \frac{17}{4} & 2 & 2 \\
1  & 2 & \frac{17}{4} & 2 \\
1  & 2 & 2 & \frac{17}{4} \end{array}\right) \ ,\label{Msax}
\eea
giving the masses squared
\beq
m^2_{sax} = \frac{(2\pi \sqrt{\alpha'})^6}{2^8} \frac{p^4 |m_0|^\frac{5}{2} }{E^\frac{3}{2}}\,   \left( \frac{3}{5} \right)^{\frac{5}{2}} 9 \ \Big( 35,9,9,9\Big)  =  \frac{|V|_0|}{M_p^2} \, \left( \frac{70}{3},6,6,6\right) \ .\label{msax}
\eeq
Using the following relation between the masses squared in the 4d anti-de Sitter bulk and the conformal weights $\Delta$ of the single trace operators in a would-be dual 3d CFT
\beq
\Delta = \frac{1}{2} \left(3 \pm \sqrt{9 + 12 m^2 \frac{M_p^2}{|V|_0|} } \right) \ ,\label{Deltam}
\eeq
rewritten for convenience, we obtain here the conformal weights
\beq
\Delta_{sax}= \Big(10,6,6,6 \Big)\ . \label{Deltasax}
\eeq

We turn to the axions, for which the only contribution comes from $\tilde{V}_2$. We first compute the Hessian at the critical point: we get
\begin{equation}
\begin{split}
\frac{E}{|m_0|}\,\frac{r^6}{2g^4}\,\del_{\tilde{\xi}}^2 \tilde{V_2}|_0 &= 1 \,,\quad
\frac{E}{|m_0|}\, \frac{r^6}{2g^4}\, \del_{\tilde{b}_i}^2 \tilde{V_2}|_0 = \frac{34}{9}\,, \\
\frac{E}{|m_0|}\, \frac{r^6}{2g^4}\, \del_{\tilde{\xi}} \del_{\tilde{b}_i} \tilde{V_2}|_0 &= s_i \,,\quad
\frac{E}{|m_0|}\, \frac{r^6}{2g^4}\, \del_{\tilde{b}_i} \del_{\tilde{b}_j} \tilde{V_2}|_0 = -\frac{5}{3} s_k + s_i s_j \,.
\end{split}
\end{equation}
We deduce the mass matrix (in non-canonical basis $\{ \tilde{\xi}, \tilde{b}_i\}$) at the critical point
\bea
\tilde{g}^{ij} \del_j \del_k V|_0 & = (2\pi \sqrt{\alpha'})^6 \frac{p^4 |m_0|^\frac{5}{2} }{E^\frac{3}{2}}\, \frac{2g^4}{r^2} \, \left( \begin{array}{cccc} \frac{1}{2g^2 r^4} & \frac{s_1}{2g^2 r^4} & \frac{s_2}{2g^2 r^4} & \frac{s_3}{2g^2 r^4} \\
s_1  & \frac{34}{9} &  (-\frac{5}{3} s_3 + s_1 s_2) &  (-\frac{5}{3} s_2 + s_1 s_3) \\
s_2  &  (-\frac{5}{3} s_3 + s_1 s_2) & \frac{34}{9}  & (-\frac{5}{3} s_1 + s_2 s_3) \\
s_3  &  (-\frac{5}{3} s_2 + s_1 s_3) &  (-\frac{5}{3} s_1 + s_2 s_3) & \frac{34}{9}  \end{array}\right) \nn\\
& = 3 \frac{|V|_0|}{M_p^2} \, \left( \begin{array}{cccc} \frac{16}{9} & \frac{16}{9}s_1 & \frac{16}{9} s_2 & \frac{16}{9} s_3 \\
s_1  & \frac{34}{9} &  (-\frac{5}{3} s_3 + s_1 s_2) &  (-\frac{5}{3} s_2 + s_1 s_3) \\
s_2  &  (-\frac{5}{3} s_3 + s_1 s_2) & \frac{34}{9}  & (-\frac{5}{3} s_1 + s_2 s_3) \\
s_3  &  (-\frac{5}{3} s_2 + s_1 s_3) &  (-\frac{5}{3} s_1 + s_2 s_3) & \frac{34}{9}  \end{array}\right) \label{Max}
\eea
Computing the eigenvalues, we conclude
\beq
m^2_{ax} = \frac{|V|_0|}{M_p^2}\, \frac{1}{3} \, \left\{ \begin{array}{cl} \Big(88, 10, 10, 10 \Big) & {\rm for}\ s_1 s_2 s_3 = -1\ ,\ {\rm i.e.}\ e_1 e_2 e_3 m_0 <0 \\[.1in]
\Big(-2, 40, 40, 40 \Big) & {\rm for}\ s_1 s_2 s_3 = 1 \ ,\ {\rm i.e.}\ e_1 e_2 e_3 m_0 >0  \end{array} \right. \label{max}
\eeq
Recall that the Breitenlohner-Freedman bound in 4d is given by $m^2 M_p^2/|V|_0| \geq -3/4$, so the negative mass squared obtained above still provides a stable anti-de Sitter solution. This spectrum gives the following conformal weights with \eqref{Deltam}
\beq
\Delta_{ax} =  \left\{ \begin{array}{cl} \Big(11,5,5,5 \Big) & {\rm for}\ s_1 s_2 s_3 = -1\ , \\[.1in]
\Big(2,8,8,8 \Big) & {\rm for}\ s_1 s_2 s_3 = 1 \ ,  \end{array} \right. \label{Deltaax}
\eeq
where $\Delta=2$ can be traded for $\Delta =1$. As a side remark, note that only $s_1 s_2 s_3 = -1$ gives a supersymmetric solution \cite[(5.17)]{DeWolfe:2005uu}.

As pointed out already, the peculiar form of the potential allows to get a mass spectrum that only depends on the flux numbers through an overall factor: the value of the potential itself at the critical point. This leads to universal conformal weights independent of the flux numbers (except for the signs $s_i$).

\section{Warping DGKT}\label{sec:warpDGKT}

The 10d DGKT solution presented in Section \ref{sec:sol} is smeared: the $O_6$-plane sources are not localized or backreacted, their contributions as well as the warp factor $e^A$ and the dilaton are constant. A more complete, backreacted solution would rather allow for varying warp factor and dilaton, and localized source contributions. An ansatz for such a solution is presented in Appendix \ref{ap:conv}. Starting with 10d massive type IIA supergravity and $O_6$-planes, we provide there the 10d equations resulting from such a compactification ansatz. We also indicate how our conventions match those of \cite{Junghans:2020acz}. In Section \ref{sec:sol} and Appendix \ref{ap:smear}, we show how taking a smeared limit on these equations (with $e^A=1$) brings us back to the setting of DGKT, with a smeared solution.

While we know the equations and ansatz for a backreacted solution, finding it is an open problem. An approximate solution, beyond the smeared limit, has however been proposed in \cite{Junghans:2020acz} (see also \cite{Marchesano:2020qvg}) using a perturbative expansion in a discretized parameter $n$. The $0^{{\rm th}}$ order in this $n$-expansion is the DGKT (smeared) solution, and the $1^{{\rm st}}$ order provides the approximate warped and localized solution.

The parameter $n$ corresponds to a common scaling parameter of the three $F_4$ flux numbers $e_i$; it can also be understood as a common integer to the three components in the quantized flux $F_4$. Importantly, the DGKT solution allows this parameter to be chosen arbitrarily large, playing a crucial role in the parametric scale separation and parametric control on classicality of the solution. This is why extensions of the DGKT solution based on a $n$-expansion can be considered. As mentioned, one such extension can be used to get an approximate warped and backreacted solution, and this is what we focus on in this section. We discuss aspects of this extension first in 10d and then in 4d. One motivation is to know how the 4d mass spectrum would be altered at $1^{{\rm st}}$ order. A different extension is then considered in the next section.\\

In this section, we consider the following $n$-expansion of the 10d fields of Appendix \ref{ap:conv}
\begin{equation}
\begin{split}
g_{mn} & = g_{mn}^{(0)} n^{1/2} + g_{mn}^{(1)} n^{-1/2} + {\cal O} (n^{-3/2})\\
g^{mn} & = g^{mn\, (0)} n^{-1/2} + g^{mn\, (1)} n^{-3/2} + {\cal O} (n^{-5/2}) \\
e^A l \equiv w & = w^{(0)} n^{3/4} + w^{(1)} n^{-1/4} + {\cal O} (n^{-5/4}) \\
e^{-\phi} \equiv \tau & = \tau^{(0)} n^{3/4} + \tau^{(1)} n^{-1/4} + {\cal O} (n^{-5/4}) \label{expansionJ}\\
F_0 & = F_0^{(0)} n^0 \\
F_2 & = F_2^{(0)} n^{1/2} + F_2^{(1)} n^{0} + {\cal O} (n^{-1/2}) \\
H & = H^{(0)} n^{0} + H^{(1)} n^{-s_h} \ ,\ 0 < s_h < 1  \\
F_4 & = F_4^{(0)} n + F_4^{(1)} n^{1-s_4}  \ ,\ 0 < s_4 < 1  \\
F_6 & = 0 + F_6^{(1)} n^{s_6} \ ,\ 1< s_6 <2-s_h \ ,\ s_6 <2-s_4 
\end{split}
\end{equation}
where $w, \tau$ are notations of \cite{Junghans:2020acz}, and $l$ is our 4d anti-de Sitter radius in 10d string frame. The $0^{{\rm th}}$ order fields should be given on-shell by the DGKT solution.
One recognises their scaling with $n$: starting with $g_{mn}, \tau$, this is seen thanks to \eqref{bcgdfields}, with $e_i \sim n$, $E \sim n^3$. For $w$, this is due to the anti-de Sitter radius $l$, whose scaling can be read from \eqref{R4EV} and related equations, with $-12 \, l^{-2}={\cal R}^S_4 \propto E g^{-2} {\cal R}_4 \propto E^{-1/2} $.
Finally, we recall that $F_2^{(0)}=F_6^{(0)}=0$ in the DGKT solution, the fluxes $F_0$ and $H$ do not scale at $0^{{\rm th}}$ order, but $F_4$ scales as $n$ by definition. To this, we could add the scaling the 4d Einstein frame metric: it can be read from ${\cal R}_4$ in the DGKT solution
\beq
g_{\mu\nu}  = g_{\mu\nu}^{(0)}\ n^{9/2} + ... \label{4dmetricscaling}
\eeq
This is however not needed for now. At $1^{{\rm st}}$ order, the scalings in \eqref{expansionJ} are those of \cite{Junghans:2020acz}, provided one picks $s_h =s_4 =\frac{1}{2}$, $s_6 =1$; we leave ourselves more freedom here on these three scalings for future purposes, and do not fix their values. Note that $s_h >0$ and $s_4 >0$ are necessary by definition of the expansion, the other bounds will be motivated below. Also, strictly speaking, the value $s_6 =1$ of \cite{Junghans:2020acz} is excluded by our bound $s_6 > 1$, but this can be ignored for now as that bound will only be required later. In addition, one can verify the following relation at $1^{{\rm st}}$ order
\beq
g^{mn \, (1)} = -g^{mp \, (0)} g_{pq}^{(1)} g^{qn \, (0)} \ ,
\eeq
consistent with the above scaling for the inverse 6d metric. Finally, note that $F_0$ is not corrected, as it is meant to be an independent quantized scalar.

We consider the $n$-expansion \eqref{expansionJ}, first in the 10d warped setting of Appendix \ref{ap:conv}, and then in a corresponding 4d warped theory.

\subsection{10d extended solution}

As verified in Section \ref{sec:sol} and Appendix \ref{ap:smear}, smearing equations and quantities of Appendix \ref{ap:conv} gives the back the DGKT compactification ansatz and solution. The $n$-expansion presented above is meant in the same way. Plugging the field expansion \eqref{expansionJ} in the 10d equations of Appendix \ref{ap:conv}, one can develop them at leading order (LO) and next-to leading order (NLO) in $n$. The LO, which only involves $0^{{\rm th}}$ order fields, should reproduce, or at least be compatible with the DGKT solution. This is verified to some extent in \cite{Junghans:2020acz}; since our conventions match, this is also true for us. At NLO, $1^{{\rm st}}$ order fields appear. This is where a solution was found in \cite{Junghans:2020acz}, that matches the expectations of what an approximate backreacted solution would be. For completeness, we give this solution in our conventions (the only difference being the RR sign, see Appendix \ref{ap:conv})
\bea
F_2^{(1)} & =  + 4 \sum_{q,t=0,1,2} *_6^{(0)} \Bigg( \d \beta \left( {\rm Re} \left( \alpha^{2a q+2t} \, z^a \right) \right) \w \d \left( {\rm Im}\left( \alpha^{2q+2t} \, z^1 \right) \right) \label{F21sol}\\
& \hspace{1.5in} \w \d \left( {\rm Im}\left( \alpha^{4q+2t} \, z^2 \right) \right) \w \d \left( {\rm Im}\left( \alpha^{2t} \, z^3 \right) \right) \Bigg) \ ,\nn\\
\frac{w^{(1)}}{w^{(0)}} & = - \frac{\tau^{(1)}}{3 \tau^{(0)}} = \frac{1}{\upnu_1 \upnu_2 \upnu_3 \tau^{(0)}} \sum_{q,t=0,1,2} \beta \left( {\rm Re} \left( \alpha^{2a q+2t} \, z^a \right) \right) \ ,\\
u_a^{(1)} &= -\frac{\upnu_a^2}{\upnu_1 \upnu_2 \upnu_3 \tau^{(0)}} \sum_{q,t=0,1,2} \alpha^{4a q+4t} \beta \left( {\rm Re} \left( \alpha^{2b q+2t} \, z^b \right) \right) \ ,\\
{\rm where}\ \beta\left( {\rm Re}\, \overrightarrow{z} \right) &= - 2 \int_0^{\infty} \d s \left[ 1- \prod_{a=1}^3 \theta_3 \left( 2 {\rm Re}\, z^a , \frac{4 \pi^2}{\upnu_a^2} s \right) \right] + {\rm constant} \ ,\\
{\rm for}\ \theta_3 (\sigma ,\tau) & = \sum_{n=-\infty}^{\infty} e^{2\pi \i n\, \sigma - \tau n^2 } \ .
\eea
The solution is expressed in terms of the $z^i$ defined in Section \ref{sec:ansatz}, and $\alpha = e^{\i \frac{\pi}{3}}$. The $\upnu_i$ are related to our $v_i$ as in \eqref{vimap} and only have a $0^{{\rm th}}$ order contribution, no $1^{{\rm st}}$ order one. The 6d metric is nevertheless corrected at $1^{{\rm st}}$ order by the $u_a^{(1)} $, with $u_a^{(0)}=0 $; we refer to \cite[(4.12)-(4.14)]{Junghans:2020acz} about it. The function $\beta$ plays the role of generalized Green's function (see \cite{Andriot:2019hay}), in such a way that $\d F_2^{(1)}$ will give a $\delta$-function, and similarly for the Laplacian of $w$ and $\tau$.

This NLO solution, expressed in terms of $1^{{\rm st}}$ order fields, does not involve $1^{{\rm st}}$ order corrections to $H$, $F_4$, $F_6$. This is because those decouple from the rest of the equations, and can consistently be set to zero. We devote the rest of this subsection to these fluxes. We allow more freedom in their scaling as indicated in \eqref{expansionJ} than in \cite{Junghans:2020acz}, and make important observations on their $1^{{\rm st}}$ order contributions.

\subsubsection{$H$, $F_4$ and $F_6$ fluxes: decoupling at NLO}\label{sec:HF4F6}

We now expand the various equations at NLO, with an interest in $H$, $F_4$ and $F_6$ at $1^{{\rm st}}$ order. We start with the Bianchi identities (BI) given in \eqref{BIs}. Using that $F_2^{(0)}=0$, those are given at NLO by
\bea
\d H^{(1)} &=0 \\
\d F_2^{(1)} - H^{(0)} \w F_0^{(0)} &= \sum_{I} \frac{T_{10}^I}{7} {\rm vol}_{\bot_I} \\
\d F_4^{(1)} n^{1-s_4} - H^{(0)}\w F_2^{(1)} &= 0
\eea
The $F_2$ BI is the one considered in \cite{Junghans:2020acz} and solved by the solution detailed above. The $F_4$ BI has two terms that we kept for illustration: since $s_4 < 1$, the second term $H^{(0)}\w F_2^{(1)}$ (which does not seem to vanish with the above solution) is subdominant and should then be dropped at NLO. This leaves us with closed $H^{(1)}, F_4^{(1)}$, and $F_6^{(1)}$ is necessarily closed.

We turn to the flux equations of motion given in \eqref{fluxeq0} - \eqref{fluxeq}. Using the metric scaling, we first get the following Hodge star action at $1^{{\rm st}}$ order on $\d y^{1..p} = \d y^1 \w ... \w \d y^p$
\beq
*_6\, \d y^{1..p} = *_6^{(0)}\, \d y^{1..p}\ n^{(3-p)/2} + *_6^{(1)}\, \d y^{1..p}\ n^{(1-p)/2} \ .
\eeq
When acting on a flux given schematically by $F_p= F_p^{(0)} n^{f_0} + F_p^{(1)} n^{f_1}$, we get at NLO
\beq
*_6 F_p = n^{(1-p)/2 + f_0} \left(*_6^{(1)} F_p^{(0)} + *_6^{(0)} F_p^{(1)} \ n^{1 + f_1 - f_0} \right) \ .
\eeq
Since in the scaling \eqref{expansionJ} we always have $1 + f_1 - f_0 > 0$, the second term dominates; for $F_6$ the first term vanishes anyway. Furthermore, we can compare $*_6^{(0)} F_p^{(1)}$ times the $0^{{\rm th}}$ order of $e^{4A} l^4$ or $e^{-2 \phi}$, to $*_6^{(0)} F_p^{(0)}$ times their $1^{{\rm st}}$ order: $e^{4A} l^4$ or $e^{-2 \phi}$ diminish by $1/n$ between the two orders, while the fluxes diminish by less than $1/n$. This means that the contribution with $*_6^{(0)} F_p^{(1)}$ is dominant and $e^{4A}$ and $e^{-2 \phi}$ can be set to their $0^{{\rm th}}$ order. We finally get the following NLO flux equations
\bea
\d( *_6^{(0)} F_2^{(1)} ) = 0 & \\
\d( *_6^{(0)} F_4^{(1)} ) n^{-s_4} + H^{(0)} \w *_6^{(0)} F_6^{(1)} n^{s_6-2} = 0 & \\
\d( *_6^{(0)} F_6^{(1)} ) = 0 & \\
\d( {\tau^{(0)}}^2 *_6^{(0)} H^{(1)} ) n^{-s_h}  - F_4^{(0)} \w *_6^{(0)} F_{6}^{(1)} n^{s_6-2} = 0 & \label{HeqNLO}
\eea
where we used that $F_2^{(0)} = 0$, $H^{(0)} \w *_6^{(0)} F_4^{(0)} = 0 $ and that $w^{(0)}$ and $\tau^{(0)}$ are constant. The LO $F_4$ and $H$ equations require that $s_6-2 < 0$, already implied by the conditions $s_6< 2- s_4$, $s_6<2-s_h$, in \eqref{expansionJ}. In addition, those two conditions on $s_6$ imply that the second terms in the $F_4$ and $H$ equations at NLO are subdominant. This makes $H^{(1)}, F_4^{(1)}, F_6^{(1)}$ co-closed. $F_2^{(1)}$ is also co-closed, which is satisfied by the above solution.

We conclude that {\sl $H^{(1)}, F_4^{(1)}, F_6^{(1)}$ are harmonic forms} (with respect to the $0^{{\rm th}}$ order metric), with the scaling constrained as in \eqref{expansionJ}. We also note that they decouple from the other equations so far, and can thus be ignored or set to zero as done in \cite{Junghans:2020acz}, with a more general scaling.\\

We now turn to the remaining equations, namely the dilaton, the 4d and 6d Einstein equations given in \eqref{eq1}-\eqref{eq3}. To compute them at NLO, let us first provide the following schematic expansion, at LO + NLO
\bea
& {\rm For} \quad F_p= F_p^{(0)} n^{f_0} + F_p^{(1)} n^{f_1} \ , \label{F2}\\
& |F_p|^2 \sim |F_p^{(0)}|^2 \, n^{2f_0 - \frac{p}{2}} + 2 F_p^{(0)}F_p^{(1)} (g^{-1\, (0)})^p \, n^{f_0 +f_1 - \frac{p}{2}} + p F_p^{(0)} F_p^{(0)} (g^{-1\, (0)})^{p-1} g^{-1\, (1)} \, n^{2 f_0  - \frac{p}{2} - 1} \ . \nn
\eea
In addition, one has
\beq
2f_0 - \frac{p}{2} > f_0 +f_1 - \frac{p}{2} > 2 f_0  - \frac{p}{2} - 1 \ ,
\eeq
which makes the last term subdominant, hence not to be considered at NLO. Indeed, by definition, $f_1 < f_0$ giving the first inequality, that guarantees that the LO is dominant. The second inequality boils down to $f_1 > f_0-1$, a property that we already noticed for our fluxes in \eqref{expansionJ} (for $F_6$, one uses $F_6^{(0)}=0$). This derivation of the NLO for $|F_p|^2$ will be useful in the following. For now, let us consider the LO: we get
\beq
{\rm LO}:\quad e^{-\phi}|H|^2 = \tau^{(0)} |H^{(0)}|^2 \, n^{-3/4} \ ,\ e^{\phi}|F_p|^2 = 1/\tau^{(0)} \, |F_p^{(0)}|^2 \, n^{-3/4} \quad {\rm for} \quad p=0,2,4 \ .
\eeq
We can compare this contribution to that of $\Delta_6 e^{-\phi}$ that appears in all equations \eqref{eq1}-\eqref{eq3}. We first establish that the Christoffel symbol goes as $\Gamma \sim \Gamma^{(0)} + \Gamma^{(1)}\, n^{-1}$. Then, using that $\tau^{(0)}$ is constant, we get at NLO
\beq
\Delta_6 e^{-\phi} = \Delta_6^{(0)} \tau^{(1)}\, n^{-3/4} \ .
\eeq
We conclude that in each of the equations \eqref{eq1}-\eqref{eq3}, the LO flux contributions have the same scaling as the NLO of $\Delta_6 e^{-\phi}$. Therefore, the NLO version of these equations only involves the $0^{{\rm th}}$ order of the fluxes, as already indicated in \cite{Junghans:2020acz}. Once again, $H^{(1)}, F_4^{(1)}, F_6^{(1)}$ decouple and can be ignored or set to zero; we also do not get any constraint on our $1^{{\rm st}}$ order scalings $s_h,s_4,s_6$ from these remaining equations at NLO. The only constraint we thus obtained for $H^{(1)}, F_4^{(1)}, F_6^{(1)}$ with the general scaling in \eqref{expansionJ}, is that they are harmonic forms.

\subsubsection{$H$, $F_4$ and $F_6$ fluxes: quantization and (N)NLO}\label{sec:F4again}

Even though these fluxes at $1^{{\rm st}}$ order decouple in equations at NLO, there are important reasons to consider them, as we will see in 4d. There is then a crucial observation to be made. The bottom line of the previous study is that $H^{(1)}, F_4^{(1)}, F_6^{(1)}$ have to be harmonic: this implies that {\sl they must be quantized}. Harmonic forms on the torus orbifold with the orientifold are constrained as in the initial flux ansatz: in other words, these fluxes have to be on the same cycles as their $0^{{\rm th}}$ order counterparts. This gives
\beq
H = -(p + p^{(1)}) \beta_0 \,,\quad
F_4= \sqrt{2}\, (e_i + e_i^{(1)} n^{-s_4}) \tilde{w}^i\,,\quad
F_6 = f^{(1)} \d^6 y \,, \label{fluxescorrect}
\eeq
where we introduced the $1^{{\rm st}}$ order flux numbers $p^{(1)}, e_i^{(1)}, f^{(1)}$. Note that those should carry (part of) the scaling in $n$, in agreement with \eqref{expansionJ}. Given that the DGKT ($0^{{\rm th}}$ order) fluxes are already quantized, the $1^{{\rm st}}$ order contributions should be quantized independently. The quantization gets written as follows
\begin{equation}
\begin{split}
& \int H =  (2\pi \sqrt{\alpha'})^2\, (h_3 + h_3^{(1)} n^{-s_h}) \ ,\ \int_i F_4 =  (2\pi \sqrt{\alpha'})^3\,  (f_{4i}\, n + f_{4i}^{(1)} n^{1-s_4}) \ , \\
& \int F_6 =  (2\pi \sqrt{\alpha'})^5\, f_6^{(1)} n^{s_6} \ , \qquad {\rm with}\ \  h_3\, ,\ h_3^{(1)} n^{-s_h}\, ,\ f_{4i}\, n\, ,\ f_{4i}^{(1)} n^{1-s_4}\, ,\ f_6^{(1)} n^{s_6} \in \mathbb{Z} \ .\label{fluxquant}
\end{split}
\end{equation}
We deduce the flux numbers quantization
\bea
p &= - \frac{(2\pi \sqrt{\alpha'})^2}{\int \beta_0}\, h_3  \,,\quad
p^{(1)} = - \frac{(2\pi \sqrt{\alpha'})^2}{\int \beta_0}\, h_3^{(1)} n^{-s_h} \ ,\\
e_i &= \frac{(2\pi \sqrt{\alpha'})^3}{\sqrt{2} \int \tilde{w}^i} \, f_{4i} \, n \,,\quad
e_i^{(1)}= \frac{(2\pi \sqrt{\alpha'})^3}{\sqrt{2} \int \tilde{w}^i} \, f_{4i}^{(1)}\, n \,,\quad
f^{(1)} = \frac{(2\pi \sqrt{\alpha'})^5}{\int \d^6 y} \, f_6^{(1)} n^{s_6} \ .
\eea
The condition $( h_3^{(1)}/ n^{s_h} ) \in \mathbb{Z}$ is constraining, because $0 < s_h < 1$. The coefficient $h_3^{(1)}$ is a fixed number but $n$ is like a variable that can be sent to arbitrary large values; in particular, the expansion considered before is valid only in the limit of large $n$. It is therefore clear that the quantization condition $h_3^{(1)} n^{-s_h} \in \mathbb{Z}$ cannot be satisfied for arbitrary $n$. To allow the large $n$ limit, we set
\beq
h_3^{(1)} = 0 \,,\quad p^{(1)}=0 \,,\quad H^{(1)} = 0 \, . \label{H1=0}
\eeq
This situation is very similar to that of $F_0$, which is not corrected beyond the $0^{{\rm th}}$ order. In addition, we recall the tadpole condition \eqref{tadpole}, giving with $ F_0= (2\pi \sqrt{\alpha'})^{-1}\, f_0 $ that $f_0 h_3 = -2 $, i.e.~$|h_3|=1$ or $2$. Getting the integer $h_3^{(1)} n^{-s_h}$ smaller than such an $|h_3|$, in order to have the $1^{{\rm st}}$ order subdominant to the $0^{{\rm th}}$ order, is very limited (and not much subdominant). So again for consistency of the expansion, we are led to \eqref{H1=0}.

We now come back to the $H$-flux equation of motion at NLO. In that equation \eqref{HeqNLO}, the $F_6$ term was subdominant thanks to the condition $s_6 < 2-s_h$ in \eqref{expansionJ}. If $H^{(1)} = 0$, one has however no access to $s_h$ so this condition cannot be checked. Another way to see the issue is to start directly with $H^{(1)} = 0$ (as one does for $F_0$): the resulting NLO equation would just be $F_4^{(0)} \w *_6^{(0)} F_6^{(1)} = 0$. Note that further terms in $F_2^{(1)}$ would appear in \eqref{HeqNLO} with a scaling $n^{-1}$; the $ F_6^{(1)}$ term remains dominant, avoiding a mixing with those, thanks to the bound $s_6>1$ in \eqref{expansionJ}. Then, the only solution to the resulting NLO equation $F_4^{(0)} \w *_6^{(0)} F_6^{(1)} = 0$ is $F_6^{(1)} = 0$. As we will see, this situation is similar to the one in the 4d scalar potential, where $F_6^{(1)}$ can generate a linear term in the axions, leading to an analogous equation; one solution will then be to make $F_6^{(1)}$ vanish. For these reasons, we set
\beq
f_6^{(1)}=0 \,,\quad f^{(1)} = 0 \,,\quad F_6^{(1)} = 0 \, .
\eeq
The NLO equations and the quantization of fluxes therefore lead to very important constraints on $H^{(1)}, F_4^{(1)}, F_6^{(1)}$, even with the general scalings of \eqref{expansionJ}, leaving for now only $F_4^{(1)}$ non-zero.\\

We finally come back to the 6d Einstein equation \eqref{eq3}. As explained in Section \ref{sec:HF4F6}, this equation at NLO only involved $0^{{\rm th}}$ order fluxes. A priori, one should not consider NNLO, since such an order can potentially involve $2^{{\rm nd}}$ order fields, that go beyond the extension one wants to consider here. The $n$-dependence for $2^{{\rm nd}}$ order fields is proposed in the expansion \eqref{expansionJ} under the symbol ${\cal O}(n^{\#})$, following \cite{Junghans:2020acz}; we did not propose any for $H,F_4, F_6$ for which we allowed more freedom in the scaling. Taking this expansion into account, let us revisit the 6d Einstein equation \eqref{eq3}, aiming to look at the NNLO. The schematic $n$-expansion for the terms appearing in that equation is as follows
\bea
\left. \begin{array}{r}
- e^{-2\phi} {\cal R}_{mn}  \\[0.1in]
+ \frac{g_{mn}}{4} \left( e^{-\phi} \Delta_6\, e^{-\phi} + (\del e^{-\phi})^2 + 4 e^{-\phi-A} \del_{p} e^{A} \del^{p} e^{-\phi} \right) \\[0.1in]
+ 4 e^{-2\phi -A} \nabla_n \del_m e^{A} + 2 e^{-\phi} \nabla_m \del_n e^{-\phi}  - 2  \del_m e^{-\phi} \del_n e^{-\phi}   \end{array}
\right| & \sim n^{3/2} \left( 1 + n^{-1} + {\cal O}(n^{-2}) \right)\nn\\[0.1in]
\left. \begin{array}{r}
+\frac{e^{-2\phi}}{4} H_{mpq}H_n^{\ \ pq} - \frac{g_{mn}}{8} e^{-2\phi} |H|^2  + \frac{e^{-\phi}}{2}T_{mn}   - g_{mn} \frac{7}{16} e^{-\phi} \frac{T_{10}}{7} \\[0.1in]
+ \frac{g_{mn}}{16} ( |F_0|^2 + 3 |F_6|^2 ) \end{array}
\right| & \sim n^{1/2} \left( 1 + n^{-1} + {\cal O}(n^{-2}) \right)\nn\\[0.1in]
\left. \begin{array}{r}
+ \frac{1}{2\times 3!} F_{4\ mpqr}F_{4\ n}^{\ \ \ pqr}  -3 \frac{g_{mn}}{16}  |F_4|^2  \end{array}
\right| & \sim n^{1/2} \left( 1 + n^{-s_4} \right)\nn\\[0.1in]
\left. \begin{array}{r}
+ \frac{1}{2} F_{2\ mp}F_{2\ n}^{\ \ \ \ p} - \frac{g_{mn}}{16} |F_2|^2 \end{array}
\right| & \sim n^{-1/2} \left( 1 + {\cal O}(n^{-1/2}) \right)\nn
\eea
where we used the argued $H^{(1)}=0$, $F_6^{(0)}=F_6^{(1)}=0$, together with $F_2^{(0)}=0$ and the definition of the source contributions $T_{10}$ and $T_{mn}$. The above is easily obtained thanks to the fact that the metric, the warp factor and the dilaton evolve by $1/n$ between each order. We recover that the LO is given by the first terms, namely the Ricci tensor and derivatives of warp factor and dilaton, at $0^{{\rm th}}$ order. The NLO is given by $1^{{\rm st}}$ order contributions of the latter, together with $0^{{\rm th}}$ order fluxes and sources, as mentioned already. The surprise comes at the next level, that we refer to as NNLO: thanks to $0< s_4 < 1$, it is given by
\beq
\left( \frac{1}{2\times 3!} F_{4\ mpqr}F_{4\ n}^{\ \ \ pqr}  -3 \frac{g_{mn}}{16}  |F_4|^2  \right)_{{\rm NNLO}} n^{1/2 -s_4} = 0 \ . \label{eqF41}
\eeq
Indeed, all other contributions, including $2^{{\rm nd}}$ order fields, appear at $n^{-1/2}$ which is a lower order. This is surprising because this gives a new constraint on $F_4^{(1)}$, which does not involve $2^{{\rm nd}}$ order fields, and therefore should a priori be obeyed. This is not mentioned in \cite{Junghans:2020acz}, even though $s_4=\frac{1}{2}$ considered there is captured by the above.\footnote{The NNLO 6d Einstein equation, as derived from the scaling of \cite{Junghans:2020acz} (included in the above analysis) would take the following form
\begin{align}
\frac{1}{2}F_{2\,mp}^{(1)}F_{2\,n}^{(0)\,p}-\frac{g_{mn}^{(0)}}{16}\vert F_2\vert^2
+\frac{e^{-2\phi^{(0)}}}{4}H_{mpq}^{(1)}H_n^{(0) \,pq }-\frac{g_{mn}^{(0)}}{8}e^{-2\phi^{(0)}}\vert H\vert^2
+\frac{1}{12}F_{4\,mpqr}^{(1)}F_{4\,n}^{(0)\,pqr}-3\frac{g_{mn}^{(0)}}{16}\vert F_4\vert^2=0 \ , \nonumber
\end{align}
where the squares of fluxes are the contraction of the $0^{{\rm th}}$ and $1^{{\rm st}}$ order components, and all terms scale as $n^0$. The DGKT solution, $F_2^{(0)}=0$, would cause the first two terms to vanish, while the flux quantization constraint would make the terms involving $H^{(1)}$ to vanish, leaving us with equation \eqref{eqF41}. This eventually gives the solution for $F_4^{(1)}$ presented in \eqref{eqF432}.} In addition, all terms entering the LO are actually vanishing (since fields at $0^{{\rm th}}$ order are constant), so the LO equation is somehow unsatisfactory. Such an LO equation could be ignored, and one may promote the above NNLO to NLO, then to be considered. This hierarchy will be made clearer in 4d with the scalar potential, where we will see that the above constraint on $F_4$ indeed has to be considered.

We rewrite \eqref{eqF41} more explicitly, using the $0^{{\rm th}}$ order metric to raise indices
\beq
\frac{1}{3!} F_{4\ mpqr}^{(0)}F_{4\ n}^{(1) \ pqr}  - \frac{3}{8} g^{(0)}_{mn} \frac{1}{4!} F_{4\ pqrs}^{(0)}F_{4}^{(1) \ pqrs}  = 0 \ . \label{eqF42}
\eeq
Contracting it with an inverse metric gives the second term in \eqref{eqF42} to vanish, which in turn requires the first term to vanish
\beq
\eqref{eqF42}\ \ \Rightarrow\ \ \frac{1}{4!} F_{4\ pqrs}^{(0)}F_{4}^{(1) \ pqrs}  = 0 \ \ \Rightarrow\ \ \frac{1}{3!} F_{4\ mpqr}^{(0)}F_{4\ n}^{(1) \ pqr} = 0 \ .\label{eqF43}
\eeq
We rewrite it more explicitly and get
\beq\label{eqF432}
\forall\, i\neq j \in \{1,2,3\} \ ,\ e_i \, e_i^{(1)}\, v_i^2 + e_j \, e_j^{(1)}\, v_j^2 = 0 \ \Rightarrow \ e_i^{(1)} = 0 \ \Rightarrow \ F_4^{(1)}=0 \ ,
\eeq
where we used that in the DGKT solution, all $F_4$ components are non-zero: $e_1 e_2 e_3 \neq 0$. The (N)NLO constraint \eqref{eqF41} therefore requires $F_4^{(1)}$ to vanish! The closer look we took at fluxes $H,F_4,F_6$ has thus set important constraints on them, starting with quantization and eventually forcing them all to vanish at $1^{{\rm st}}$ order. We will revisit these constraints from a 4d perspective.

\subsection{4d theory and solution}

In Appendix \ref{ap:conv}, we have provided the equations of 10d type IIA supergravity with sources in a warped compactification ansatz. In this section, we consider a 4d effective theory for such a compactification. We derive in full generality in Appendix \ref{ap:4dwarp} such a ``warped 4d theory'' from 10d type IIA supergravity, and repeat the result here. We use conventions of \cite{Andriot:2022bnb}, slightly different than those of \cite{Junghans:2020acz}, and our 4d theory is more general. It is obtained using the following 10d string frame metric
\beq
\d s^2 = {\tau(x)}^{-2}\, e^{2A(y)}\, g_{\mu\nu}(x)\, \d x^{\mu} \d x^{\nu} + g_{mn}(x,y)\, \d y^m \d y^n \ , \label{10dourmetric}
\eeq
with the 4d field
\beq
\tau^{2} = {\cal V}_{6\phi} \equiv (2\pi \sqrt{\alpha'})^{-6}  \int \d^6 y \sqrt{|g_6|}\, e^{2A-2 \phi} \ ,
\eeq
and $g_{\mu\nu}$ ends up being the 4d Einstein frame metric. The action for the resulting 4d theory is given by
\bea
\int \d^{4} x \sqrt{|g_{4}|} & \Bigg( \frac{M_p^2}{2} {\cal R}_{4} - V_{{\rm part}} \label{4dactionfullywarped}\\
& - \frac{M_p^2}{4}\bigg( {\cal V}_{6\phi}^{-2}\, (\del {\cal V}_{6\phi} )^2  - \frac{1}{2} \del_{\mu} g_{mn}\, \del^{\mu} g^{mn} + |\del B_2|^2 + |\del C_{1,3}|^2 \ \frac{\int \d^6 y \sqrt{|g_6|} e^{2A}}{ (2\pi \sqrt{\alpha'})^{6}\, {\cal V}_{6\phi} }
\bigg)  \Bigg) + {\cal S}_{F_6}  \ . \nn
\eea
The terms captured by ${\cal S}_{F_6}$ are given in \eqref{SF6}; they cannot be better detailed in all generality without some knowledge of the fields dependence on $y$. These terms eventually contribute to the scalar potential. The other part of the scalar potential is given by
\bea
V_{{\rm part}} = \frac{M_p^2}{2}  (2\pi \sqrt{\alpha'})^{-6}\, {\cal V}_{6\phi}^{-2}  \int \d^6 y & \sqrt{|g_6|} \, e^{4A-2 \phi} \, \bigg( - {\cal R}_6 + 12 e^{-2A} (\del e^{A} )^2 + 8 e^{-A} \Delta_6 e^{A} - 4(\del \phi)^2 \nn\\
& \ - e^{\phi} \sum_I\frac{T_{10}^I}{7} + \frac{1}{2} |H|^2 +  \frac{e^{2\phi}}{2}  \bigg[ F_0^2 +\big|F_2 + F_0 B_2\big|^2  \label{potw} \\
& \ +\left|F_4 +C_1 \w H + F_2 \w B_2 +\frac{1}{2}F_0\ B_2 \w B_2\right|^2  \bigg] \bigg) \ .\nn
\eea
We recall from \cite{Andriot:2022bnb} that we consider here only background fluxes, while the axions $B_2, C_3$ are 4d fluctuations. The source term in the potential can also be rewritten in general using the Bianchi identity, as explained around \eqref{potBI}.

With this general warped 4d theory at hand, we can apply the $n$-expansion \eqref{expansionJ} of the fields to get a corresponding 4d theory order by order. At LO, we will recover the theory used for the DGKT solution, and we will then consider the higher orders corrections. In this expansion, we use for simplicity that $F_2^{(0)}=0 , F_6^{(0)}=0, C_1=0$, and that $w^{(0)}, \tau^{(0)}, g_{mn}^{(0)}$ are independent of $y$. In the above general 4d theory, we also see a new ingredient w.r.t.~10d: the axions. In order to perform an $n$-expansion of the 4d theory, we consider
\bea
B_2 & = B_2^{(0)} n^{1/2} + B_2^{(1)} n^{b} \nn\\
C_3 & = C_3^{(0)} n^{3/2} + C_3^{(1)} n^{c} \label{axionscaling}\\
{\rm with}\ &\ b< 0 \ ,\ c < 1 \ . \nn
\eea
As we will see, their $0^{{\rm th}}$ order scalings are fixed by requiring that their kinetic terms, as well as their scalar potential contributions, scale in the same manner as the rest of the action at LO.\footnote{One may be bothered by the fact that the axions do not scale at $0^{{\rm th}}$ order as their internal flux counterparts. Those should however be considered as independent quantities, because the above axions are 4d fluctuations. This can be seen through their associated field strength, which has components given by $\del_{\mu} B_2$ or $\del_{\mu} C_3$, i.e.~with one 4d leg, contrary to the 6d $H$ and $F_4$. This emphasizes the independence of those fields and of their $n$-scaling.} In the 4d DGKT solution, $B_2^{(0)}=C_3^{(0)}=0$, so one could as well ignore these fields. It will still be useful to have a 4d off-shell notion for this scaling, in particular in the potential. Regarding the $1^{{\rm st}}$ order, it would a priori be enough to require $b< \frac{1}{2} \ ,\ c < \frac{3}{2}$. The more stringent scalings above is a choice to have their $1^{{\rm st}}$ order contributions subdominant at NLO, as we will see. We are now ready to expand the warped 4d theory order by order.

\subsubsection{4d theory at LO}\label{sec:4dLO}

We first discuss $0^{{\rm th}}$ order and then $1^{{\rm st}}$ order fields contributions.

\begin{itemize}
  \item \bf{$0^{{\rm th}}$ order and LO}
\end{itemize}

We first consider the $0^{{\rm th}}$ order contributions of the various fields: those will contribute to the LO of the 4d action. We start with the kinetic terms in \eqref{4dactionfullywarped}: at $0^{{\rm th}}$ order in the fields, they boil down to
\beq
- \frac{M_p^2}{4}\bigg( {{\cal V}_{6\phi}^{(0)}}^{-2}\, (\del {\cal V}_{6\phi}^{(0)} )^2  - \frac{1}{2} \del_{\mu} g_{mn}^{(0)}\, \del^{\mu} g^{mn\, (0)} + |\del B_2^{(0)}|^2 + |\del C_{3}^{(0)}|^2 \, e^{2 \phi^{(0)}} \bigg) \ .
\eeq
Note that we neglected to consider the 4d metric. As mentioned around \eqref{4dmetricscaling}, the 4d metric in Einstein frame scales at $0^{{\rm th}}$ order as $n^{9/2}$. We deduce the following first part of the 4d action in terms of $0^{{\rm th}}$ order fields
\bea
 \int \d^{4} x \sqrt{|g_{4}|}\, \frac{M_p^2}{2} & \Bigg( {\cal R}_{4}^{(0)} \\
 - & \frac{1}{2}\bigg( {{\cal V}_{6\phi}^{(0)}}^{-2}\, (\del {\cal V}_{6\phi}^{(0)} )^2  - \frac{1}{2} \del_{\mu} g_{mn}^{(0)}\, \del^{\mu} g^{mn\, (0)} + |\del B_2^{(0)}|^2 + |\del C_{3}^{(0)}|^2 \, e^{2 \phi^{(0)}} \bigg) \Bigg) n^{-9/2} \ .\nn
\eea
This is the LO expression of this part of the action, as there is no other contribution. Note that here and in the following, it will not be necessary to expand the overall 4d volume given by $\sqrt{|g_4|}$.

We turn to the scalar potential. Using the independence  of $w^{(0)}, \tau^{(0)}, g_{mn}^{(0)}$ with respect to $y$, and the same for $F_6 + \dots$, the $F_6$ contribution \eqref{SF6} becomes at LO (given by $0^{{\rm th}}$ order fields)
\beq
{\cal S}_{F_6}= -\frac{M_p^2}{4} \int \d^4 x \sqrt{|g_4|}\, \Bigg[\frac{e^{4\phi}}{(2\pi \sqrt{\alpha'})^{-6} \int \d^6 y \sqrt{|g_6|}} \left| C_3 \w H + F_4 \w B_2 + \frac{1}{6}F_0\ B_2 \w B_2\w B_2\right|^2 \Bigg]^0\ ,\nn
\eeq
where the superscript ${}^0$ denotes the $0^{{\rm th}}$ order (with $n$ included). From this and \eqref{potw}, we deduce the following potential at LO
\bea
V = \frac{M_p^2}{2}  & \frac{e^{2\phi^{(0)}}}{(2\pi \sqrt{\alpha'})^{-6} \int \d^6 y \sqrt{|g_6^{(0)}|}} \, \Bigg( - e^{\phi} \sum_I\frac{T_{10}^I}{7} + \frac{1}{2} |H|^2 +  \frac{e^{2\phi}}{2}  \bigg[ F_0^2 +\big|F_0 B_2\big|^2\\
& +\left|F_4 +\frac{1}{2}F_0\ B_2 \w B_2\right|^2  + \left| C_3 \w H + F_4 \w B_2 + \frac{1}{6}F_0\ B_2 \w B_2\w B_2\right|^2 \bigg] \Bigg)^{(0)} n^{-9/2}\ .\nn
\eea
Contrary to the other pieces of the action, we will see that $1^{{\rm st}}$ order fields could have contributed to this LO scalar potential, but they actually do not; $0^{{\rm th}}$ order fields are thus enough. We verify that the complete 4d theory at LO scales as $n^{-9/2}$.

The LO 4d theory matches perfectly the one used in DGKT. Indeed, the kinetic terms correspond to the ones obtained for DGKT, given in \eqref{kinsax1} or \eqref{kinsax2} for the saxions and \eqref{kinax1} for the axions. The scalar potential agrees with the DGKT one given in \eqref{potgen0}. This was already mentioned in \cite{Junghans:2020acz}, even though we provided a more general analysis, including axions. As emphasized, it was enough to consider $0^{{\rm th}}$ order fields; we turn to $1^{{\rm st}}$ order contributions at LO, the fate of which is still very instructive.

\begin{itemize}
  \item \bf{$1^{{\rm st}}$ order and LO}
\end{itemize}

Some terms in the scalar potential of the warped 4d theory \eqref{potw} turn out to give the same scaling as above, namely $n^{-9/2}$, while contributing a priori with $1^{{\rm st}}$ order fields. In that case, $1^{{\rm st}}$ order fields can also be said to contribute at LO. There are three such terms: we show here that they all vanish, because they are total derivatives that integrate to zero, as briefly mentioned in \cite{Junghans:2020acz}.

The first one is the term in ${\cal R}_6$. Since $g_{mn}^{(0)}$ is independent of $y$, the $0^{{\rm th}}$ order of ${\cal R}_6$ vanishes, since it is built on derivatives of the metric. The next order will come from having a single $g_{mn}^{(1)}$ or $g^{mn \, (1)}$, acted upon by the two derivatives of ${\cal R}_6$; the other (inverse) metrics involved being $0^{{\rm th}}$ order. Using the definition of ${\cal R}_6$ and Christoffel symbols, we get schematically at $1^{{\rm st}}$ order
\beq
{\cal R}_6 \sim g^{mn \, (0)} \del \left( g^{pq \, (0)} \del g_{rs}^{(1)} \right) \, n^{-3/2} = \del \left( g^{mn \, (0)}  g^{pq \, (0)} \del g_{rs}^{(1)} \right) \, n^{-3/2} \ .
\eeq
The complete term in the potential \eqref{potw} then goes as follows
\beq
{\cal V}_{6\phi}^{-2}  \int \d^6 y \sqrt{|g_6|} \, e^{4A-2 \phi}\ {\cal R}_6 \sim \left[\frac{(2\pi \sqrt{\alpha'})^{12}\ e^{2\phi}}{\left( \int \d^6 y \sqrt{|g_6|}\right)^2} \right]^{(0)} \!\!\!\! \int \d^6 y \sqrt{|g_6^{(0)}|}\  \del \left( g^{mn \, (0)}  g^{pq \, (0)} \del g_{rs}^{(1)} \right) \, n^{-9/2} \ ,
\eeq
where the extra factors are all necessarily at $0^{{\rm th}}$ order. The scaling is indeed the LO one in $n^{-9/2}$, with a $1^{{\rm st}}$ order field. Since $|g_6^{(0)}|$ is independent of $y$, this becomes a total derivative, that we thus integrate to zero, so this term eventually does not contribute.

The second term is in $e^{-A} \Delta_6 e^{A}$. The same reasoning applies, where this time we use that $w^{(0)}$ is independent of $y$: we get
\beq
e^{-A} \Delta_6 e^{A} \sim \frac{1}{w^{(0)}} g^{mn \, (0)}\, \del \del w^{(1)} \ n^{-3/2} \ .
\eeq
This complete term in the potential has the same prefactor as before, giving the same overall scaling in $n^{-9/2}$. As above, it becomes a total derivative that we set to zero.

The third occurrence of a  $1^{{\rm st}}$ order field is through the source term. One may replace the $T_{10}^I$ or $T_{10} = \sum_I T_{10}^I $ (at $0^{{\rm th}}$ order) using the Bianchi identity, as in \eqref{potBI}. We then get the following LO term
\bea
& {\cal V}_{6\phi}^{-2} \int  e^{4A- \phi} \sum_{I} {\rm vol}_{||_I} \w \left( \d F_{2} - H \w F_{0} \right) \\
\sim & \left[\frac{(2\pi \sqrt{\alpha'})^{12}\ e^{3\phi}}{\left( \int \d^6 y \sqrt{|g_6|}\right)^2} \right]^{(0)} \!\!\!\! \int \sum_{I} {\rm vol}_{||_I}^{(0)} \w \left( \d F_{2}^{(1)} - H^{(0)} \w F_{0}^{(0)} \right) \ n^{-9/2} \ .
\eea
Since sources are wrapping cycles, $\d {\rm vol}_{||_I}^{(0)}=0$, so the $1^{{\rm st}}$ order contribution of $ F_{2}^{(1)} $ becomes a total derivative, which then integrates to zero.\\

It is important to note that these appearances of $1^{{\rm st}}$ order fields at LO coincide with 10d equations considered in \cite{Junghans:2020acz}. For instance the $F_2$ Bianchi identity that is solved involves $ F_{2}^{(1)} $. Similarly, as discussed in Section \ref{sec:HF4F6}, the Einstein equations involve $1^{{\rm st}}$ order contributions from the Ricci tensor and the warp factor, while fluxes contribute at $0^{{\rm th}}$ order. We mentioned a doubt in Section \ref{sec:F4again} on whether those equations should be considered as NLO or actually LO; from the 4d perspective here, we would rather choose the latter. Consistently with this choice, turning to the 4d NLO in the following, we will see appearing the next order equation that constrains the $F_4$-flux \eqref{eqF42}.

\subsubsection{4d theory at NLO and critical point}\label{sec:4dNLO}

We now develop the warped 4d theory at NLO in the $n$-expansion \eqref{expansionJ}; we only consider its scalar potential. Because of flux quantization discussed in Section \ref{sec:F4again}, we consider $H^{(1)}=0$; we leave at first the possibility of having $F_6^{(1)}$. We also consider $F_2^{(0)}=0$. We start with the first terms of the scalar potential \eqref{potw}, and provide the following schematic LO and NLO expansion
\beq
\left. \begin{array}{r}
{\cal V}_{6\phi}^{-2}  \int \d^6 y  \sqrt{|g_6|} \, e^{4A-2 \phi} \, \bigg( - {\cal R}_6 + 12 e^{-2A} (\del e^{A} )^2 + 8 e^{-A} \Delta_6 e^{A} - 4(\del \phi)^2 \\[0.1in]
 - e^{\phi} \sum_I\frac{T_{10}^I}{7} + \frac{1}{2} |H|^2 +  \frac{e^{2\phi}}{2}  \big[ F_0^2 +|F_2|^2  \big] \bigg) \end{array}
 \right|  \sim n^{-9/2} \left( 1 + n^{-1} \right) \ . \label{4dNLOcorr}
\eeq
This can be understood as follows. Most LO contributions come from $0^{{\rm th}}$ order fields; going to $1^{{\rm st}}$ order fields to get to NLO diminishes the scaling by $1/n$ (for the metric, the warp factor and the dilaton, which is what contributes here), hence the NLO scaling above. The other LO contributions come from $1^{{\rm st}}$ order fields, as total derivatives. For those terms (${\cal R}_6, e^{-A} \Delta_6 e^{A}$ and $\d F_2$),\footnote{For the source contributions, one may either use the definition of $T_{10}^I$ in terms of metrics, or replace it by the Bianchi identity; in both cases, the reasoning leads to the same scaling.} we have to be more cautious: there are two possibilities for the NLO. First it can be the product of two $1^{{\rm st}}$ order fields; the new $1^{{\rm st}}$ order field w.r.t. to LO is either a metric, a warp factor or a dilaton, so we reach the scaling indicated above. Second, the NLO can be due to a single $2^{{\rm nd}}$ order field times $0^{{\rm th}}$ order fields. In that case, the single $2^{{\rm nd}}$ order field has to be under a derivative, leading eventually to a total derivative contribution, as explained previously at LO. So such NLO contributions are set to zero. We conclude that the terms considered above have NLO contributions that scale as $n^{-11/2}$.\\

The remaining terms in the scalar potential depend on $F_4$, $F_6$ and the axions. For those, a threshold will rather be $n^{-10/2}$. For now on, we therefore restrict further the scaling of $F_4$ given in \eqref{expansionJ} towards
\beq
F_4 = F_4^{(0)} n + F_4^{(1)} n^{1-s_4}  \ ,\quad 0 < s_4 < \frac{1}{2} \ . \label{s4restr}
\eeq
This will allow us to get $F_4$ contributions dominant, and higher than $n^{-10/2}$. An important motivation for this restriction will be the topic of the next subsubsection, $\alpha'$-corrections, as those are argued in \cite{Junghans:2020acz} to appear at $n^{-10/2}$. Note that the following discussion of $F_4$, $F_6$ and axionic contributions is not done in \cite{Junghans:2020acz}; the above scaling restriction also does not include the value $s_4=\frac{1}{2}$ of \cite{Junghans:2020acz}, so we deviate from now on from that reference.

Let us start with the following term from \eqref{potw}, that we develop at LO and NLO schematically as follows
\bea
& {\cal V}_{6\phi}^{-2}  \int \d^6 y \sqrt{|g_6|} \, e^{4A-2 \phi} e^{2\phi}\ |F_2 + F_0 B_2|^2 \label{VaxF2}\\
& \sim \dots \times \left(  (F_0 B_2^{(0)})^2 \, n^{-9/2} +  2 F_2^{(1)} F_0 B_2^{(0)} \, n^{-10/2} + 2 F_0^2 B_2^{(0)} B_2^{(1)} \, n^{-10/2 + b} \right) + {\rm subdominant}  \ ,\nn
\eea
where omitted prefactors are $0^{{\rm th}}$ order fields. As anticipated, the restriction $b<0$ avoids contributions of $1^{{\rm st}}$ order axions, leaving as NLO contribution the term in $F_2^{(1)} F_0 B_2^{(0)} \, n^{-10/2}$. Interestingly, this term is linear in the axion.\footnote{Less schematically, this term can be written as being proportional to the 6-form $B_2^{(0)} \w *_6^{(0)} F_2^{(1)}$. It would be interesting to check whether this vanishes, given the solution \eqref{F21sol}.} As mentioned, this term will be subdominant to the $F_4$ terms, so we do not consider it.

We turn to the following $F_4$ terms, read from the warped 4d effective action \eqref{potw}: we develop it schematically at NLO and on our compactification ansatz
\bea
& {\cal V}_{6\phi}^{-2}  \int \d^6 y \sqrt{|g_6|} \, e^{4A-2 \phi} e^{2\phi}\ \left|F_4 +C_1 \w H + F_2 \w B_2 +\frac{1}{2}F_0\ B_2 \w B_2\right|^2 \label{VaxF4}
\\
& \sim \dots \times n^{-10/2} \left(  F_4^{(0)}  + \frac{1}{2}(F_0\, B_2^2)^{(0)} \right) \left(  F_4^{(1)} n^{1/2-s_4}  + F_2^{(1)}\, B_2^{(0)} n^{0} + F_0^{(0)}\, B_2^{(0)} B_2^{(1)} n^{ b} \right) + {\rm subd.}  \ ,\nn
\eea
where omitted prefactors are $0^{{\rm th}}$ order fields. The last two terms are of order $n^{-10/2}$ or lower, while the $F_4^{(1)}$ term is of higher level thanks to the restriction \eqref{s4restr}, as anticipated.

We are left with the contributions from ${\cal S}_{F_6}$ in \eqref{SF6}. We first get the scaling
\beq
{\cal S}_{F_6} \sim n^{-9/2-3} (1 + n^{-1}) \times \text{scaling of}\ (F_6+C_3\w H +F_4 \w B_2 + \frac{1}{2} F_2 \w B_2 \w B_2 + \frac{1}{6} F_0 B_2\w  B_2 \w B_2 )^2 \ ,\nn
\eeq
where the square does not involve the metric. At NLO, we get schematically
\beq
n^{-10/2-1} \left( C_3\! \w H + F_4 \!\w B_2 + \frac{1}{6}F_0\ B_2^3\right)^{\!(0)} \left( F_6^{(1)}\, n^{s_6}  + C_3^{(1)}\! \w H^{(0)} \, n^{c} + F_4^{(1)} \! \w B_2^{(0)}\, n^{3/2-s_4} \right) +\, {\rm subd.} \ ,\nn
\eeq
where several terms were shown to be subdominant (to the $F_4^{(1)}$ one) thanks to $b<0$ and the restriction on $s_4$ \eqref{s4restr}. As anticipated in \eqref{axionscaling}, the scaling $c<1$ also makes the $C_3^{(1)}$ term subdominant, and with a lower scaling than $n^{-10/2}$. On the contrary, \eqref{s4restr} allows the $F_4^{(1)}$ term to scale higher than $n^{-10/2}$. Finally, let us focus on the $F_6^{(1)}$ term, which also scales higher than $n^{-10/2}$ thanks to $s_6>1$ in \eqref{expansionJ}. This term is problematic because it generates a linear term for both $0^{{\rm th}}$ order axions, the rest of their potential being quadratic. This would change the $0^{{\rm th}}$ order axions in the solution, by giving them a non-zero value, contrary to \eqref{bx}. To avoid this, a first requirement is $s_6 < \frac{3}{2}$, making the term subdominant to LO; note this requirement can also be seen from a $(F_6^{(1)})^2$ term not written above. We further want this contribution to be lower at our NLO, so that the $0^{{\rm th}}$ order axions are left unchanged. This leads us to pick the more stringent scaling (with respect to $\frac{3}{2}$ and to \eqref{expansionJ})
\beq
s_6 < \frac{3}{2} -s_4 \ .
\eeq
This effectively makes $F_6^{(1)}$ disappear from the 4d theory at NLO, which amounts to set $F_6^{(1)}=0$; this is consistent with the discussion in Section \ref{sec:F4again}. The only dominant term in ${\cal S}_{F_6}$ is finally the one in $F_4^{(1)} $; considering no dependence on $y$ for $F_4^{(1)} $ as in \eqref{fluxescorrect}, ${\cal S}_{F_6}$ gets simplified as at $0^{{\rm th}}$ order.

To summarize, starting from the warped 4d theory \eqref{4dactionfullywarped}, using the scalings \eqref{expansionJ} and \eqref{axionscaling}, restricting further to
\beq
0 < s_4 < \frac{1}{2} \ ,\quad 1 < s_6 < \frac{3}{2} -s_4 \ ,
\eeq
we get the following scalar potential at NLO
\bea
& \frac{2}{M_p^2} \left[\frac{ \int \d^6 y \sqrt{|g_6|}}{(2\pi \sqrt{\alpha'})^{6}\ e^{4\phi}} \right]^{(0)} \ V_{{\rm NLO}} \nn\\
& = \left(F_4^{(0)} +\frac{1}{2}F_0^{(0)}\ B_2^{(0)} \w B_2^{(0)}\right)\cdot F_4^{(1)} \ n^{-9/2-s_4} \nn\\
& + \left( C_3\! \w H + F_4 \!\w B_2 + \frac{1}{6}F_0\ B_2\! \w B_2 \! \w B_2\right)^{\!(0)} \!\!\! \cdot \left(  F_4^{(1)} \! \w B_2^{(0)} \right) \ n^{-9/2-s_4}   \ . \label{potwNLO}
\eea
The dot is the contraction of forms with the $0^{{\rm th}}$ order 6d metric, from which the scaling has been extracted. We get an NLO scaling at $n^{-9/2-s_4}$, which as argued is above $n^{-10/2}$.\\

The axion terms in $V_{{\rm NLO}}$ are at least quadratic, so their critical point can consistently be set at $B_2^{(0)}= C_3^{(0)}=0$. The first derivative of $V_{{\rm NLO}}$ with respect to the saxions, at the critical point, therefore only involves the term without axion, that we denote $V_{{\rm NLO}}^1$
\beq
 \frac{2}{M_p^2}  V_{{\rm NLO}}^1 = \left[\frac{(2\pi \sqrt{\alpha'})^{6}\ e^{4\phi}}{ \int \d^6 y \sqrt{|g_6|}} \right]^{(0)}  F_4^{(0)}\cdot F_4^{(1)} \ n^{-9/2-s_4} \propto \frac{e^{4\phi}}{{\rm vol}^3} \sum_{i=1}^{3} e_i e_i^{(1)} v_i^2 \ ,
\eeq
where the last expression captures the dependence on the saxions. It is then straightforward to show that
\beq
\del_{\phi} V_{{\rm NLO}} = \del_{v_i} V_{{\rm NLO}} =0 \ \ \Rightarrow \ \ \forall i,\ e_i^{(1)}=0 \ \ \Rightarrow \ \ F_4^{(1)} =0 \ ,
\eeq
where we used that $\forall i,\ e_i \neq 0 $. As in 10d with the (N)NLO equations discussed in Section \ref{sec:F4again}, we obtain that the solution requires $F_4^{(1)} =0$ at this $n$-level.

So in the end, $V_{{\rm NLO}} = 0$ and there is no correction to the potential, the critical point or the mass spectrum above $n^{-10/2}$. We now turn to this $n$-level and the related $\alpha'$ corrections.

\subsubsection{Hitting the $\alpha'$-corrections}\label{sec:alpha}

Bulk $\alpha'$-corrections to the 4d theory are said in \cite{Junghans:2020acz} to arise at order $n^{-5}= n^{-10/2}$, while we recall from above that the LO order 4d theory is at $n^{-9/2}$. Once one reaches $\alpha'$-corrections, the 4d theory gets modified, so the perturbative corrections to the mass spectrum due to the $n$-expansion alone are not meaningful anymore. Those could still be computed, but $\alpha'$-corrections should as well be taken into account and this would then change the result. In addition, $\alpha'$-corrections are not fully known, so it seems out of sight to be able to fully compute corrections to the mass spectrum once these corrections appear. Interestingly, we tried above to get NLO corrections above the $\alpha'$-correction level of $n^{-10/2}$, thanks to a specific scaling $F_4^{(1)}$, but we eventually showed that such a correction to the solution is not admissible and enforces $F_4^{(1)} =0$. The 4d theory, the critical point and the mass spectrum are therefore robust against corrections, at least until that level.

It is interesting to revisit the arguments presented in \cite{Junghans:2020acz} regarding the bulk $\alpha'$-corrections. As is well-known in type II supergravities, those arise at ${\alpha'}^3$, corresponding to 8-derivative terms. They could involve a mixture of supergravity fluxes and metric contractions; however, fluxes typically lead to subdominant terms because of the extra inverse metrics necessary to contract their indices, and the dilaton factors on RR fluxes. A dominant $m$-derivative term is then rather given by considering only metric contractions and derivatives (e.g.~Riemann or Ricci tensors). An estimate of its scaling is schematically given by
\beq
(g^{-1})^{\frac{m}{2}}\, g^{-1}\, (\del)^m g \sim n^{-\frac{m}{4}} \ ,\label{gdelg}
\eeq
where the scaling is obtained from $0^{{\rm th}}$ order metrics. Such terms come with a $0^{{\rm th}}$ order prefactor $e^{2 \phi}/ \sqrt{|g_6|} \sim n^{-3}$. The overall scaling of a dominant $m$-derivative term is then given by $n^{-\frac{m+12}{4}} = n^{-\frac{9}{2}} \times n^{-\frac{m-6}{4}}$. This leads to the mentioned $n^{-10/2}$ for 8-derivatives, i.e.~the first $\alpha'$-correction. To this, let us add the string loop corrections, which as argued in \cite{Junghans:2020acz}, grow with $g_s^2 = \tau^{-2} \sim n^{-\frac{3}{2}}$. The $l$-loop then adds a factor $n^{-\frac{3l}{2}}$ to the LO scaling, bringing us here lower than $n^{-10/2}$ for the first correction at $l=1$.

While the above arguments prevent us from pursuing the computation of corrections to the 4d theory, we note however that the evaluation of the $\alpha'$-correction scalings could be refined. Indeed, the 2-derivatives term, given e.g.~by ${\cal R}_6^{(0)}$ should appear at $n^{-\frac{9}{2} + 1}$. We recall however that this does not happen because this contribution vanishes, since $0^{{\rm th}}$ order metrics are constant. This allows the LO to be at $n^{-\frac{9}{2}}$. The above argument, counting the scaling in \eqref{gdelg}, would face the same issue. To get a non-vanishing term, the metric acted upon by the derivatives should be non-constant, which only happens at $1^{{\rm st}}$ order (note that this point and the following reasoning hold true for the torus orbifold considered in this paper, but things could be different on a general Calabi-Yau manifold). Considering $\del g^{(1)}$ instead of $\del g^{(0)}$ lowers the scaling by $n^{-1}$. This is what happens for us at LO, where ${\cal R}_6^{(1)}$ appears instead of ${\cal R}_6^{(0)}$. But as argued in Section \ref{sec:4dLO}, this term becomes a total derivative that gets integrated to 0. The same would be true here, if all other fields are the $0^{{\rm th}}$ order metric and dilaton, which are constant. To get a contributing term, one needs a second $1^{{\rm st}}$ order (inverse) metric, which further reduces the scaling by an extra $n^{-1}$. We deduce the actual scaling to be
\beq
{\alpha'}^3 \int \frac{e^{2 \phi^{(0)}}}{\sqrt{|g_6^{(0)}|}}\ \left({g^{(0)}}^{-1}\right)^{4}\, {g^{(1)}}^{-1}\, \del^8 g^{(1)} \sim n^{-7} = n^{-\frac{9}{2}} \times n^{-\frac{5}{2}} \ ,
\eeq
much lower than $n^{-10/2}$. It then becomes relevant to compare this to the scaling of flux terms, or of string loop corrections.

\begin{figure}[H]
\centering
\vspace{-0.7in}
\includegraphics[width=0.7\textwidth]{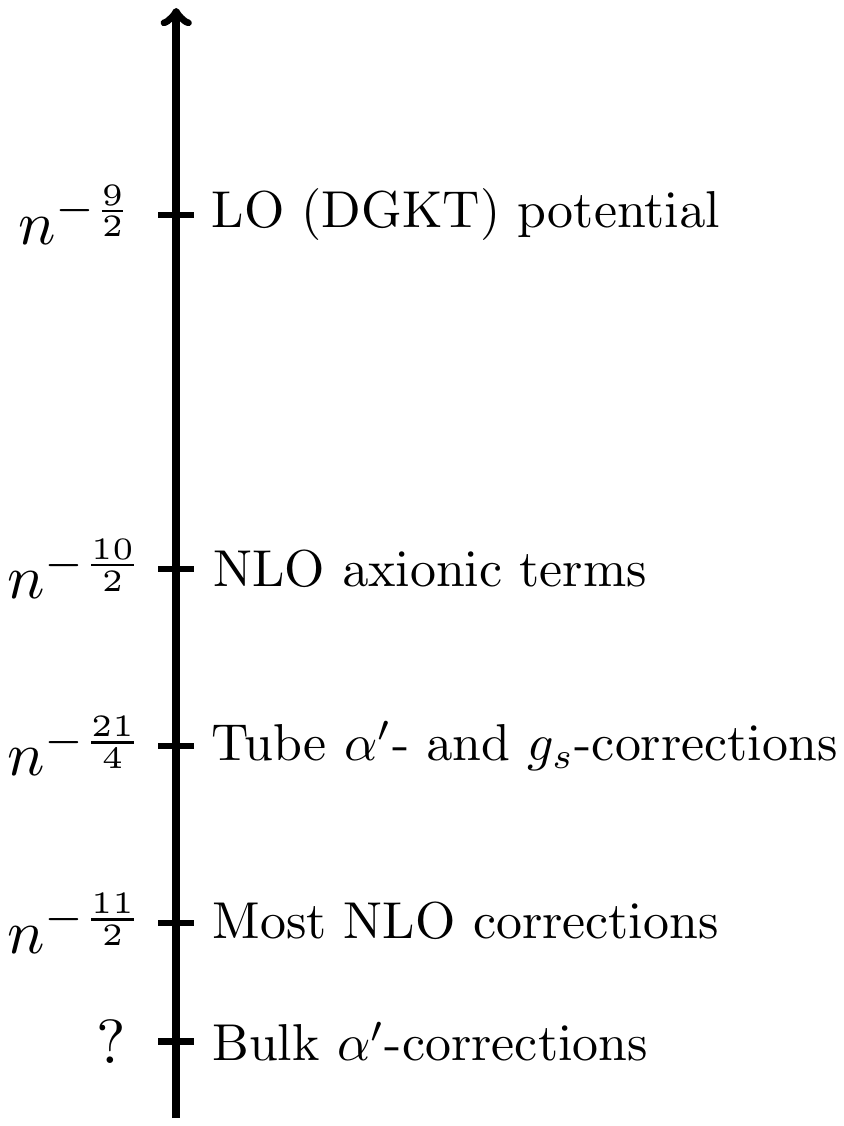}
\vspace{-0.6in}
\caption{Analogue to Figure \ref{fig:scalingintro}, with the level of bulk $\alpha'$-corrections reassessed for the torus orbifold. We refer to the main text for more details.}\label{fig:scalingmodif}
\end{figure}

If $\alpha'$-corrections appear at a lower level than $n^{-10/2}$, as indicated in Figure \ref{fig:scalingmodif}, then the previous perturbative corrections could be considered. Indeed, we recall that most of them were appearing at $n^{-11/2}$ as in \eqref{4dNLOcorr}. We also found few axionic terms appearing in \eqref{VaxF2}, \eqref{VaxF4} and possibly coming from ${\cal S}_{F_6}$, that would scale as $n^{-10/2}$ or lower; it would be interesting to reexamine those in detail. However, another type of corrections were discussed in \cite{Junghans:2020acz}: the $\alpha'$- and $g_s$-corrections in tube regions around the sources, where the backreaction cannot be neglected. It was argued that those would arise at $n^{-21/4}$, so very close to the previously mentioned $n^{-10/2}$; we refer to the discussion in \cite{Junghans:2020acz} about these corrections. A more advanced study then seems required to determine the precise $n$-level of the various corrections, including ours. The 4d theory, the critical point and the mass spectrum appear at first to be robust against corrections until they hit $\alpha'$- or $g_s$-corrections, but as suggested, a more thorough analysis of the various corrections involved would be welcome.

\section{New DGKT extension}\label{sec:newextension}

As explained at the beginning of Section \ref{sec:warpDGKT}, the 10d DGKT solution admits an unbounded discretized parameter $n$, common to the three components of $F_4$. This parameter allows to consider extensions of this solution as a perturbative expansion in $n$: the $0^{{\rm th}}$ order is nothing but the 10d DGKT solution, and corrections to the 10d fields appear at $1^{{\rm st}}$ order. A first $n$-expansion was given in \eqref{expansionJ}, generalising the one of \cite{Junghans:2020acz}, and was studied in the previous section. We consider in this section a new $n$-expansion, given as follows in terms of the fields defined in Appendix \ref{ap:conv}
\bea
g_{\mu\nu} & = g_{\mu\nu}^{(0)}\ n^{9/2} + g_{\mu\nu}^{(1)}\ n^{9/2-g} \ ,\ 0 < g < 1 \nn\\
g^{\mu\nu} & = g^{\mu\nu\, (0)}\ n^{-9/2} + g^{\mu\nu\, (1)}\ n^{-9/2-g} \ ,\nn\\
g_{mn} & = g_{mn}^{(0)} n^{1/2} + g_{mn}^{(1)} n^{1/2-g} \ , \nn\\
g^{mn} & = g^{mn\, (0)} n^{-1/2} + g^{mn\, (1)} n^{-1/2-g} \nn\\
e^A l= w & = w^{(0)} n^{3/4} + w^{(1)} n^{3/4-w} \ ,\  0 < w <1 \nn\\
e^{-\phi} = \tau & = \tau^{(0)} n^{3/4} + \tau^{(1)} n^{3/4-t} \ ,\  0 < t <1 \label{expansionnew}\\
F_0 & = F_0^{(0)} \nn\\
F_2 & = F_2^{(1)} n^{0} \ , \nn\\
H & = H^{(0)} n^{0} + H^{(1)} n^{-s_h} \ ,\ 0 < s_h < 1/2 \ ,  \nn \\
F_4 & = F_4^{(0)} n + F_4^{(1)} n^{1-s_4}  \ ,\ 0 < s_4 < 1/2 \ ,\nn \\
F_6 & = F_6^{(1)} n^{s_6} \ ,\ 1< s_6 <2-s_h \ ,\ s_6 <2-s_4 \nn
\eea
Once again, $0^{{\rm th}}$ order fields match the DGKT solution, with $F_2^{(0)}=0, F_6^{(0)}=0 $. We recall that $g_{\mu\nu}^{(0)}, g_{mn}^{(0)}, w^{(0)}, \tau^{(0)}$ are independent of $y$. The same holds for $F_0^{(0)}, H^{(0)},  F_4^{(0)}$, which are quantized. The scalings chosen at $1^{{\rm st}}$ order make the corresponding fields and corrections dominant over the scalings of \cite{Junghans:2020acz}, except for $F_2$. The $1^{{\rm st}}$ order solution of \cite{Junghans:2020acz} might then be recovered at a subdominant order, but we will not investigate this. Further motivations for the scalings chosen will appear later in this section. We will use this new $n$-expansion in the warped compactification 10d setting of Appendix \ref{ap:conv}, and then in the corresponding 4d theory. We will see how the solution and the mass spectrum get corrected.\\

Before doing so, let us specify more the compactification ansatz for the $1^{{\rm st}}$ order fields. We first take $g_{\mu\nu}^{(1)}, g_{mn}^{(1)}, w^{(1)}, \tau^{(1)}$ to be independent of $y$. Regarding the metric, we consider a much more restricted ansatz, respecting the orbifold and orientifold constraints: similarly to the $0^{{\rm th}}$ order metric in \eqref{6dmetric}, we choose at $1^{{\rm st}}$ order
\beq
g_{11}^{(1)}= g_{22}^{(1)} \ ,\ \ g_{33}^{(1)}= g_{44}^{(1)} \ , \ \ g_{55}^{(1)}= g_{66}^{(1)}  \ ,\ \ g_{mn}^{(1)}=0 \ {\rm for}\ m\neq n \ .
\eeq
We introduce accordingly the quantities $v_1^{(1)}, \ v_2^{(1)},\ v_3^{(1)}$, independent of $y$, together with the $0^{{\rm th}}$ order ones $v_i^{(0)}=v_i/n^{1/2}$; none of those carry any $n$. They obey
\beq
i=1,2,3:\quad \frac{v_i^{(1)}}{v_i^{(0)}} \equiv \frac{g_{2i-1,2i-1}^{(1)}}{g_{2i-1,2i-1}^{(0)}} = \frac{g_{2i,2i}^{(1)}}{g_{2i,2i}^{(0)}} \ .
\eeq
We finally introduce $v^{(0)}= \frac{|e_i|}{n} v_i^{(0)}  = v / n^{3/2}$ (without sum on $i$), which also carries no $n$. We introduce as well for convenience $v^{(1)}$
\beq
v^{(1)} = \frac{|e_i|}{n}\, v_i^{(1)} \ \text{(no sum)} \ \Rightarrow \ \frac{v_i^{(1)}}{v_i^{(0)}} = \frac{v^{(1)}}{v^{(0)}} \ . \label{vdef}
\eeq
Introducing such a $v^{(1)}$ is a priori a further restriction on the possible solutions, as it imposes an isotropy among the $i$, but we choose to look here for such ``simple'' solutions. We comment more on this ansatz at the end on the Section \ref{sec:interpret}.

The fact $g_{mn}^{(1)}$ is diagonal has several technical implications. First, any component $g_{mn}^{(1)}$ can be viewed as proportional to $g_{mn}^{(0)}$ with a factor independent of $y$, namely a ratio $\frac{v_i^{(1)}}{v_i^{(0)}}$. As a consequence, the Hodge star at $1^{{\rm st}}$ order $*_6^{(1)}$ (with one $g_{mn}^{(1)}$ or $g^{mn\, (1)}$) on any form is proportional to $*_6^{(0)}$. Second, we can rewrite the metric expansion as follows, knowing that the components are diagonal and using that $g^{mn \, (1)} = -g^{mp \, (0)} g_{pq}^{(1)} g^{qn \, (0)}$
\beq
g_{mn} = g_{mn}^{(0)}\, n^{1/2} \left( 1 + \frac{g_{mn}^{(1)}}{g_{mn}^{(0)}}\, n^{-g} \right) \ ,\quad g^{mn} = g^{mn\, (0)}\, n^{-1/2} \left( 1 - \frac{g_{mn}^{(1)}}{g_{mn}^{(0)}} \, n^{-g} \right) \ ,
\eeq
for fixed $m=n$. Finally, the following quantity will often appear, and it gets simplified
\beq
\sum_{m=1}^6 \frac{g_{mm}^{(1)}}{g_{mm}^{(0)}} = 2 \sum_{i=1}^3 \frac{v_i^{(1)}}{v_i^{(0)}} = 6\, \frac{v^{(1)}}{v^{(0)}}\ . \label{sumg1g0}
\eeq
We now have all tools to consider the $n$-expansion \eqref{expansionnew} with the specified compactification ansatz in 10d and in 4d.

\subsection{10d extended solution}

\subsubsection{Equations at NLO and flux quantization}

We consider here the complete 10d equations presented in Appendix \ref{ap:conv} and expand them order by order. As mentioned in Section \ref{sec:warpDGKT}, $0^{{\rm th}}$ order fields should contribute to LO equations, and satisfy them as being the DGKT solution. We are then interested in NLO with $1^{{\rm st}}$ order fields. We start with the flux equations of motion \eqref{fluxeq0} - \eqref{fluxeq} and focus first on $H,F_4, F_6$. Thanks to the above ansatz, in particular the independence on $y$, and the fact that $H^{(0)}, F_4^{(0)}$ are harmonic, we get at NLO
\bea
\d(  *_6^{(0)} F_4^{(1)} )\, n^{2-s_4} + H^{(0)} \w *_6^{(0)} F_6^{(1)}\, n^{s_6} = 0  &\\
\d(  *_6^{(0)} F_6^{(1)} ) = 0 &\\
e^{- 2\phi^{(0)}} \d(*_6^{(0)} H^{(1)} )\, n^{2-s_h} - F_4^{(0)} \w *_6^{(0)} F_6^{(1)} \, n^{s_6}  = 0 &
\eea
where we used $s_6>1$ from \eqref{expansionnew}. Using as well $s_6 < 2-s_4,\, s_6 < 2-s_h$, the $F_6$ terms are then subdominant, making the fluxes $H^{(1)}, F_4^{(1)}, F_6^{(1)}$ co-closed w.r.t.~the $0^{{\rm th}}$ order metric. Looking now at the Bianchi identities \eqref{BIs} at NLO, it is easy to see that they require these forms to be closed. We conclude that {\sl $H^{(1)}, F_4^{(1)}, F_6^{(1)}$ are harmonic w.r.t.~the $0^{{\rm th}}$ order metric}.

This situation has already been encountered with the previous extension, and discussed in Section \ref{sec:F4again}. These harmonic fluxes have to be quantized. An ansatz for them, compatible with the orbifold and orientifold, was given in \eqref{fluxescorrect}, followed by the quantization of their flux numbers. As argued there, this together with the large $n$ expansion led to the requirement in \eqref{H1=0} of a vanishing $H^{(1)}$. The $H$-flux e.o.m.~additionally led, given the scalings chosen, to having a vanishing $F_6^{(1)}$. The reasoning holds true here so we consider from now on
\beq
H^{(1)} = 0 \ ,\quad F_6^{(1)} = 0 \ .
\eeq

We finally turn to the $F_2$ equations. Simply because both $H$ and $F_4$ components are along the same directions at $0^{{\rm th}}$ and $1^{{\rm st}}$ order (see \eqref{fluxescorrect}), one has $H \w *_6 F_4 =0 $. So one gets at NLO the equations
\bea
\d(  *_6^{(0)} F_2^{(1)} ) &= 0 \\
\d F_{2}^{(1)} - H^{(0)} \w F_{0}^{(0)} &= \sum_{I} \frac{T_{10}^I}{7} {\rm vol}_{\bot_I}
\eea
None of our corrections (except $F_2^{(1)}$) are $y$ dependent, so we will not be able to solve the dilaton or Einstein equations with a localized source contribution, defined in Appendix \ref{ap:conv}. So we have to smear our source contributions, and the solution to be obtained will be a smeared correction to DGKT. Note that smeared or not, $\frac{T_{10}^I}{7} {\rm vol}_{\bot_I}$ is not metric dependent and is only counting the number of $O_6$, so this quantity does not get corrected by our $1^{{\rm st}}$ order fields. A first way to solve the above equations is to take this source contribution as smeared and then consider a harmonic $F_2^{(1)}$ (possibly vanishing). Another option could be to maintain a localized Bianchi identity, and replace $H^{(0)} \w F_{0}^{(0)}$ for the smeared source contribution as done in \cite{Junghans:2020acz}; in that case, the above equations completely decouple from the rest of the equations. As we will see, there is no other $F_2^{(1)}$ contribution to the other equations at NLO, so in the end we do not need to bother about this field. In the 4d theory at NLO, it will also not appear.\\

Coming back to the source contributions $T_{10}^I$ appearing in the dilaton and Einstein equations, those only get corrected due to corrections of the metric. At $0^{{\rm th}}$ order, one could trade $\sum_I T_{10}^I$ for $T_{10}$ and use a universal dependence on the metric: those quantities all go as $1/\sqrt{g_{\bot}}$, where $\sqrt{g_{\bot}} \sim |g_6|^{1/4}$. Since corrections are only those of the metric, then the $1^{{\rm st}}$ order correction to this source contribution is easily obtained from this metric dependence: we compute using \eqref{sumg1g0}
\beq
T_{10} = T_{10}^{(0)}\, n^{-3/4} \left( 1 -\frac{3}{2}\, \frac{v^{(1)}}{v^{(0)}}\, n^{-g} \right) \ , \label{T10corr}
\eeq
where the $0^{{\rm th}}$ order DGKT value is $T_{10}^{(0)} n^{-3/4}$. Similarly, following the discussion around \eqref{Tijdef}, we get the relation $T_{mn}= \frac{1}{2}g_{mn}\frac{T_{10}}{7}$. From this, we get the following corrected expression
\beq
T_{mn}= \frac{1}{2}\, g_{mn}^{(0)}\, \frac{T_{10}^{(0)}}{7}\, n^{-1/4} \left( 1 - \frac{1}{2} \frac{v^{(1)}}{v^{(0)}} n^{-g}  \right) \ . \label{Tmncorr}
\eeq
These expressions will provide a solution at $1^{{\rm st}}$ order to the remaining equations.\\

We turn to the dilaton and the Einstein equations given in \eqref{eq1}-\eqref{eq3}, in view of developing them at NLO and solving them. First, using the independence of the metric, dilaton and warp factor w.r.t.~$y$, true for us both at $0^{{\rm th}}$ and $1^{{\rm st}}$ order, the equations simplify to the following ones
\bea
0 = &\, 2 e^{-2A} {\cal R}_4^S - |H|^2 + e^{\phi } \frac{T_{10}}{7} \ ,\\
0 = &\, - 4 |H|^2 - e^{2\phi } \sum_{q=0}^6 (q-1) |F_q|^2   +3 e^{\phi}  \frac{T_{10}}{7} \ ,\\
0=  & \, \frac{1}{2} H_{mpq}H_n^{\ \ pq} + \left(e^{2\phi } F_{2\ mp}F_{2\ n}^{\ \ \ \ p} +\frac{e^{2\phi }}{3!} F_{4\ mpqr}F_{4\ n}^{\ \ \ pqr} \right)  + e^{\phi}T_{mn} \nn\\
&   +  \frac{g_{mn}}{8} (\sum_{q=0}^6 (1-q) e^{2\phi }|F_q|^2 + 8  e^{2\phi }|F_6|^2 - 2 |H|^2  - e^{\phi} T_{10} ) \ .
\eea
Those match the equations \eqref{dilatoneom} - \eqref{6dEinstein} solved in 10d by DGKT, showing once again the compatibility when using only $0^{{\rm th}}$ order quantities, corresponding to LO. When developing these equations at NLO, $1^{{\rm st}}$ order fields must appear.

The first equation above, the dilaton e.o.m., need not be solved as it only defines $e^{-2A}{\cal R}_4^S$. The 4d cosmological constant is related to ${\cal R}_{4}$ the Ricci scalar of the 4d Einstein frame metric. On-shell, the two Ricci scalars are related by ${\cal R}_4^S = {\cal V}_{6\phi}\, {\cal R}_{4}$, where ${\cal V}_{6\phi}$ is defined in \eqref{V6p}. We then rewrite the above equation as follows
\beq
{\cal R}_{4} = (2\pi \sqrt{\alpha'})^{6} \  \frac{ e^{2 \phi}}{\int \d^6 y \sqrt{|g_6|}}\, \left( \frac{1}{2} |H|^2 - \frac{1}{2} e^{\phi } \frac{T_{10}}{7} \right) \ , \label{R4Ecorrect}
\eeq
and will consider later its expansion at NLO.

We turn to the other two equations to solve. Using that $F_6=0$ and verifying that $F_2$ contributions are subdominant (to the $F_4$ ones), one is left with the following equations to expand and solve
\bea
0 = &\, - 4 |H|^2 + e^{2\phi } ( |F_0|^2 - 3|F_4|^2)   +3 e^{\phi}  \frac{T_{10}}{7} \ , \label{eq11}\\
0=  &\, \frac{1}{2} H_{mpq}H_n^{\ \ pq} + \frac{e^{2\phi }}{3!} F_{4\ mpqr}F_{4\ n}^{\ \ \ pqr}  + e^{\phi}T_{mn} \nn\\
&   +  \frac{g_{mn}}{8} ( e^{2\phi }|F_0|^2 - 3e^{2\phi }|F_4|^2 - 2 |H|^2  - e^{\phi} T_{10} ) \ . \label{eq12}
\eea
This leaves three $1^{{\rm st}}$ order field contributions, from the metric, the dilaton and $F_4$. Having them all to contribute at the same level at NLO requires to fix their scalings in \eqref{expansionnew} as
\beq
t=g=s_4 \ , \label{sameexpo}
\eeq
which constrains the range of $g,t$. With this choice, we are now ready to expand and solve these remaining equations at NLO.

\subsubsection{Solution}

We have proposed in \eqref{expansionnew} a new $n$-expansion at $1^{{\rm st}}$ order beyond the DGKT solution. After \eqref{expansionnew}, we have specified further our 10d solution ansatz. We have started developing and solving the 10d equations at NLO, restricting further our solution. We are left with the Einstein equations \eqref{eq11} and \eqref{eq12} to expand at NLO and solve, which we do in the following. Our variables are three $1^{{\rm st}}$ order fields: the metric, the dilaton and $F_4$, whose scaling are now related as in \eqref{sameexpo}. Let us make already the following observation on the $F_4$-flux \eqref{fluxescorrect}: due to equation \eqref{eq12}, a solution with the above ansatz will need to satisfy
\beq
\forall \, i:\quad \frac{e_i^{(1)}}{e_i} \equiv e^{(1)} \ ,
\eeq
meaning that $e_i^{(1)}$ is proportional to $e_i$ with the same proportionality factor $\forall\, i$, given by the symbol $e^{(1)}$. This isotropy is related to the isotropy of the metric components, and the symbol $v^{(1)}$ introduced in \eqref{vdef}. We now find a solution with this simplified solution ansatz.\\

We start with equation \eqref{eq11}. We consider it at NLO, making $1^{{\rm st}}$ order fields appear. We reconstruct $0^{{\rm th}}$ order quantities that we denote with a superscript ${}^0$ (i.e.~containing the $n$ dependence). We get at NLO the following quantities
\bea
& e^{-2\phi} |H|^2 = \left( e^{-2\phi} |H|^2 \right)^0 \left(2 \frac{\tau^{(1)}}{\tau^{(0)}} - 3 \frac{v^{(1)}}{v^{(0)}} \right) n^{-g} \ ,\ e^{-\phi} T_{10} = \left( e^{-\phi} T_{10} \right)^0 \left(\frac{\tau^{(1)}}{\tau^{(0)}} - \frac{3}{2} \frac{v^{(1)}}{v^{(0)}} \right) n^{-g} \nn\\
& |F_4|^2 =  \left( |F_4|^2 \right)^0 \left( - 4 \frac{v^{(1)}}{v^{(0)}} + 2\, e^{(1)} \right) n^{-g} \ ,
\eea
where we used \eqref{T10corr}. Note that $F_0$ does not contribute at NLO when considering $e^{-2\phi}\times \eqref{eq11}$. Focusing then on $\left(e^{2\phi}\right)^0 e^{-2\phi}\times \eqref{eq11}$, we use the $0^{{\rm th}}$ order results \eqref{H2} and \eqref{Fq2} to eventually rewrite \eqref{eq11} at NLO as
\beq
-\frac{1}{2}\frac{\tau^{(1)}}{\tau^{(0)}} + 21 \frac{v^{(1)}}{v^{(0)}} - \frac{81}{8} e^{(1)}  = 0 \ .\label{1stordereq1}
\eeq

We turn to \eqref{eq12} and proceed similarly. We get the following quantities at NLO
\bea
& e^{-2\phi} \frac{1}{2} H_{mpq}H_n^{\ pq}  = \left( e^{-2\phi} \frac{1}{2} |H|^2 g_{mn} \right)^0 \left(2 \frac{\tau^{(1)}}{\tau^{(0)}}  - 2 \frac{v^{(1)}}{v^{(0)}} \right) n^{-g}  \\
&  \frac{1}{3!} F_{4\ mpqr}F_{4\ n}^{\ \ \ pqr} = \left( \frac{2}{3} |F_4|^2 g_{mn} \right)^0 \left( 2\, e^{(1)} - 3 \frac{v^{(1)}}{v^{(0)}}  \right) n^{-g} \\
& e^{-\phi} T_{mn} = \left( \frac{1}{2}\, g_{mn}\,  e^{-\phi}\, \frac{T_{10}}{7} \right)^0 \left( \frac{\tau^{(1)}}{\tau^{(0)}} - \frac{1}{2} \frac{v^{(1)}}{v^{(0)}} \right) n^{-g} \\
& \frac{g_{mn}}{8} \left( - e^{-\phi} T_{10} - 2e^{-2\phi} |H|^2 + |F_0|^2 - 3 |F_4|^2 \right) \nn\\
& = \left( \frac{g_{mn}}{8} \right)^0 \Big[\frac{\tau^{(1)}}{\tau^{(0)}} \left( - e^{-\phi} T_{10} - 4 e^{-2\phi} |H|^2 \right)^0 + \frac{v^{(1)}}{v^{(0)}} \left( \frac{1}{2} e^{-\phi} T_{10} + 4 e^{-2\phi} |H|^2 + F_0^2 + 9 |F_4|^2 \right)^0 \nn\\
& \hspace{3.5in} + e^{(1)} \left( - 6 |F_4|^2 \right)^0 \Big] n^{-g}
\eea
where we used \eqref{Tmncorr}. These are all ingredients needed to rewrite $\left(e^{2\phi}\right)^0 e^{-2\phi}\times \eqref{eq12}$. To that end, we use again the $0^{{\rm th}}$ order results \eqref{H2} and \eqref{Fq2}, and rewrite \eqref{eq12} at NLO as
\beq
\frac{\tau^{(1)}}{\tau^{(0)}} + 3 \frac{v^{(1)}}{v^{(0)}} - \frac{9}{4} e^{(1)}  = 0 \ .\label{1stordereq2}
\eeq
Combining \eqref{1stordereq2} with \eqref{1stordereq1}, we get the following 10d solution
\beq
\frac{\tau^{(1)}}{\tau^{(0)}} = \frac{3}{4} e^{(1)} \ ,\quad \frac{v^{(1)}}{v^{(0)}} = \frac{1}{2} e^{(1)} \ .\label{solution}
\eeq

We use the above to compute the correction to the cosmological constant. From \eqref{R4Ecorrect}, we first obtain at LO and NLO
\beq
\frac{{\cal R}_{4} - {\cal R}_{4}^0}{(2\pi \sqrt{\alpha'})^{6}} = \left(\frac{ e^{2 \phi}}{\int \d^6 y \sqrt{|g_6|}} \right)^0\, \Big[\frac{\tau^{(1)}}{\tau^{(0)}} \left(- |H|^2 + \frac{3}{2} e^{\phi} \frac{T_{10}}{7} \right)^0 + \frac{v^{(1)}}{v^{(0)}} \left( -3 |H|^2 + \frac{9}{4} e^{\phi} \frac{T_{10}}{7} \right)^0 \Big] n^{-g} \ .\nn
\eeq
Using the $0^{{\rm th}}$ order results \eqref{H2}, \eqref{Fq2}, as well as the expressions for the 4d Ricci scalar \eqref{R4s}, \eqref{relR4}, and the solution \eqref{solution}, we finally obtain
\beq
{\cal R}_{4} = {\cal R}_{4}^0 \left( 1 - \frac{9}{2}\, n^{-g}\ e^{(1)} \right) \ .
\eeq
Note that this is compatible with having a $1^{{\rm st}}$ order correction to the 4d metric in $n^{-g}$, as indicated with the initial ansatz \eqref{expansionnew}.\\

We summarize the 10d solution obtained at NLO: it is given by the expansion \eqref{expansionnew} with all $1^{{\rm st}}$ order fields independent of $y$, except $F_2^{(1)}$, and
\bea
& g_{11}^{(1)}= g_{22}^{(1)} \ ,\ \ g_{33}^{(1)}= g_{44}^{(1)} \ , \ \ g_{55}^{(1)}= g_{66}^{(1)}  \ ,\ \ g_{mn}^{(1)}=0 \ \ {\rm for} \ \ m\neq n \ ,\nn \\
& i=1,2,3:\quad \frac{g_{2i-1,2i-1}^{(1)}}{g_{2i-1,2i-1}^{(0)}} = \frac{g_{2i,2i}^{(1)}}{g_{2i,2i}^{(0)}}= \frac{v_i^{(1)}}{v_i^{(0)}} = \frac{v^{(1)}}{v^{(0)}}  \ \ {\rm with}\ \ v^{(0,1)} = v_i^{(0,1)} \frac{|e_i|}{n} \ \ \text{(no sum)}\ ,\ v_i^{(0)}= \frac{v_i}{n^{1/2}} \nn\\
& g^{mn (1)}= - g^{mn (0)}\, \frac{v^{(1)}}{v^{(0)}} \ , \nn\\
& T_{10} = \left( T_{10} \right)^0 \left( 1 -\frac{3}{2}\, \frac{v^{(1)}}{v^{(0)}}\, n^{-g} \right) \ ,\ T_{mn}= \left( \frac{1}{2}\, g_{mn}\, \frac{T_{10}}{7} \right)^0 \left( 1 - \frac{1}{2} \frac{v^{(1)}}{v^{(0)}} n^{-g}  \right)  \ ,\nn\\
& H^{(1)}=0 \ ,\ F_6^{(1)}=0 \ ,\ F_4 = \sqrt{2}\, e_i\, \left(1 + e^{(1)}\, n^{-s_4} \right)\, \tilde{w}^i \ \ {\rm with}\ \  e_i \sim n \ ,\nn\\
& 0 < g=t=s_4 < \frac{1}{2} \ ,\\
& \frac{\tau^{(1)}}{\tau^{(0)}} = \frac{3}{4} e^{(1)} \ ,\quad \frac{v^{(1)}}{v^{(0)}} = \frac{1}{2} e^{(1)} \ ,\nn\\
& {\cal R}_{4} = {\cal R}_{4}^0 \left( 1 - \frac{9}{2}\, n^{-g}\ e^{(1)} \right) \ . \label{R410dsol}
\eea
In addition, $F_2^{(1)}$ and $w^{(1)}$ drop out of the equations at NLO; they can equivalently be set to zero at this order. Contrary to the $n$-expansion \eqref{expansionJ} studied in the previous section, we obtain here a non-vanishing correction in $F_4$ that gets related to that of the metric and the dilaton. The scalings allow precisely here such a relation. Before commenting more on this solution, we now turn to its 4d description.

\subsection{4d description}

We consider here the new $n$-expansion \eqref{expansionnew} and compactification ansatz previously detailed within a 4d effective theory. We start from the general warped 4d effective theory given in \eqref{4dactionfullywarped} and will expand it accordingly at NLO. To that end, we need, as for the other expansion, the scaling of the 4d axions. We give it as follows
\bea
B_2 & = B_2^{(0)} n^{1/2} + B_2^{(1)} n^{1/2-b} \ , \ b>0 \ ,\label{axionscalingnew}\\
C_3 & = C_3^{(0)} n^{3/2} + C_3^{(1)} n^{3/2-c} \ ,\ c>0 \ .\nn
\eea
We then have all ingredients to expand the 4d theory at NLO. From the resulting 4d theory, we will recover the previous 10d solution \eqref{solution} as a 4d critical point. We will finally investigate the corrections to the mass spectrum.

\subsubsection{4d theory}

We start with the warped 4d effective theory in \eqref{4dactionfullywarped} and expand it at LO and NLO. We use the expansion \eqref{expansionnew} as well as the above axion scalings, together with the independence of most of the fields w.r.t.~$y$. We perform as explained in Section \ref{sec:4dLO} an integration to zero of the $F_2$ total derivative, and use $F_6=0$. We get at first the following 4d theory, to be further expanded at LO and NLO
\bea
\int \d^{4} x \sqrt{|g_{4}|} & \Bigg( \frac{M_p^2}{2} {\cal R}_{4} - V \\
& - \frac{M_p^2}{4}\bigg( {\cal V}_{6\phi}^{-2}\, (\del {\cal V}_{6\phi} )^2  - \frac{1}{2} \del_{\mu} g_{mn}\, \del^{\mu} g^{mn} + |\del B_2|^2 + e^{2\phi} |\del C_{3}|^2
\bigg)  \Bigg)  \ , \nn
\eea
with
\bea
& \frac{2}{M_p^2} \frac{1}{(2\pi \sqrt{\alpha'})^{6}} V \\
= &\,   \frac{e^{2\phi} }{ \int \d^6 y \sqrt{|g_6|} } \, \Bigg(  \frac{1}{2} |H|^2 +  \frac{e^{\phi} }{ \int \d^6 y \sqrt{|g_6|} } \int  {\rm vol}_{||} \w  H \w F_{0} + \frac{e^{2\phi}}{2}  \bigg[ F_0^2 +\big|F_0 B_2\big|^2  \nn \\
& \qquad \qquad\quad  +\left|F_4  +\frac{1}{2}F_0\ B_2 \w B_2\right|^2 + \left| C_3\w H +F_4 \w B_2 + \frac{1}{6} F_0 B_2\w  B_2 \w B_2 \right|^2 \bigg] \Bigg) \nn
\eea
where ${\rm vol}_{||}$ is meant here generally, to possibly include metric corrections. The quantity $\frac{e^{\phi} }{ \int \d^6 y \sqrt{|g_6|} } \int  {\rm vol}_{||} \w  H \w F_{0} $ could also be traded for $-e^{\phi}\,  \frac{T_{10}}{7}$. Note that as explained in Section \ref{sec:4dNLO}, the $F_2$ NLO contributions are subdominant to the $F_4$ ones, so they have been already dropped.

We further consider the compactification ansatz, allowing us to introduce the 4d fields. Respecting the orbifold and orientifold projections, we take the following ansatz (in line with 10d considerations)
\bea
i=1,2,3:\quad & g_{2i-1,2i-1} = g_{2i,2i} = 2 (\kappa \sqrt{3})^{1/3}\, v_i \ ,\\
& B_2 = \sum_{i=1}^3 b_i w^i \ , \quad C_3 = \sqrt{2}\, \xi\, \alpha_0 \ , \nn
\eea
where together with $\phi$, the above $v_i, b_i, \xi$ are our 4d fields. We slightly abuse of notations: those fields were used already for DGKT in 4d, but we now consider them to contain both $0^{{\rm th}}$ and $1^{{\rm st}}$ order contributions. In agreement with the above $n$-expansions and notations, they can be written as
\bea
& v_i = v_i^{(0)} n^{1/2} + v_i^{(1)} n^{1/2-g} \ ,\ 0 < g < 1/2 \label{viexp}\\
& e^{-\phi} = e^{-\phi^{(0)}} n^{3/4} + e^{-\phi^{(1)}} n^{3/4-g} \nn\\
& b_i = b_i^{(0)} n^{1/2} + b_i^{(1)} n^{1/2-b} \nn\\
& \xi = \xi^{(0)} n^{1/2} + \xi^{(1)} n^{1/2-c} \ .\nn
\eea
We also detail the background fluxes: as in the 10d solution ansatz, $F_0, H$ only have a $0^{{\rm th}}$ order contribution, contrary to $F_4$
\beq
F_0 = -\sqrt{2}\,m_0 \,,\quad
H = -p \beta_0 \,,\quad
F_4 = \sqrt{2}\,e_i\,\left(1 + e^{(1)}\, n^{-g} \right)\, \tilde{w}^i \,,
\eeq
with $e_i \sim n$. Using these ingredients in the above 4d theory, we rewrite it as follows
\bea
\int \d^{4} x \sqrt{|g_{4}|} & \Bigg( \frac{M_p^2}{2} {\cal R}_{4} - V \\
& - \frac{M_p^2}{4}\bigg( \left(\del \ln \frac{\kappa v_1v_2v_3}{e^{2\phi}} \right)^2  + \sum_{i=1}^3 (\del \ln v_i)^2 + \sum_{i=1}^3 \frac{1}{v_i^2} (\del b_i)^2 + 2 \frac{e^{2\phi}}{\kappa v_1v_2v_3} (\del \xi)^2
\bigg)  \Bigg)  \ , \nn
\eea
with
\bea
& \frac{2}{M_p^2} \frac{1}{(2\pi \sqrt{\alpha'})^{6}} V \\
& =  \Bigg( \frac{p^2 e^{2\phi}}{2 (\kappa v_1v_2v_3)^2 } + 2\sqrt{2} p m_0 \frac{e^{3\phi} }{ (\kappa v_1v_2v_3)^\frac{3}{2} }  + \frac{e^{4\phi}}{\kappa v_1v_2v_3}  \bigg[ m_0^2 +  \left(1 + e^{(1)}\, n^{-g} \right)^2 \sum_i \frac{e_i^2 v_i^2}{(\kappa v_1v_2v_3)^2}  \nn \\
& \qquad \qquad\quad + m_0^2 \sum_{i=1}^3 \frac{b_i^2}{v_i^2}  -2 m_0 \kappa\, \left(1 + e^{(1)}\, n^{-g} \right) \frac{b_1b_2b_3}{(\kappa v_1v_2v_3)^2} \sum_{i=1}^3 \frac{e_i v_i^2}{b_i} \nn\\
& \qquad \qquad\quad + \frac{1}{(\kappa v_1v_2v_3)^2} \left( - \xi p + \left(1 + e^{(1)}\, n^{-g} \right)  \sum_{i=1}^3  e_i b_i \right)^2 + {\cal O}(b_i^4,\xi b_i^3)\bigg] \Bigg) \ . \nn
\eea
Remarkably, the effective 4d theory just obtained matches formally the DGKT 4d theory given in \eqref{pot}, and \eqref{kinsax2}, \eqref{kinax1}, up to the map
\beq
e_i \rightarrow e_i \left(1 + e^{(1)}\, n^{-g} \right) \,,\quad  v_i, \phi, b_i , \xi \ {\rm (DGKT)} \rightarrow v_i, \phi, b_i , \xi \ {\rm (here)}  \ .\label{eicorr}
\eeq
Studying the extremum, as well as the mass spectrum, will then be straightforward.

We have not developed the theory at LO and NLO. The previous identification with DGKT makes such a development unnecessary. As we will see, if one nevertheless performs such a development, we recover the 10d solution obtained via such an expansion. Note also that the NLO correction to the theory is in $n^{-g}$ with respect to the LO, because of the correction of $F_4$, of the fields and of ${\cal R}_4$. Since $0<g< 1/2$, we get the NLO at order $n^{-9/2-g} > n^{-10/2}$, i.e.~above the level where $\alpha'$- and $g_s$-corrections may appear, as discussed in Section \ref{sec:alpha}. This motivates us further to consider the new $n$-expansion \eqref{expansionnew}.

\subsubsection{Critical point, mass spectrum and interpretation}\label{sec:interpret}

Given the formal matching of our 4d theory with the DGKT one, indicated in \eqref{eicorr}, determining the critical point and the resulting mass spectrum is straightforward: we will follow Section \ref{sec:4dV} and \ref{sec:4dkinmass}. We start with the critical point. As in \eqref{bx}, we get here that the axions vanish
\beq
b_i = \xi = 0 \ ,
\eeq
which comes from the fact that the potential is at least quadratic in them. Turning to the saxions, we can proceed as for DGKT: we first introduce some fields $r_i, g$ as in \eqref{var}, including in their definition the $e_i$ correction of \eqref{eicorr}. The potential then gets similarly simplified, and is formally the same as \eqref{V1} and \eqref{V2}, up to overall factors. The extremum values for $r_i, g$ can be read from \eqref{r0g0}. Translating back to $v_i, \phi$, we obtain at the critical point
\beq
v_i |e_i|  = (v_i |e_i|)^0\,  \left(1 + e^{(1)}\, n^{-g} \right)^{1/2} \ ,\qquad \frac{e^{\phi}}{\sqrt{\kappa v_1v_2v_3}} = \left(\frac{e^{\phi}}{\sqrt{\kappa v_1v_2v_3}}\right)^0\, \left(1 + e^{(1)}\, n^{-g} \right)^{-3/2} \ ,
\eeq
where the numerics are absorbed within the $0^{{\rm th}}$ order field values. Using the $v_i$ expansion \eqref{viexp}, with $(v_i |e_i|)^0 = v_i^{(0)} n^{1/2} |e_i| $, we expand and rewrite the first equality as
\beq
1 + \frac{v_i^{(1)}}{v_i^{(0)}} n^{-g} = 1+ \frac{1}{2} e^{(1)}\, n^{-g} \ .
\eeq
We note that the ratio is independent of $i$ and can thus be written as $\frac{v_i^{(1)}}{v_i^{(0)}} = \frac{v^{(1)}}{v^{(0)}} $, as with the 10d notations. The second equality above can be rewritten into
\beq
e^{-\phi} = \left(e^{-\phi}\right)^0\, \left(1 + e^{(1)}\, n^{-g} \right)^{3/4} \ ,
\eeq
then further expanded and compared to the $\phi$ expansion in \eqref{viexp}. Eventually, we get the following critical point values
\beq
\frac{v^{(1)}}{v^{(0)}} = \frac{1}{2} e^{(1)} \ ,\qquad \frac{e^{-\phi^{(1)}}}{e^{-\phi^{(0)}}} = \frac{3}{4} e^{(1)} \ .
\eeq
Those reproduce precisely the 10d solution \eqref{solution}, that was obtained by solving the NLO expansion of the 10d equations.

The value of the potential at the extremum and the 4d Ricci scalar can be obtained from the DGKT ones \eqref{R4EV}, with the mapping \eqref{eicorr}
\beq
{\cal R}_{4} = 4\, \frac{V|_0}{M_p^2} = \left( {\cal R}_{4} \right)^0 \left(1 + e^{(1)}\, n^{-g} \right)^{-9/2} \simeq \left( {\cal R}_{4} \right)^0 \left(1 - \frac{9}{2} e^{(1)}\, n^{-g} \right) \ . \label{V0corr}
\eeq
The last expression, obtained by expanding, matches precisely the value of the 10d solution \eqref{R410dsol}.\\

We turn to the mass spectrum, for which we follow again the DGKT derivation of Section \ref{sec:4dkinmass}. As above, we introduce the fields $r_i, g, \tilde{b}_i, \xi$ of \eqref{var}, including in their definition the $e_i$ correction of \eqref{eicorr}. Note that the signs $s_i$ are left unchanged by the correction of \eqref{eicorr} for sufficiently large $n$. As mentioned above, each part of the potential (for the saxions and the axions) only depends on the fluxes by an overall factor, so the correction \eqref{eicorr} only affects the potential in this way. The same is true for the kinetic terms: the fluxes only appear though an overall factor for the axions, and not at all for the saxions. The mass spectrum computation could then only be altered by such a correction to overall factors. As can be seen in the mass matrices \eqref{Msax}, \eqref{Max}, or in the masses \eqref{msax}, \eqref{max}, the overall factor is in the end the same for the saxions and axions, and is in addition nothing but the vacuum value of the scalar potential $V|_0$. The only change in the mass spectrum is thus the correction to $V|_0$, that can be read in \eqref{V0corr}. In other words, the masses obtained are
\bea
m^2 =  \left( m^2 \right)^0 \left(1 + e^{(1)}\, n^{-g} \right)^{-9/2} \simeq \left( m^2 \right)^0 \left(1 - \frac{9}{2} e^{(1)}\, n^{-g} \right) \ ,\\
\frac{m^2}{|V|_0|} = \left( \frac{m^2}{|V|_0|} \right)^0 \ \text{given in}\ \eqref{msax}\ {\rm and}\ \eqref{max}\ .
\eea
We conclude that the mass spectrum is barely changed at $1^{{\rm st}}$ order by the new $n$-expansion \eqref{expansionnew}. Indeed, since the ratio $m^2/|V|_0|$ is not corrected, the conformal dimensions are unchanged. They equate the same integers as given in \eqref{Deltasax} and \eqref{Deltaax}.\\

The 4d description finally allows us to reinterpret the new $n$-expansion \eqref{expansionnew}, together with the compactification ansatz considered. We have seen that the 4d theory was formally the same as DGKT up to the map \eqref{eicorr}, that redefines the 4d fields and the $F_4$-flux. The change of the latter can be viewed as
\beq
n \rightarrow n + e^{(1)}\, n^{1-g} \equiv n' \ ,
\eeq
where we recall that $e^{(1)}$ is a constant and $0 < g < 1/2$. As shown around \eqref{fluxquant}, the quantization of the $F_4$-flux is still possible with such a correction, contrary to the $H$-flux. In the end, the $n$-expansion \eqref{expansionnew} can then be understood as being only a redefinition of the discretized parameter $n$ of DGKT to $n'$, making use of the fact that it is still possible to satisfy the quantization conditions and get flux integers with a lower power of $n$, namely $n^{1-g}$. Another interpretation, that we followed, is to view it as a correction to (or extension of) DGKT with a parameter $n$. In that case, the corrections are however very specific as they only alter the masses through the cosmological constant. Let us emphasize that interpreting the new expansion \eqref{expansionnew} as a simple redefinition $n\rightarrow n'$ of DGKT was not obvious from the start, especially not from 10d, where we solved equations perturbatively in the parameter $n$; this is more easily seen in the 4d theory where the treatment was not perturbative. Getting to this interpretation also required constraints from the 10d equations and the flux quantization to set some fluxes to zero and refine the scalings.\footnote{One may argue that provided a smeared and isotropic ansatz, the only solution to 10d equations is DGKT. The ansatz considered in this section being close to this, one may not be surprised that we eventually obtain as a solution a repackaging of DGKT. We however have at $1^{{\rm st}}$ order a few differences to a purely smeared, isotropic, DGKT-like ansatz. First, we allowed for a coordinate dependent $F_2^{(1)}$; its scaling eventually made it drop out of relevant equations though. Second, flux quantization sets $H^{(1)}=0$, while the DGKT ansatz contains $H^{(0)}\neq 0$; the repackaging is still possible thanks to the $n^0$ scaling of the latter. These points help understanding why a repackaging into DGKT is possible despite few initial ansatz differences. We thank D.~Junghans for exchanges on this matter.} Let us also stress that taking rather the point of view of a correction to DGKT, then not only the flux is corrected, but also the metric, the dilaton and the cosmological constant. It remains a remarkable property of DGKT that the conformal dimensions are eventually not changed, making again its spectrum robust under possible extensions.

\vspace{0.4in}

\subsection*{Acknowledgements}

We deeply thank E.~Plauschinn, M.~Rajaguru and T.~Wrase for helpful discussions during the completion of this project.

\newpage

\begin{appendix}

\section{10d elements}

\subsection{Warped 10d type IIA supergravity}\label{ap:conv}

In this appendix, we detail our conventions for 10d massive type IIA supergravity. For them, we follow \cite[App. A]{Andriot:2016xvq}, and \cite[App. A]{Andriot:2017jhf} for the multiple source sets. In addition, we consider a compactification ansatz on a 6d manifold, with a warp factor in the 4d part of the metric, and give here the resulting 10d equations split on 4d and 6d. We finally compare for future purposes our equations to those of \cite{Junghans:2020acz}, finding a perfect match upon translation.

We start with the following 10d string frame metric
\beq
\d s^2 = e^{2A(y)} \, g^S_{\mu\nu}(x) \d x^{\mu} \d x^{\nu} + g_{mn}(y) \d y^m \d y^n \ ,\label{metric10d}
\eeq
split as a warped product of a 4d spacetime and a 6d compact space. The 4d volume form associated to $g^S_{\mu\nu}$ is ${\rm vol}^S_4$, and ${\cal R}^S_4$ is the corresponding Ricci scalar. We denote with an $S$ these 4d quantities, to be later distinguished from those in 4d Einstein frame. In addition, we consider the dilaton to have only internal (6d) dependence: $\phi(y)$.

Prior to giving the 10d supergravity equations, split on 4d and 6d, let us recall our notations for the (space-filling) $O_6$-plane contribution. We first introduce various notations with the following rewriting of DBI action, where we consider no world-volume $B$-field
\bea
\left.\frac{{\cal S}_{{\rm DBI}}}{- c_p T_p}\right|_{O_6} = \int_{\Sigma_{7}} \d^{7} \xi\, e^{-\phi} \sqrt{|i^*[g_{10}]|} & = \int_{10} e^{-\phi} {\rm vol}^S_4\w {\rm vol}_{||} \w \delta_3^{\bot} \label{DBIrewrite}\\
& = \int \d^{10} x\, e^{-\phi} \sqrt{|P[g_{10}]|} \, \delta(\bot)  \ , \nn
\eea
referring to \cite[(A.7)]{Andriot:2016xvq} for more details. We also introduce the transverse volume form ${\rm vol}_{\bot}$ such that ${\rm vol}_{||} \w {\rm vol}_{\bot} = {\rm vol}_6$. We further recall the concept of source set labeled by $I$: one set is made of the sources that wrap the same internal submanifold. We consider here one $O_6$ source (i.e.~one DBI action) for each set $I$ and no $D_6$, and then use the notation \cite{Andriot:2016xvq, Andriot:2017jhf}
\beq
\frac{T_{10}^I}{7} = 2 \times 2\pi \sqrt{\alpha'}\, \sqrt{\frac{|P[g_{10}]|}{|g_{10}|}} \, \delta({\bot}) \ .\label{T10def}
\eeq
We then have the following relation between previously introduced quantities for one set $I$
\beq
2\pi \sqrt{\alpha'} \times 2 \delta_3^{\bot_I} = \frac{T_{10}^I}{7} {\rm vol}_{\bot_I} \ .\label{BIRHS}
\eeq
We also denote
\beq
 T_{10} = \sum_{I} T_{10}^I \ .
\eeq

Finally, we complete our compactification ansatz as follows: we consider all flux-forms $H,F_0,F_2,F_4,F_6$ to be internal. The form $F_6$ comes from the 10d 4-form RR flux, split as $F_{4}^{10} = e^{4A} {\rm vol}^S_4 *_6 F_6+ F_4$. This gives the square $|F_{4}^{10}|^2 = |F_4|^2 - |F_6|^2$. With all these ingredients, we are ready to give the 10d equations in string frame, split on 4d and 6d.\\

To start with, the flux Bianchi identities are given by
\begin{equation}
\begin{split}
\d H &=0 \\
\d F_0 &=0 \label{BIs}\\
\d F_{2} - H \w F_{0} &= \sum_{I} \frac{T_{10}^I}{7} {\rm vol}_{\bot_I} \\
\d F_4 - H\w F_2 &= 0 
\end{split}
\end{equation}
and the flux equations of motion are given by
\bea
e^{-4A} \d( e^{4A} *_6 F_2 ) + H \w *_6 F_4 &= 0 \, \label{fluxeq0}\\
e^{-4A} \d( e^{4A} *_6 F_4 ) + H \w *_6 F_6 &= 0 \,\\
\d( e^{4A} *_6 F_6 ) &= 0 \,\\
e^{-4A} \d( e^{4A- 2\phi} *_6 H )  - \sum_{q=0}^{4} F_q \w *_6 F_{q+2} &= 0 \ .\label{fluxeq}
\eea
The 10d version of the dilaton equation of motion is given in \cite[(A.14)]{Andriot:2016xvq}. We combine it here with \cite[(C.5)]{Andriot:2016xvq} and \cite[(2.28)]{Andriot:2017oaz} to write it as follows
\bea
0 = &\, 2 e^{-2A} {\cal R}^S_4 + 2 {\cal R}_6 - |H|^2 + e^{\phi}  \frac{T_{10}}{7} \\
& - 2 e^{-4A} (\del e^{2A} )^2 - 8 e^{-2A} \Delta_6 e^{2A} +2 e^{-4\phi} (\del e^{2\phi})^2 - 4  e^{2\phi} \Delta_6 e^{-2\phi} -8 e^{2\phi - 2A} \del_m e^{2A} \del^m e^{-2\phi}    \ .\nn
\eea
We finally turn to the Einstein equations. The trace-reversed 10d Einstein equation is obtained using \cite[(A.15), (2.14)]{Andriot:2016xvq}
\bea
{\cal R}_{MN} & = \frac{1}{4} H_{MPQ}H_N^{\ \ PQ}+\frac{e^{2\phi}}{2}\left(F_{2\ MP}F_{2\ N}^{\ \ \ \ P} +\frac{1}{3!} F^{10}_{4\ MPQR}F_{4\ N}^{10 \ \ PQR} \right)  + \frac{e^{\phi}}{2}T_{MN} -2\nabla_M \del_N{\phi} \nn\\
& +\frac{g_{MN}}{4} \left(-\frac{e^{\phi}}{4} T_{10} -\frac{1}{2} |H|^2 + \frac{e^{2\phi}}{4} (|F_0|^2 - |F_2|^2 -3 |F_4^{10}|^2 ) - \Delta \phi + 2 |\del \phi|^2 \right) \ ,
\eea
and we give in the bulk of the paper the definition of the source energy-momentum tensor $T_{MN}$. From there, we first get the 4d Einstein equation, proportional to its trace: we obtain it using \cite[App. C]{Andriot:2016xvq}
\begin{equation}
\begin{split}
0 = &\, 2e^{-2A} {\cal R}^S_4  +|H|^2 + \frac{e^{2\phi}}{2} \sum_{q=0}^6 (q-1) |F_q|^2   -\frac{e^{\phi}}{2} \frac{T_{10}}{7} \\
&  -4 e^{-2A} \Delta_6\, e^{2A} -4 e^{-4A} (\del e^{2A})^2 - e^{2\phi} \Delta_6\, e^{-2\phi}  - 6 e^{2\phi-2A}  \del_{m} e^{2A} \del^{m} e^{-2\phi}  \ .
\end{split}
\end{equation}
The 6d (trace-reversed) Einstein equation is obtained using in addition \cite[(2.23c)]{Andriot:2017oaz}
\bea
0={\cal R}_{mn}  & -\frac{1}{4} H_{mpq}H_n^{\ \ pq} - \frac{e^{2\phi}}{2}\left(F_{2\ mp}F_{2\ n}^{\ \ \ \ p} +\frac{1}{3!} F_{4\ mpqr}F_{4\ n}^{\ \ \ pqr} \right)  - \frac{e^{\phi}}{2}T_{mn} \nn\\
& - \frac{g_{mn}}{4} \Bigg(-\frac{e^{\phi}}{4} T_{10} -\frac{1}{2} |H|^2 + \frac{e^{2\phi}}{4} (|F_0|^2 - |F_2|^2 -3 |F_4^{10}|^2 ) +\frac{1}{2} e^{2\phi} \Delta_6\, e^{-2\phi} \\
&+ e^{2\phi-2A} \del_{p} e^{2A} \del^{p} e^{-2\phi} \Bigg)
- 2 e^{-2A} \nabla_n \del_m e^{2A} + e^{-4A} \del_m e^{2A} \del_n e^{2A} + 2\nabla_m \del_n{\phi} \ . \nn
\eea
For convenience, we rewrite the above dilaton, 4d and 6d Einstein equations as follows
\begin{equation}
\begin{split}
0 = &\, 2 e^{-\phi -2A} {\cal R}^S_4 + 2 e^{-\phi } {\cal R}_6 - e^{-\phi } |H|^2 + \frac{T_{10}}{7} \label{eq1}\\
& - 24 e^{-\phi -2A} (\del e^{A} )^2 - 16 e^{-\phi -A} \Delta_6 e^{A}  - 8 \Delta_6 e^{-\phi} - 32 e^{- A} \del_m e^{A} \del^m e^{-\phi} \,,
\end{split}
\end{equation}
\bea
0 = &\, -e^{-2\phi -2A} {\cal R}^S_4  - \frac{1}{2} e^{-2\phi}|H|^2 - \frac{1}{4} \sum_{q=0}^6 (q-1) |F_q|^2   +\frac{e^{-\phi}}{4} \frac{T_{10}}{7} \label{eq2}\\
& +4 e^{-2\phi-A} \Delta_6\, e^{A} + 12 e^{-2\phi -2A} (\del e^{A})^2 + e^{-\phi} \Delta_6\, e^{-\phi} + (\del e^{-\phi})^2  +12 e^{-\phi-A}  \del_{m} e^{A} \del^{m} e^{-\phi}  \ .\nn
\eea
\begin{equation}
\begin{split}
0=- e^{-2\phi} {\cal R}_{mn}  & +\frac{e^{-2\phi}}{4} H_{mpq}H_n^{\ \ pq} + \frac{1}{2}\left(F_{2\ mp}F_{2\ n}^{\ \ \ \ p} +\frac{1}{3!} F_{4\ mpqr}F_{4\ n}^{\ \ \ pqr} \right)  + \frac{e^{-\phi}}{2}T_{mn} \\
&  - \frac{g_{mn}}{8} e^{-2\phi} |H|^2 + \frac{g_{mn}}{16} (\sum_{q=0}^6 (1-q) |F_q|^2 + 8 |F_6|^2 )   - g_{mn} \frac{7}{16} e^{-\phi} \frac{T_{10}}{7} \\
& + \frac{g_{mn}}{4} \left( e^{-\phi} \Delta_6\, e^{-\phi} + (\del e^{-\phi})^2 + 4 e^{-\phi-A} \del_{p} e^{A} \del^{p} e^{-\phi} \right) \\
&  + 4 e^{-2\phi -A} \nabla_n \del_m e^{A} + 2 e^{-\phi} \nabla_m \del_n e^{-\phi}  - 2  \del_m e^{-\phi} \del_n e^{-\phi}   \ .\label{eq3}
\end{split}
\end{equation}

Having presented the 10d type IIA supergravity equations split as 4d and 6d on a warped compactification ansatz, in our conventions, let us compare them to those of \cite{Junghans:2020acz}. Starting with the dilaton and Einstein equations \eqref{eq1}-\eqref{eq3}, we find a perfect agreement with \cite[(2.9)-(2.11)]{Junghans:2020acz} thanks to the following translation
\bea
& e^{-\phi} \equiv \tau \ ,\qquad e^A \times {\rm constant} \equiv w \ ,\\
& \delta_3^{\bot_I} \equiv \delta_{i3} \ ,\quad \frac{1}{2\pi \sqrt{\alpha'}} \frac{T_{10}^I}{7} \equiv  2 \delta(\pi_i) \ ,\quad \frac{1}{2\pi \sqrt{\alpha'}} T_{mn} \equiv 2 \ssum \Pi_{i,mn} \delta(\pi_i) \ ,\label{sourcematch}
\eea
where here and in the following the left-hand side are our notations and the right-hand side those of \cite{Junghans:2020acz}. One should also note that $2\pi \sqrt{\alpha'} = 1$ in \cite{Junghans:2020acz}. This translation is based on the definitions of quantities in each set of conventions, and one then verifies the matching of the equations.

Let us add more details on the warp factor and the map of conventions. The 4d metric (in 10d string frame) in \cite{Junghans:2020acz} is $w^2 g^s_{\mu\nu}$ where $g^s_{\mu\nu}$ is said to be the AdS metric with unit radius, so that $w$ effectively carries the AdS scale. In general, the AdS metric is proportional to $l^2$ where $l$ is the AdS radius, and in 4d, one has for the AdS Ricci scalar ${\cal R}^s_4 = -12/ l^2$; $g^{\mu\nu\, s} {\cal R}^s_{\mu\nu} = -12$ is indeed used in \cite{Junghans:2020acz} (we use a label $s$, which is absent in that reference; this is to avoid confusion with our metrics). Our 4d metric in \eqref{metric10d} is denoted $e^{2A} g^S_{\mu\nu}$, where $g^S_{\mu\nu}$ denotes the complete AdS metric. We conclude on the matching
\beq
e^A \, l \equiv w \ ,
\eeq
recalling that $l$ is here our 4d anti-de Sitter radius in 10d string frame, and verify the matching of the ${\cal R}_4^S$ terms in the above 10d equations.

Regarding flux equations, we get a matching of equations \eqref{BIs} and \eqref{fluxeq0} - \eqref{fluxeq} with \cite[(2.3)-(2.8)]{Junghans:2020acz},\footnote{The sign between the two terms in the $H$-flux e.o.m.~seems at first sight different. This is due to a difference in the Hodge star conventions: in $d$ dimensions on a $p$-form, they differ by a sign $(-1)^{p(d-p)}$. This leads to a minus sign only on the $H$-flux, eventually giving the same equation.} except for a sign difference in the non-trivial Bianchi identity in \eqref{BIs}. That sign has no physical impact and can be compensated by changing the sign of RR fluxes (which leaves all other equations unchanged): we get
\beq
F_{q\, {\rm here}} \equiv -F_{q\, {\rm there}} \ .\label{signconv}
\eeq
Having mapped our conventions to those of \cite{Junghans:2020acz}, we can for instance import the analysis and solutions worked-out there.

\subsection{Smearing source contributions}\label{ap:smear}

In \cite{Junghans:2020acz} is considered a smearing procedure, mentioned in Section \ref{sec:sol}. We use it here to reproduce (to some extent) the smeared source contributions in the smeared DGKT solution of Section \ref{sec:sol}, starting from their localised definitions. Note that the smeared 10d solution is not fully explicited nor verified in \cite{Junghans:2020acz}, and we bridge here this gap.

The map between our conventions and those of \cite{Junghans:2020acz} on source contributions is given in \eqref{sourcematch}. Those involve 6d metric components so let us first map those. Our notations follow \cite{DeWolfe:2005uu}: the metric is given in \eqref{6dmetric}. In \cite{Junghans:2020acz}, a similar symbol $v_i^{(0)2}$ is unfortunately used with a different meaning. We then change that notation as follows: $v_i^{(0)2}$ in \cite{Junghans:2020acz} $\rightarrow \upnu_i^{2}$, and map the notations
\beq
2 (\kappa \sqrt{3} )^{1/3}\, v_i \equiv \upnu_i^2 \, n^{1/2} \ . \label{vimap}
\eeq
We verify the matching of the 6d volume computation, recalling that the work in \cite{Junghans:2020acz} is done on the covering torus $T^6$ instead of the orbifold $T^6/\mathbb{Z}_3^2$
\beq
9\, {\rm vol} = \int_{T^6} \d^6 y \sqrt{g_6} = 9 \, \kappa v_1 v_2 v_3 \equiv \left( \frac{\sqrt{3}}{2} \right)^3 \upnu_1^{2} \upnu_2^{2} \upnu_3^{2} \, n^{3/2}  \ .
\eeq
This can be found in \cite[(2.14),(4.18)]{Junghans:2020acz}, and here in \eqref{vol}; the orbifold divides this volume by 9.

We now start considering the source contributions. From the definitions \eqref{DBIrewrite}-\eqref{BIRHS} giving the source quantization, we reached in Section \ref{sec:sol} the integral $\int \frac{ T_{10} }{7}\, {\rm vol}_{\bot} =  2 \times 2\pi \sqrt{\alpha'}$. Using in addition the (smeared) solution value for ${\rm vol}_{\bot} $ \eqref{volbp}, we deduce the following value of the smeared source contribution in the DGKT solution
\beq
\frac{1}{2\pi \sqrt{\alpha'}} \frac{ T_{10} }{7} =  4 \left( \sqrt{{\rm vol}} \ \int_{T^6/\mathbb{Z}_3^2} \beta_0 \right)^{-1} \ .
\eeq
We then need to compute the above integral. It is unclear to us how to do so, and on which 3d space to perform it. We still propose the following result, which will pass various consistency check
\beq
 \int_{T^6/\mathbb{Z}_3^2} \beta_0  = \frac{1}{2} \sqrt{2}\, 3^{\frac{1}{4}} \ ,
\eeq
where one recognises the factor $\sqrt{2}\, 3^{\frac{1}{4}}$ in $\beta_0$ \eqref{beta0}. Going to the covering torus brings a factor 9 to this integral. $T_{10}$ becomes the sum of the different set contributions, $ T_{10}= \sum_I  T_{10}^I$. We deduce that the smeared contribution on the covering torus is
\beq
\frac{1}{2\pi \sqrt{\alpha'}} \sum_I \frac{ T_{10}^I }{7} = 18 \times \frac{ 4}{\sqrt{2}\, 3^{\frac{1}{4}} \sqrt{{\rm vol}} } \ .\label{T10ap}
\eeq
According to the map \eqref{sourcematch}, the above should be the smeared version of $2 \sum_i \delta(\pi_i)$ in \cite{Junghans:2020acz}, where the sum is done on the 9 sources on the covering torus. The smearing procedure described in \cite[(2.13),(2.14),(4.18)]{Junghans:2020acz} leads to the following replacement
\beq
\forall i \ ,\ \delta(\pi_i) \rightarrow \frac{8}{\upnu_1 \upnu_2 \upnu_3 \, n^{3/4}} = \frac{4}{\sqrt{2}\, 3^{\frac{1}{4}} \, \sqrt{\rm vol}} \ . \label{deltasmear}
\eeq
Since the result is independent of $i$, we get that
\beq
2 \sum_i \delta(\pi_i) = 18 \times \frac{4}{\sqrt{2}\, 3^{\frac{1}{4}} \, \sqrt{{\rm vol}} } \ ,
\eeq
which matches the result \eqref{T10ap}. In Section \ref{sec:sol}, the value of $T_{10}$ was rather fixed through the Bianchi identity. From the above expression, the value of Section \ref{sec:sol} can now be recovered from its definition and source quantization, or the flux quantization and tadpole condition \eqref{tad}.

It would also be interesting to recover ${\rm vol}_{\bot}$ in the smeared solution \eqref{volbp}, or $T_{10} {\rm vol}_{\bot}$, from the smearing procedure of \cite{Junghans:2020acz}. $ {\rm vol}_{\bot}$ would then appear as an effective transverse volume built from the sum of all those on the covering torus. Doing so would require the volumes ${\rm vol}_{\tilde{\pi}_i}$, where $\tilde{\pi}_i$ is the dual cycle to the wrapped one $\pi_i$. Those are however not made explicit in \cite{Junghans:2020acz}, and we refrain from doing so.

Nevertheless, another related quantity can be obtained through smearing: the source energy-momentum tensor $T_{mn}$. While we have its value in the smeared solution, let us reproduce it as an effective quantity from smearing, as done above for $T_{10}$. We start in \cite{Junghans:2020acz} with $\ssum_i \Pi_{i,mn} \delta(\pi_i)$: in the smeared limit, since $\delta(\pi_i)$ becomes independent of $i$, that quantity becomes proportional to $\ssum_i \Pi_{i,mn}$. We read that sum from \cite[Tab. 2]{Junghans:2020acz} and get
\beq
\ssum \Pi_{i,11}^{(0)} = \ssum \Pi_{i,22}^{(0)} = \frac{9}{2} \upnu_1^2 \ ,\ \ssum \Pi_{i,33}^{(0)} = \ssum \Pi_{i,44}^{(0)} = \frac{9}{2} \upnu_2^2 \ ,\ \ssum \Pi_{i,55}^{(0)} = \ssum \Pi_{i,66}^{(0)} = \frac{9}{2} \upnu_3^2 \ ,
\eeq
while $\ssum \Pi_{i,mn}^{(0)} = 0$ for $m\neq n$. In other words, we deduce in the smeared limit for $m,n=2k-1,2k-1$ and $2k ,2k$
\beq
\ssum_i \Pi_{i,mn} \delta(\pi_i) =  \frac{9}{2} \upnu_k^2\, n^{1/2} \ \delta(\pi_i) = \frac{\upnu_k^2\, n^{1/2}}{2} \ssum_i\, \delta(\pi_i) \ .
\eeq
Using the mapping to our notations \eqref{sourcematch} (having verified above the match between $ \delta(\pi_i)$ and $T_{10}^I$), we get
\beq
T_{mn} = (\kappa \sqrt{3} )^{1/3}\, v_k\ \frac{T_{10}}{7} \ .
\eeq
One verifies that this is precisely the smeared value of the DGKT solution \eqref{F4F4}, thanks to \eqref{relphivol}. We conclude that the smearing procedure, as described in \cite{Junghans:2020acz}, is able to reproduce the values of the smeared source contributions of the DGKT solution given in Section \ref{sec:sol}.

\section{4d elements}

\subsection{DGKT saxions kinetic terms}\label{ap:kin}

We derive here the saxions kinetic terms from 10d supergravity for the DGKT setting, allowing us to have consistent conventions. This result is used in Section \ref{sec:4dkinmass}. To get these kinetic terms, we use known results; a more general and complete derivation is provided in Appendix \ref{ap:4dwarp}. In \cite{Andriot:2020wpp} was considered the following 10d string frame metric split in 4d and 6d
\bea
\d s_{10}^2 & = \tau^{-2} g_{\mu \nu} \d x^{\mu} \d x^{\nu} \label{metric10dtaurho}\\
& + \rho \Big ( \sigma_1^{-4} \sigma_2^{2} ((\d y^1)^2 + (\d y^2)^2) + \sigma_1^{2} \sigma_2^{-4} ((\d y^3)^2 + (\d y^4)^2) + \sigma_1^{2} \sigma_2^{2} ((\d y^5)^2 + (\d y^6)^2)  \Big) \ ,\nn
\eea
with 4d fluctuations $\tau, \rho, \sigma_1, \sigma_2$ and the 4d dilaton $\tau = e^{-\phi} \rho^{\frac{3}{2}}$ used to go to 4d Einstein frame. Then, the 4d kinetic terms were found in \cite[(D.16)]{Andriot:2020wpp} to be given by
\beq
{\cal S}_{{\rm kin}} =- \frac{M_p^2}{2} \int \d^4 x \sqrt{|g_4|} \left( 2 (\del \ln \tau)^2 + \frac{3}{2} (\del \ln \rho )^2 + 12 \left((\del \ln \sigma_1 )^2 + (\del \ln \sigma_2 )^2 - \del \ln \sigma_1 \del \ln \sigma_2  \right) \right) \ .\nn
\eeq
The following 6d metric
\beq
\d s_6^2 = R_1^2 \left((\d y^1)^2 + (\d y^2)^2\right) + R_2^2 \left((\d y^3)^2 + (\d y^4)^2\right)
+ R_3^2 \left((\d y^5)^2 + (\d y^6)^2\right) \ ,\label{6dmetricRi}
\eeq
corresponds to the previous one with
\beq
\rho^{\frac{3}{2}} = R_1 R_2 R_3 \,,\quad
\sigma_1= \left(\frac{R_3}{R_1} \right)^{\frac{1}{3}}\,,\quad
\sigma_2= \left(\frac{R_3}{R_2} \right)^{\frac{1}{3}} \ .
\eeq
From these results, we obtain the following kinetic terms
\beq
{\cal S}_{{\rm kin}} =- \frac{M_p^2}{2} \int \d^4 x \sqrt{|g_4|} \ 2 \left(  (\del \ln \tau)^2 +  (\del \ln R_1 )^2 + (\del \ln R_2 )^2 +  (\del \ln R_3 )^2 \right) \,. \label{kinsax1}
\eeq
We now apply the above results to the 10d metrics of interest. The relation between the 10d string frame 4d metric and the 4d Einstein frame one is given in \eqref{relR4}. Comparing to \eqref{metric10dtaurho}, we can identify $\tau^{-2}$ to $(2\pi \sqrt{\alpha'})^6\ \frac{e^{2\phi}}{{\rm vol}}$. Given the definition of $g$ in \eqref{var}, we deduce $\tau^{-1} \propto g$. Turning to the 6d metric, comparing \eqref{6dmetricRi} to \eqref{6dmetric}, and using the definition of $r_i$ in \eqref{var}, we get $R_i^2 \propto v_i \propto r_i^2$. We conclude on the following kinetic terms for the 4d fields of interest
\beq
{\cal S}_{{\rm kin}} =- \frac{1}{2} \int \d^4 x \sqrt{|g_4|}\ 2 M_p^2\  \left( (\del \ln\, g)^2 + (\del \ln\, r_1 )^2 + (\del \ln \, r_2 )^2 +  (\del \ln\, r_3 )^2 \right) \ .\label{kinsax2ap}
\eeq

\subsection{Warped 4d theory: kinetic terms and scalar potential}\label{ap:4dwarp}

We work out here in full generality a 4d effective theory obtained by dimensional reduction of 10d type IIA supergravity with sources, on a warped compactification. We will use it in the main text for the extensions of DGKT.

We start by considering the following 10d string frame metric
\beq
\d s^2 = {\tau(x)}^{-2}\, e^{2A(y)}\, g_{\mu\nu}(x)\, \d x^{\mu} \d x^{\nu} + \rho(x)\, g_{mn}(x,y)\, \d y^m \d y^n \ , \label{10dfullmetric}
\eeq
which is more general than \eqref{metric10d}. The 4d Einstein frame metric will eventually be given by $g_{\mu\nu}$. We first compute the corresponding 10d Ricci scalar: it will contribute to the kinetic terms and the potential for this warped compactification. It is given as follows
\begin{equation}
\begin{split}
{\cal R}_{10} & = \tau^2\, e^{-2A}\, {\cal R}_4 + \rho^{-1}\, \left({\cal R}_6 - 12 e^{-2A} (\del e^{A} )^2 - 8  e^{-A} \Delta_6 e^{A} \right) \\
 & - e^{-2A}\, \nabla_{\mu} \left( 3 \tau^4 \del^{\mu} \tau^{-2} + \tau^2 \rho^{-1} g^{mn} \del^{\mu} (\rho g_{mn}) \right) \\
& - \frac{9}{2} e^{-2A}\, \tau^6 (\del \tau^{-2})^2 - 2 e^{-2A}\, \tau^4 \del_{\mu} \tau^{-2} \del^{\mu} (\rho g_{mn}) \rho^{-1} g^{mn} \\
& - \frac{\tau^2}{4} e^{-2A}\, \del_{\mu} (\rho g_{mn}) \rho^{-1} g^{mn} \del^{\mu} (\rho g_{pq}) \rho^{-1} g^{pq} + \frac{\tau^2}{4} e^{-2A}\, \del_{\mu} (\rho g_{mn}) \del^{\mu} (\rho^{-1} g^{mn}) \ .
\end{split}
\end{equation}
In the above, ${\cal R}_4$ and ${\cal R}_6$ are respectively built from $g_{\mu\nu}$, 4d derivatives, and $g_{mn}$, 6d derivatives. The 4d squares and covariant derivatives are built with $g_{\mu\nu}$, while the 6d squares are built with $g_{mn}$. Note that the 10d Ricci scalar for the purely warped 10d metric \eqref{metric10d} can also be read from \cite[(2.28)]{Andriot:2017oaz}, and the one for the metric \eqref{10dfullmetric} without warp factor can be read from \cite[(D.2)]{Andriot:2020wpp}, and \cite[(4.8)]{Andriot:2022xjh} in $d$ dimensions.

The above formula can have interesting applications, given its generality. For our purposes however, we set $\rho=1$: the resulting metric from \eqref{10dfullmetric}, namely
\beq
\d s^2 = {\tau(x)}^{-2}\, e^{2A(y)}\, g_{\mu\nu}(x)\, \d x^{\mu} \d x^{\nu} + g_{mn}(x,y)\, \d y^m \d y^n \ , \label{10dourmetricap}
\eeq
will be enough for the 4d fields to be considered. We then simplify and rewrite the above
\bea
{\cal R}_{10} & ={\cal R}_6 - 12 e^{-2A} (\del e^{A} )^2 - 8  e^{-A} \Delta_6 e^{A}  \label{R10full} \\
 &\  + \tau^2\, e^{-2A}\, \Big( {\cal R}_4  - \nabla_{\mu} \left( 3 \tau^2 \del^{\mu} \tau^{-2} +  g^{mn} \del^{\mu} g_{mn} \right) - \frac{3}{2} \tau^{-4} (\del \tau^{2})^2 + \tau^{-2} \del_{\mu} \tau^{2}\, \del^{\mu} (g_{mn}) g^{mn}  \nn\\
& \  \phantom{+ \tau^2\, e^{-2A} } \ \ - \frac{1}{4} \del_{\mu} (g_{mn}) g^{mn} \del^{\mu} (g_{pq}) g^{pq} + \frac{1}{4} \del_{\mu} g_{mn} \del^{\mu} g^{mn} \Big) \nn\ .
\eea
The first line will contribute to the 4d scalar potential, while the others will give the kinetic terms.

We now write the 10d Ricci scalar \eqref{R10full} as the first line plus $\tau^2\, e^{-2A}\, ( {\cal R}_4 + \dots )$ for simplicity, and denote the determinant of the 10d metric \eqref{10dourmetricap} by $|G|$. We obtain
\bea
& \frac{1}{2 \kappa_{10}^2} \int \d^{10} x \sqrt{|G|}\, e^{-2\phi}\ {\cal R}_{10}  \\
= &\ \frac{M_p^2}{2} \int \d^{4} x \sqrt{|g_4|}\, \tau^{-2}\, {\cal V}_{6\phi} \Big(( {\cal R}_4 + \dots) \nn\\
 & \phantom{\ M_p^2 \d^{4} x }  + (2\pi \sqrt{\alpha'})^{-6}\, {\cal V}_{6\phi}^{-1}\, \tau^{-2}\, \int \d^6 y \sqrt{|g_6|}\, e^{4A-2 \phi} \, \left( {\cal R}_6 - 12 e^{-2A} (\del e^{A} )^2 - 8 e^{-A} \Delta_6 e^{A} \right) \Big) \nn\\
&\ \ {\rm with}\ {\cal V}_{6\phi} = (2\pi \sqrt{\alpha'})^{-6}  \int \d^6 y \sqrt{|g_6|}\, e^{2A-2 \phi} \label{V6p}\\
= &\ \int \d^{4} x \sqrt{|g_{4}|} \left(\frac{M_p^2}{2}  ( {\cal R}_{4} + \dots ) - V_{{\cal R}} \right) \quad {\rm with}\ \tau^{-2} = {\cal V}_{6\phi}^{-1} \ ,\nn\\
{\rm and} &\ V_{{\cal R}}= - \frac{M_p^2}{2}  (2\pi \sqrt{\alpha'})^{-6}\, {\cal V}_{6\phi}^{-2} \int \d^6 y \sqrt{|g_6|}\, e^{4A-2 \phi} \, \left( {\cal R}_6 - 12 e^{-2A} (\del e^{A} )^2 - 8 e^{-A} \Delta_6 e^{A} \right) \ ,\nn
\eea
with $2 \kappa_{10}^2= (2\pi)^7 {\alpha'}^4$ and $M_p^2 = (\pi \alpha')^{-1}$. Note that $\tau^{2} = {\cal V}_{6\phi}$ allows to reach the 4d Einstein frame, with metric $g_{\mu\nu}$. We follow conventions of \cite{Andriot:2022bnb}, adapting them to the fact that $A$ and $\phi$ depend on internal coordinates. We have minor differences with conventions of \cite{Junghans:2020acz}.

In the following, we assume (which will be true in our applications) that the dependence on 4d versus 6d coordinates, in $|g_6|$ and in $e^{\phi}$, factorizes. As a consequence, $\del_{\mu} \ln |g_6|$ and $\del_{\mu} \ln e^{\phi}$ are independent of 6d coordinates. We are then able to factorize out a 6d integral as follows
\begin{equation}
\begin{split}
\frac{\del_{\mu} {\cal V}_{6\phi}}{ (2\pi \sqrt{\alpha'})^{-6}} & = \int \d^6 y \, e^{2A} \, \del_{\mu}(\sqrt{|g_6|} e^{-2 \phi} ) \\
& = \int \d^6 y \, e^{2A} \, \sqrt{|g_6|} e^{-2 \phi} \left( \frac{1}{2} \del_{\mu} \ln |g_6| + \del_{\mu} \ln \, e^{-2 \phi} \right) \\
& = \frac{{\cal V}_{6\phi}}{ (2\pi \sqrt{\alpha'})^{-6}} \left( \frac{1}{2} g^{mn} \del_{\mu} g_{mn}  -2 \del_{\mu} \phi \right) \ . 
\end{split}
\end{equation}
In other words, we get
\beq
2 \del_{\mu} \phi = - {\cal V}_{6\phi}^{-1}  \del_{\mu} {\cal V}_{6\phi} + \frac{1}{2} g^{mn} \del_{\mu} g_{mn} =  - \tau^{-2} \del_{\mu} \tau^{2} + \frac{1}{2} g^{mn} \del_{\mu} g_{mn} \ ,
\eeq
where all terms are only 4d dependent. We use this to compute the contribution from the 10d dilaton kinetic term, minus the 6d square on 6d derivatives
\beq
4(\del \phi)_{10}^2 - 4(\del \phi)_{6}^2 = \tau^2\, e^{-2A}\, \Big(  \tau^{-4} (\del \tau^{2})^2 + \frac{1}{4} \del_{\mu} (g_{mn}) g^{mn} \del^{\mu} (g_{pq}) g^{pq} - \tau^{-2} \del_{\mu} \tau^{2}\, \del^{\mu} (g_{mn}) g^{mn}  \Big) \ .\nn
\eeq
This allows us to conclude
\begin{equation}
\begin{split}
\frac{1}{2 \kappa_{10}^2}&\int \d^{10} x \sqrt{|G|}\, e^{-2\phi}\ \left( {\cal R}_{10} + 4(\del \phi)_{10}^2 \right)  \\
= &\ \int \d^{4} x \sqrt{|g_{4}|} \left(\frac{M_p^2}{2} \bigg( {\cal R}_{4} - \frac{1}{2} {\cal V}_{6\phi}^{-2}\, (\del {\cal V}_{6\phi} )^2  + \frac{1}{4} \del_{\mu} g_{mn}\, \del^{\mu} g^{mn}  \bigg) - V_{{\cal R}\phi} \right) \ ,
\end{split}
\end{equation}
where we integrated to zero the 4d total derivative in \eqref{R10full}, and with
\bea
& V_{{\cal R}\phi} = - \frac{M_p^2}{2}  (2\pi \sqrt{\alpha'})^{-6}\, {\cal V}_{6\phi}^{-2} \int \d^6 y \sqrt{|g_6|}\, e^{4A-2 \phi} \, \bigg( {\cal R}_6 - 12 e^{-2A} (\del e^{A} )^2 - 8 e^{-A} \Delta_6 e^{A} + 4(\del \phi)^2 \bigg) \ ,\nn
\eea
in which we have 6d derivatives and metric.\\

We turn to the fluxes and their contribution to the scalar potential. For those we follow the procedure described in \cite{Andriot:2022bnb} which takes care of the bulk terms as well as the Chern-Simons ones, and the need to distinguish the background fluxes from the axion fluctuations. For the contribution of the 6d fluxes to the scalar potential, we first get the following expression, with a 6d integral as above already for ${\cal R}_6$
\bea
V_f= \frac{M_p^2}{2}  (2\pi \sqrt{\alpha'})^{-6}\, {\cal V}_{6\phi}^{-2} \int \d^6 y \sqrt{|g_6|}\, e^{4A-2 \phi} \, \bigg( & \frac{1}{2} |H|^2 +  \frac{e^{2\phi}}{2}  \bigg[ F_0^2 +\big|F_2 + F_0 B_2\big|^2\\
& +\left|F_4 +C_1 \w H + F_2 \w B_2 +\frac{1}{2}F_0\ B_2 \w B_2\right|^2  \bigg] \bigg) \ .\nn
\eea
In addition, one has to consider possible 4d fluxes: the 4d component of $F_4$ gives rise to the $F_6$ term. Following \cite[(2.13)]{Andriot:2022bnb}, we get for it the term
\bea
& {\cal S}_{F_6}  = -\frac{M_p^2}{4} \int \d^4 x \sqrt{|g_4|}\ \frac{(2\pi \sqrt{\alpha'})^{6}}{ {\cal V}_{6\phi}^{2} \int \d^6 y \sqrt{|g_6|} e^{-4A} } \left( (2\pi \sqrt{\alpha'})^{-6}\,  \int_6 (F_6 + \dots ) \right)^2 \ , \label{SF6}\\
& F_6 + \dots = F_6+C_3\w H +F_4 \w B_2 + C_1\w H \w B_2 + \frac{1}{2} F_2 \w B_2 \w B_2 + \frac{1}{6} F_0 B_2\w  B_2 \w B_2 \ .\nn
\eea
Trading the 6d integral squared for a flux square would require to know more on the $y$-dependence of the fluxes and the 6d metric, a simplification we do not consider for now.

From the 10d flux terms, we can also obtain the kinetic terms for the axions, that we take to be only dependent on 4d coordinates (and we neglect for simplicity here the possible $y$ dependence of the 6d metric entering the squares). Proceeding as before, we obtain
\beq
{\cal S}_{{\rm kin}_{axw}}  = \int \d^4 x \sqrt{|g_4|}\, \frac{M_p^2}{2}  \left(-\frac{1}{2} |\del B_2|^2 -\frac{1}{2} |\del C_{1,3}|^2 \ \frac{\int \d^6 y \sqrt{|g_6|} e^{2A}}{ (2\pi \sqrt{\alpha'})^{6}\, {\cal V}_{6\phi} }  \right) \ .
\eeq

Finally, we consider the sources contributions to the scalar potential. Starting from the DBI action for one $O_6$ per set $I$, we eventually obtain
\beq
V_{{\rm sources}} = - \frac{M_p^2}{2}  (2\pi \sqrt{\alpha'})^{-6}\, {\cal V}_{6\phi}^{-2} \int \d^6 y \sqrt{|g_6|}\, e^{4A-2 \phi} \, e^{\phi} \sum_I\frac{T_{10}^I}{7} \ ,
\eeq
where the definition of $\frac{T_{10}^I}{7}$ is given in \eqref{T10def}.

Putting all pieces together, we conclude with the full action for the warped 4d theory, obtained from type IIA supergravity with sources, with the 10d metric $\eqref{10dourmetricap}$. There, we took $\tau^2= {\cal V}_{6\phi}$, the latter being defined in \eqref{V6p}. The action is
\bea
\int \d^{4} x \sqrt{|g_{4}|} & \Bigg( \frac{M_p^2}{2} {\cal R}_{4} - V_{{\rm part}} \label{4dactionfullywarpedap}\\
& - \frac{M_p^2}{4}\bigg( {\cal V}_{6\phi}^{-2}\, (\del {\cal V}_{6\phi} )^2  - \frac{1}{2} \del_{\mu} g_{mn}\, \del^{\mu} g^{mn} + |\del B_2|^2 + |\del C_{1,3}|^2 \, \frac{\int \d^6 y \sqrt{|g_6|} e^{2A}}{ (2\pi \sqrt{\alpha'})^{6}\, {\cal V}_{6\phi} }
\bigg)  \Bigg) + {\cal S}_{F_6}  \ , \nn
\eea
with
\bea
V_{{\rm part}} = \frac{M_p^2}{2}  (2\pi \sqrt{\alpha'})^{-6}\, {\cal V}_{6\phi}^{-2}  \int \d^6 y & \sqrt{|g_6|} \, e^{4A-2 \phi} \, \bigg( - {\cal R}_6 + 12 e^{-2A} (\del e^{A} )^2 + 8 e^{-A} \Delta_6 e^{A} - 4(\del \phi)^2 \nn\\
& \  - e^{\phi} \sum_I\frac{T_{10}^I}{7} + \frac{1}{2} |H|^2 +  \frac{e^{2\phi}}{2}  \bigg[ F_0^2 +\big|F_2 + F_0 B_2\big|^2  \label{potwap} \\
& \ \ \   +\left|F_4 +C_1 \w H + F_2 \w B_2 +\frac{1}{2}F_0\ B_2 \w B_2\right|^2  \bigg] \bigg) \ ,\nn
\eea
and where the $F_6$ term \eqref{SF6} provides the rest of the potential.

Let us add that the source term in the potential can be rewritten using the Bianchi identity, here given by \eqref{BIs}, which should hold in any case. Combining relations around \eqref{T10def}, we get the rewriting
\bea
\int \d^6 y \sqrt{|g_6|} \, e^{4A- \phi} \sum_I\frac{T_{10}^I}{7} & = \int  e^{4A- \phi} \sum_{I} \frac{T_{10}^I}{7} {\rm vol}_{||_I} \w {\rm vol}_{\bot_I} \nn \\
& = \int  e^{4A- \phi} \sum_{I} {\rm vol}_{||_I} \w \left( \d F_{2} - H \w F_{0} \right) \ ,\label{potBI}
\eea
where one may project $\d F_{2} - H \w F_{0} $ on each transverse space $I$. For our applications, it won't be necessary since the ${\rm vol}_{||_I}$ will be enough for this projection, so that $\d F_{2} - H \w F_{0} $ can be factorized as above.

\end{appendix}

\newpage

\providecommand{\href}[2]{#2}\begingroup\raggedright\endgroup


\begin{thebibliography}{10}

\bibitem{DeWolfe:2005uu}
O.~DeWolfe, A.~Giryavets, S.~Kachru and W.~Taylor, \emph{{Type IIA moduli
  stabilization}},
  \href{https://doi.org/10.1088/1126-6708/2005/07/066}{\emph{JHEP} {\bfseries
  07} (2005) 066} [\href{https://arxiv.org/abs/hep-th/0505160}{{\ttfamily
  hep-th/0505160}}].

\bibitem{Lust:2004ig}
D.~Lust and D.~Tsimpis, \emph{{Supersymmetric AdS(4) compactifications of IIA
  supergravity}},
  \href{https://doi.org/10.1088/1126-6708/2005/02/027}{\emph{JHEP} {\bfseries
  02} (2005) 027} [\href{https://arxiv.org/abs/hep-th/0412250}{{\ttfamily
  hep-th/0412250}}].

\bibitem{Camara:2005dc}
P.G.~Camara, A.~Font and L.E.~Ibanez, \emph{{Fluxes, moduli fixing and
  MSSM-like vacua in a simple IIA orientifold}},
  \href{https://doi.org/10.1088/1126-6708/2005/09/013}{\emph{JHEP} {\bfseries
  09} (2005) 013} [\href{https://arxiv.org/abs/hep-th/0506066}{{\ttfamily
  hep-th/0506066}}].

\bibitem{Acharya:2006ne}
B.S.~Acharya, F.~Benini and R.~Valandro, \emph{{Fixing moduli in exact type IIA
  flux vacua}},
  \href{https://doi.org/10.1088/1126-6708/2007/02/018}{\emph{JHEP} {\bfseries
  02} (2007) 018} [\href{https://arxiv.org/abs/hep-th/0607223}{{\ttfamily
  hep-th/0607223}}].

\bibitem{Andriot:2022yyj}
D.~Andriot, L.~Horer and P.~Marconnet, \emph{{Exploring the landscape of
  (anti-) de Sitter and Minkowski solutions: group manifolds, stability and
  scale separation}},
  \href{https://doi.org/10.1007/JHEP08(2022)109}{\emph{JHEP} {\bfseries 08}
  (2022) 109} [\href{https://arxiv.org/abs/2204.05327}{{\ttfamily
  2204.05327}}].

\bibitem{Gautason:2018gln}
F.F.~Gautason, V.~Van~Hemelryck and T.~Van~Riet, \emph{{The Tension between 10D
  Supergravity and dS Uplifts}},
  \href{https://doi.org/10.1002/prop.201800091}{\emph{Fortsch. Phys.}
  {\bfseries 67} (2019) 1800091}
  [\href{https://arxiv.org/abs/1810.08518}{{\ttfamily 1810.08518}}].

\bibitem{Lust:2019zwm}
D.~L\"ust, E.~Palti and C.~Vafa, \emph{{AdS and the Swampland}},
  \href{https://doi.org/10.1016/j.physletb.2019.134867}{\emph{Phys. Lett. B}
  {\bfseries 797} (2019) 134867}
  [\href{https://arxiv.org/abs/1906.05225}{{\ttfamily 1906.05225}}].

\bibitem{Blumenhagen:2019vgj}
R.~Blumenhagen, M.~Brinkmann and A.~Makridou, \emph{{Quantum Log-Corrections to
  Swampland Conjectures}},
  \href{https://doi.org/10.1007/JHEP02(2020)064}{\emph{JHEP} {\bfseries 02}
  (2020) 064} [\href{https://arxiv.org/abs/1910.10185}{{\ttfamily
  1910.10185}}].

\bibitem{Apruzzi:2019ecr}
F.~Apruzzi, G.~Bruno De~Luca, A.~Gnecchi, G.~Lo~Monaco and A.~Tomasiello,
  \emph{{On AdS$_{7}$ stability}},
  \href{https://doi.org/10.1007/JHEP07(2020)033}{\emph{JHEP} {\bfseries 07}
  (2020) 033} [\href{https://arxiv.org/abs/1912.13491}{{\ttfamily
  1912.13491}}].

\bibitem{Font:2019uva}
A.~Font, A.~Herr\'aez and L.E.~Ib\'a\~nez, \emph{{On scale separation in type
  II AdS flux vacua}},
  \href{https://doi.org/10.1007/JHEP03(2020)013}{\emph{JHEP} {\bfseries 03}
  (2020) 013} [\href{https://arxiv.org/abs/1912.03317}{{\ttfamily
  1912.03317}}].

\bibitem{Emelin:2020buq}
M.~Emelin, \emph{{Effective Theories as Truncated Trans-Series and Scale
  Separated Compactifications}},
  \href{https://doi.org/10.1007/JHEP11(2020)144}{\emph{JHEP} {\bfseries 11}
  (2020) 144} [\href{https://arxiv.org/abs/2005.11421}{{\ttfamily
  2005.11421}}].

\bibitem{Buratti:2020kda}
G.~Buratti, J.~Calderon, A.~Mininno and A.M.~Uranga, \emph{{Discrete
  Symmetries, Weak Coupling Conjecture and Scale Separation in AdS Vacua}},
  \href{https://doi.org/10.1007/JHEP06(2020)083}{\emph{JHEP} {\bfseries 06}
  (2020) 083} [\href{https://arxiv.org/abs/2003.09740}{{\ttfamily
  2003.09740}}].

\bibitem{Emelin:2021gzx}
M.~Emelin, F.~Farakos and G.~Tringas, \emph{{Three-dimensional flux vacua from
  IIB on co-calibrated G2 orientifolds}},
  \href{https://doi.org/10.1140/epjc/s10052-021-09261-y}{\emph{Eur. Phys. J. C}
  {\bfseries 81} (2021) 456}
  [\href{https://arxiv.org/abs/2103.03282}{{\ttfamily 2103.03282}}].

\bibitem{DeLuca:2021mcj}
G.B.~De~Luca and A.~Tomasiello, \emph{{Leaps and bounds towards scale
  separation}}, \href{https://doi.org/10.1007/JHEP12(2021)086}{\emph{JHEP}
  {\bfseries 12} (2021) 086}
  [\href{https://arxiv.org/abs/2104.12773}{{\ttfamily 2104.12773}}].

\bibitem{DeLuca:2021ojx}
G.B.~De~Luca, N.~De~Ponti, A.~Mondino and A.~Tomasiello, \emph{{Cheeger bounds
  on spin-two fields}},
  \href{https://doi.org/10.1007/JHEP12(2021)217}{\emph{JHEP} {\bfseries 12}
  (2021) 217} [\href{https://arxiv.org/abs/2109.11560}{{\ttfamily
  2109.11560}}].

\bibitem{Cribiori:2022trc}
N.~Cribiori and G.~Dall'Agata, \emph{{Weak gravity versus scale separation}},
  \href{https://doi.org/10.1007/JHEP06(2022)006}{\emph{JHEP} {\bfseries 06}
  (2022) 006} [\href{https://arxiv.org/abs/2203.05559}{{\ttfamily
  2203.05559}}].

\bibitem{Lust:2022lfc}
S.~L\"ust, C.~Vafa, M.~Wiesner and K.~Xu, \emph{{Holography and the KKLT
  scenario}}, \href{https://doi.org/10.1007/JHEP10(2022)188}{\emph{JHEP}
  {\bfseries 10} (2022) 188}
  [\href{https://arxiv.org/abs/2204.07171}{{\ttfamily 2204.07171}}].

\bibitem{Andriot:2022brg}
D.~Andriot, L.~Horer and G.~Tringas, \emph{{Negative scalar potentials and the
  swampland: an Anti-Trans-Planckian Censorship Conjecture}},
  \href{https://doi.org/10.1007/JHEP04(2023)139}{\emph{JHEP} {\bfseries 04}
  (2023) 139} [\href{https://arxiv.org/abs/2212.04517}{{\ttfamily
  2212.04517}}].

\bibitem{DeLuca:2023kjj}
G.B.~De~Luca, N.~De~Ponti, A.~Mondino and A.~Tomasiello, \emph{{Harmonic
  functions and gravity localization}},
  \href{https://doi.org/10.1007/JHEP09(2023)127}{\emph{JHEP} {\bfseries 09}
  (2023) 127} [\href{https://arxiv.org/abs/2306.05456}{{\ttfamily
  2306.05456}}].

\bibitem{Caviezel:2008ik}
C.~Caviezel, P.~Koerber, S.~Kors, D.~Lust, D.~Tsimpis and M.~Zagermann,
  \emph{{The Effective theory of type IIA AdS(4) compactifications on
  nilmanifolds and cosets}},
  \href{https://doi.org/10.1088/0264-9381/26/2/025014}{\emph{Class. Quant.
  Grav.} {\bfseries 26} (2009) 025014}
  [\href{https://arxiv.org/abs/0806.3458}{{\ttfamily 0806.3458}}].

\bibitem{Tsimpis:2012tu}
D.~Tsimpis, \emph{{Supersymmetric AdS vacua and separation of scales}},
  \href{https://doi.org/10.1007/JHEP08(2012)142}{\emph{JHEP} {\bfseries 08}
  (2012) 142} [\href{https://arxiv.org/abs/1206.5900}{{\ttfamily 1206.5900}}].

\bibitem{McOrist:2012yc}
J.~McOrist and S.~Sethi, \emph{{M-theory and Type IIA Flux Compactifications}},
  \href{https://doi.org/10.1007/JHEP12(2012)122}{\emph{JHEP} {\bfseries 12}
  (2012) 122} [\href{https://arxiv.org/abs/1208.0261}{{\ttfamily 1208.0261}}].

\bibitem{Petrini:2013ika}
M.~Petrini, G.~Solard and T.~Van~Riet, \emph{{AdS vacua with scale separation
  from IIB supergravity}},
  \href{https://doi.org/10.1007/JHEP11(2013)010}{\emph{JHEP} {\bfseries 11}
  (2013) 010} [\href{https://arxiv.org/abs/1308.1265}{{\ttfamily 1308.1265}}].

\bibitem{Gautason:2015tig}
F.F.~Gautason, M.~Schillo, T.~Van~Riet and M.~Williams, \emph{{Remarks on scale
  separation in flux vacua}},
  \href{https://doi.org/10.1007/JHEP03(2016)061}{\emph{JHEP} {\bfseries 03}
  (2016) 061} [\href{https://arxiv.org/abs/1512.00457}{{\ttfamily
  1512.00457}}].

\bibitem{Marchesano:2019hfb}
F.~Marchesano and J.~Quirant, \emph{{A Landscape of AdS Flux Vacua}},
  \href{https://doi.org/10.1007/JHEP12(2019)110}{\emph{JHEP} {\bfseries 12}
  (2019) 110} [\href{https://arxiv.org/abs/1908.11386}{{\ttfamily
  1908.11386}}].

\bibitem{Ishiguro:2021csu}
K.~Ishiguro and H.~Otsuka, \emph{{Sharpening the boundaries between flux landscape and swampland by tadpole charge}}, \href{https://doi.org/10.1007/JHEP12(2021)017}{\emph{JHEP} {\bfseries 12}
  (2021) 017} [\href{https://arxiv.org/abs/2104.15030}{{\ttfamily
  2104.15030}}].

\bibitem{Cribiori:2021djm}
N.~Cribiori, D.~Junghans, V.~Van~Hemelryck, T.~Van~Riet and T.~Wrase,
  \emph{{Scale-separated AdS4 vacua of IIA orientifolds and M-theory}},
  \href{https://doi.org/10.1103/PhysRevD.104.126014}{\emph{Phys. Rev. D}
  {\bfseries 104} (2021) 126014}
  [\href{https://arxiv.org/abs/2107.00019}{{\ttfamily 2107.00019}}].

\bibitem{Bardzell:2022jfh}
J.~Bardzell, E.~Gonzalo, M.~Rajaguru, D.~Smith and T.~Wrase, \emph{{Type IIB
  flux compactifications with h$^{1,1}$ = 0}},
  \href{https://doi.org/10.1007/JHEP06(2022)166}{\emph{JHEP} {\bfseries 06}
  (2022) 166} [\href{https://arxiv.org/abs/2203.15818}{{\ttfamily
  2203.15818}}].

\bibitem{Becker:2022hse}
K.~Becker, E.~Gonzalo, J.~Walcher and T.~Wrase, \emph{{Fluxes, vacua, and
  tadpoles meet Landau-Ginzburg and Fermat}},
  \href{https://doi.org/10.1007/JHEP12(2022)083}{\emph{JHEP} {\bfseries 12}
  (2022) 083} [\href{https://arxiv.org/abs/2210.03706}{{\ttfamily
  2210.03706}}].

\bibitem{Carrasco:2023hta}
R.~Carrasco, T.~Coudarchet, F.~Marchesano and D.~Prieto, \emph{{New families of
  scale separated vacua}}, [\href{https://arxiv.org/abs/2309.00043}{{\ttfamily
  2309.00043}}].

\bibitem{Tringas:2023vzn}
G.~Tringas, \emph{{Anisotropic scale-separated AdS$_4$ flux vacua}},
 [\href{https://arxiv.org/abs/2309.16542}{{\ttfamily 2309.16542}}].

\bibitem{Farakos:2020phe}
F.~Farakos, G.~Tringas and T.~Van~Riet, \emph{{No-scale and scale-separated
  flux vacua from IIA on G2 orientifolds}},
  \href{https://doi.org/10.1140/epjc/s10052-020-8247-5}{\emph{Eur. Phys. J. C}
  {\bfseries 80} (2020) 659}
  [\href{https://arxiv.org/abs/2005.05246}{{\ttfamily 2005.05246}}].

\bibitem{VanHemelryck:2022ynr}
V.~Van~Hemelryck, \emph{{Scale-Separated AdS3 Vacua from G2-Orientifolds Using
  Bispinors}}, \href{https://doi.org/10.1002/prop.202200128}{\emph{Fortsch.
  Phys.} {\bfseries 70} (2022) 2200128}
  [\href{https://arxiv.org/abs/2207.14311}{{\ttfamily 2207.14311}}].

\bibitem{Farakos:2023nms}
F.~Farakos, M.~Morittu and G.~Tringas, \emph{{On/off scale separation}},
 [\href{https://arxiv.org/abs/2304.14372}{{\ttfamily 2304.14372}}].

\bibitem{Conlon:2020wmc}
J.P.~Conlon and F.~Revello, \emph{{Moduli Stabilisation and the Holographic
  Swampland}}, \href{https://doi.org/10.31526/lhep.2020.171}{\emph{LHEP}
  {\bfseries 2020} (2020) 171}
  [\href{https://arxiv.org/abs/2006.01021}{{\ttfamily 2006.01021}}].

\bibitem{Conlon:2021cjk}
J.P.~Conlon, S.~Ning and F.~Revello, \emph{{Exploring the holographic
  Swampland}}, \href{https://doi.org/10.1007/JHEP04(2022)117}{\emph{JHEP}
  {\bfseries 04} (2022) 117}
  [\href{https://arxiv.org/abs/2110.06245}{{\ttfamily 2110.06245}}].

\bibitem{Apers:2022zjx}
F.~Apers, M.~Montero, T.~Van~Riet and T.~Wrase, \emph{{Comments on classical
  AdS flux vacua with scale separation}},
  \href{https://doi.org/10.1007/JHEP05(2022)167}{\emph{JHEP} {\bfseries 05}
  (2022) 167} [\href{https://arxiv.org/abs/2202.00682}{{\ttfamily
  2202.00682}}].

\bibitem{Apers:2022tfm}
F.~Apers, J.P.~Conlon, S.~Ning and F.~Revello, \emph{{Integer conformal
  dimensions for type IIa flux vacua}},
  \href{https://doi.org/10.1103/PhysRevD.105.106029}{\emph{Phys. Rev. D}
  {\bfseries 105} (2022) 106029}
  [\href{https://arxiv.org/abs/2202.09330}{{\ttfamily 2202.09330}}].

\bibitem{Ning:2022zqx}
S.~Ning, \emph{{Holographic perspectives on models of moduli stabilization in
  M-theory}}, \href{https://doi.org/10.1007/JHEP09(2022)042}{\emph{JHEP}
  {\bfseries 09} (2022) 042}
  [\href{https://arxiv.org/abs/2206.13332}{{\ttfamily 2206.13332}}].

\bibitem{Apers:2022vfp}
F.~Apers, \emph{{Aspects of AdS flux vacua with integer conformal dimensions}},
  \href{https://doi.org/10.1007/JHEP05(2023)040}{\emph{JHEP} {\bfseries 05}
  (2023) 040} [\href{https://arxiv.org/abs/2211.04187}{{\ttfamily
  2211.04187}}].

\bibitem{Quirant:2022fpn}
J.~Quirant, \emph{{Noninteger conformal dimensions for type IIA flux vacua}},
  \href{https://doi.org/10.1103/PhysRevD.106.066017}{\emph{Phys. Rev. D}
  {\bfseries 106} (2022) 066017}
  [\href{https://arxiv.org/abs/2204.00014}{{\ttfamily 2204.00014}}].

\bibitem{Andriot:2019hay}
D.~Andriot and D.~Tsimpis, \emph{{Gravitational waves in warped
  compactifications}},
  \href{https://doi.org/10.1007/JHEP06(2020)100}{\emph{JHEP} {\bfseries 06}
  (2020) 100} [\href{https://arxiv.org/abs/1911.01444}{{\ttfamily
  1911.01444}}].

\bibitem{Junghans:2020acz}
D.~Junghans, \emph{{O-Plane Backreaction and Scale Separation in Type IIA Flux
  Vacua}}, \href{https://doi.org/10.1002/prop.202000040}{\emph{Fortsch. Phys.}
  {\bfseries 68} (2020) 2000040}
  [\href{https://arxiv.org/abs/2003.06274}{{\ttfamily 2003.06274}}].

\bibitem{Marchesano:2020qvg}
F.~Marchesano, E.~Palti, J.~Quirant and A.~Tomasiello, \emph{{On supersymmetric
  AdS$_{4}$ orientifold vacua}},
  \href{https://doi.org/10.1007/JHEP08(2020)087}{\emph{JHEP} {\bfseries 08}
  (2020) 087} [\href{https://arxiv.org/abs/2003.13578}{{\ttfamily
  2003.13578}}].

\bibitem{Plauschinn:2022ztd}
E.~Plauschinn, \emph{{Mass spectrum of type IIB flux compactifications
  \textemdash{} comments on AdS vacua and conformal dimensions}},
  \href{https://doi.org/10.1007/JHEP02(2023)257}{\emph{JHEP} {\bfseries 02}
  (2023) 257} [\href{https://arxiv.org/abs/2210.04528}{{\ttfamily
  2210.04528}}].

\bibitem{Andriot:2022bnb}
D.~Andriot, P.~Marconnet, M.~Rajaguru and T.~Wrase, \emph{{Automated consistent
  truncations and stability of flux compactifications}},
  \href{https://doi.org/10.1007/JHEP12(2022)026}{\emph{JHEP} {\bfseries 12}
  (2022) 026} [\href{https://arxiv.org/abs/2209.08015}{{\ttfamily
  2209.08015}}].

\bibitem{Giddings:2005ff}
S.B.~Giddings and A.~Maharana, \emph{{Dynamics of warped compactifications and
  the shape of the warped landscape}},
  \href{https://doi.org/10.1103/PhysRevD.73.126003}{\emph{Phys. Rev. D}
  {\bfseries 73} (2006) 126003}
  [\href{https://arxiv.org/abs/hep-th/0507158}{{\ttfamily hep-th/0507158}}].

\bibitem{Shiu:2008ry}
G.~Shiu, G.~Torroba, B.~Underwood and M.R.~Douglas, \emph{{Dynamics of Warped
  Flux Compactifications}},
  \href{https://doi.org/10.1088/1126-6708/2008/06/024}{\emph{JHEP} {\bfseries
  06} (2008) 024} [\href{https://arxiv.org/abs/0803.3068}{{\ttfamily
  0803.3068}}].

\bibitem{Douglas:2008jx}
M.R.~Douglas and G.~Torroba, \emph{{Kinetic terms in warped
  compactifications}},
  \href{https://doi.org/10.1088/1126-6708/2009/05/013}{\emph{JHEP} {\bfseries
  05} (2009) 013} [\href{https://arxiv.org/abs/0805.3700}{{\ttfamily
  0805.3700}}].

\bibitem{Frey:2008xw}
A.R.~Frey, G.~Torroba, B.~Underwood and M.R.~Douglas, \emph{{The Universal
  Kahler Modulus in Warped Compactifications}},
  \href{https://doi.org/10.1088/1126-6708/2009/01/036}{\emph{JHEP} {\bfseries
  01} (2009) 036} [\href{https://arxiv.org/abs/0810.5768}{{\ttfamily
  0810.5768}}].

\bibitem{Martucci:2009sf}
L.~Martucci, \emph{{On moduli and effective theory of N=1 warped flux
  compactifications}},
  \href{https://doi.org/10.1088/1126-6708/2009/05/027}{\emph{JHEP} {\bfseries
  05} (2009) 027} [\href{https://arxiv.org/abs/0902.4031}{{\ttfamily
  0902.4031}}].

\bibitem{Martucci:2014ska}
L.~Martucci, \emph{{Warping the K\"ahler potential of F-theory/IIB flux
  compactifications}},
  \href{https://doi.org/10.1007/JHEP03(2015)067}{\emph{JHEP} {\bfseries 03}
  (2015) 067} [\href{https://arxiv.org/abs/1411.2623}{{\ttfamily 1411.2623}}].

\bibitem{Grimm:2014efa}
T.W.~Grimm, T.G.~Pugh and M.~Weissenbacher, \emph{{The effective action of
  warped M-theory reductions with higher derivative terms \textemdash{} part
  I}}, \href{https://doi.org/10.1007/JHEP01(2016)142}{\emph{JHEP} {\bfseries
  01} (2016) 142} [\href{https://arxiv.org/abs/1412.5073}{{\ttfamily
  1412.5073}}].

\bibitem{Grimm:2015mua}
T.W.~Grimm, T.G.~Pugh and M.~Weissenbacher, \emph{{The effective action of
  warped M-theory reductions with higher-derivative terms - Part II}},
  \href{https://doi.org/10.1007/JHEP12(2015)117}{\emph{JHEP} {\bfseries 12}
  (2015) 117} [\href{https://arxiv.org/abs/1507.00343}{{\ttfamily
  1507.00343}}].

\bibitem{Andriot:2017jhf}
D.~Andriot, \emph{{On classical de Sitter and Minkowski solutions with
  intersecting branes}},
  \href{https://doi.org/10.1007/JHEP03(2018)054}{\emph{JHEP} {\bfseries 03}
  (2018) 054} [\href{https://arxiv.org/abs/1710.08886}{{\ttfamily
  1710.08886}}].

\bibitem{Andriot:2020wpp}
D.~Andriot, P.~Marconnet and T.~Wrase, \emph{{New de Sitter solutions of 10d
  type IIB supergravity}},
  \href{https://doi.org/10.1007/JHEP08(2020)076}{\emph{JHEP} {\bfseries 08}
  (2020) 076} [\href{https://arxiv.org/abs/2005.12930}{{\ttfamily
  2005.12930}}].

\bibitem{Andriot:2016xvq}
D.~Andriot and J.~Bl\r{a}b\"ack, \emph{{Refining the boundaries of the
  classical de Sitter landscape}},
  \href{https://doi.org/10.1007/JHEP03(2017)102}{\emph{JHEP} {\bfseries 03}
  (2017) 102} [\href{https://arxiv.org/abs/1609.00385}{{\ttfamily
  1609.00385}}].

\bibitem{Andriot:2017oaz}
D.~Andriot and G.~Lucena~G\'omez, \emph{{Signatures of extra dimensions in
  gravitational waves}},
  \href{https://doi.org/10.1088/1475-7516/2017/06/048}{\emph{JCAP} {\bfseries
  06} (2017) 048} [\href{https://arxiv.org/abs/1704.07392}{{\ttfamily
  1704.07392}}].

\bibitem{Andriot:2022xjh}
D.~Andriot and L.~Horer, \emph{{(Quasi-) de Sitter solutions across dimensions
  and the TCC bound}},
  \href{https://doi.org/10.1007/JHEP01(2023)020}{\emph{JHEP} {\bfseries 01}
  (2023) 020} [\href{https://arxiv.org/abs/2208.14462}{{\ttfamily
  2208.14462}}].

\end{thebibliography}
\end{document}